\documentclass{aa}

\usepackage{graphicx}
\usepackage[varg]{txfonts}
\usepackage{stfloats}
\usepackage{balance}
\begin{document}

   \title{\textit{RadioAstron} discovery of a mini-cocoon around the restarted parsec-scale jet in 3C\,84}

   \author{T.~Savolainen
          \inst{1,2,3}\thanks{\email{tuomas.k.savolainen@aalto.fi}}
          \and
          G.~Giovannini
          \inst{4,13}
          \and
          Y.~Y.~Kovalev
          \inst{3,5,6}
          \and
          M.~Perucho
          \inst{7,8}
          \and
          J.~M.~Anderson
          \inst{9}
          \and
          G.~Bruni
          \inst{10}
          \and
          P.~G.~Edwards
          \inst{11}
          \and
          A.~Fuentes
          \inst{12}
          \and
          M.~Giroletti
          \inst{13}
          \and
          J.~L.~G\'omez
          \inst{12}
          \and
          K.~Hada
          \inst{14}
          \and
          S.-S.~Lee
          \inst{15,16}
          \and
          M.~M.~Lisakov
          \inst{3,5}
          \and
          A.~P.~Lobanov
          \inst{3,6}
          \and
          J.~L\'opez-Miralles
          \inst{7}
          \and
          M.~Orienti
          \inst{13}
          \and
          L.~Petrov
          \inst{17}
          \and
          A.~V.~Plavin
          \inst{5,6}
          \and
          B.~W.~Sohn
          \inst{15,16}
          \and
          K.~V.~Sokolovsky
          \inst{18,19}
          \and
          P.~A.~Voitsik
          \inst{5}
          \and
          J.~A.~Zensus
          \inst{3}
          }

        \institute{Aalto University Department of Electronics and Nanoengineering, 
             PL 15500, FI-00076 Aalto, Finland
         \and
             Aalto University Mets\"ahovi Radio Observatory,
             Mets\"ahovintie 114, FI-02540 Kylm\"al\"a, Finland
         \and
             Max-Planck-Institut f\"ur Radioastronomie,
             Auf dem H\"ugel 69, D-53121 Bonn, Germany
         \and
             Dipartimento di Fisica e Astronomia, Universit\'a di Bologna,
             via Gobetti 93/2, I-40129 Bologna, Italy 
         \and
             Lebedev Physical Institute of the Russian Academy of Sciences,
             Leninsky prospekt 53, 119991 Moscow, Russia
         \and
             Moscow Institute of Physics and Technology, 
             Institutsky per.~9, Dolgoprudny, Moscow region, 141700, Russia
         \and
             Departament d'Astronomia i Astrof\'{\i}sica, Universitat de Val\`encia,
             C/ Dr. Moliner, 50, E-46100, Burjassot, Val\`encia, Spain
         \and
             Observatori Astron\`omic, Universitat de Val\`encia,
             C/ Catedr\`atic Jos\'e Beltr\'an 2, E-46980, Paterna, Val\`encia, Spain
         \and
             Helmholtz Centre Potsdam, GFZ German Research Centre for Geosciences,
             Telegrafenberg, D-14473 Potsdam, Germany
         \and
             INAF -- Istituto di Astrofisica e Planetologia Spaziali,
             via del Fosso del Cavaliere 100, I-00133 Roma, Italy
         \and 
             CSIRO Space and Astronomy,
             Epping, NSW, 1710, Australia
         \and
             Instituto de Astrof\'isica de Andaluc\'ia-CSIC,
             Glorieta de la Astronom\'ia s/n, E-18008 Granada, Spain
         \and 
             INAF -- Istituto di Radio Astronomia, 
             via P. Gobetti 101, I-40129 Bologna, Italy
         \and
             Mizusawa VLBI Observatory, National Astronomical Observatory of Japan,
             2-12 Hoshigaoka, Mizusawa, Oshu, Iwate 023-0861, Japan
         \and    
             Korea Astronomy and Space Science Institute,
             Yuseong-gu, Daejeon, Korea
         \and
             Korea University of Science and Technology,
             Yuseong-gu, Daejeon, Korea
         \and
             NASA Goddard Space Flight Center,
             Greenbelt, ND 20771, USA
         \and
             Center for Data Intensive and Time Domain Astronomy, Department of Physics and Astronomy,
             Michigan State University, 567 Wilson Rd, East Lansing, MI 48824, USA
         \and
             Sternberg Astronomical Institute, Moscow State University,
             Universitetskii~pr.~13, 119992~Moscow, Russia
}

   \date{Received 8.11.2021; accepted 11.4.2023}

 

\abstract{
  We present \emph{RadioAstron} space-based very long baseline interferometry (VLBI) observations of the nearby radio galaxy 3C\,84 (NGC\,1275) at the centre of the Perseus cluster. The observations were carried out during a perigee passage of the \emph{Spektr-R} spacecraft on September 21$-$22, 2013 and involved a global array of 24 ground radio telescopes observing at 5\,GHz and 22\,GHz, together with the Space Radio Telescope (SRT). Furthermore, the Very Long Baseline Array (VLBA) and the phased Very Large Array (VLA) observed the source quasi-simultaneously at 15\,GHz and 43\,GHz. Fringes between the ground array and the SRT were detected on baseline lengths up to $8.1$ times the Earth's diameter, providing unprecedented resolution for 3C\,84 at these wavelengths. We note that the corresponding fringe spacing is 125\,$\mu$as at 5\,GHz and 27\,$\mu$as at 22\,GHz. Our space-VLBI images reveal a previously unseen sub-structure inside the compact $\sim 1$\,pc long jet that was ejected about ten\,years earlier. In the 5\,GHz image, we detected, for the first time, low-intensity emission from a cocoon-like structure around the restarted jet. Our results suggest that the increased power of the young jet is inflating a bubble of hot plasma as it carves its way through the ambient medium of the central region of the galaxy. Here, we estimate the minimum energy stored in the mini-cocoon, along with its pressure, volume, expansion speed, and the ratio of heavy particles to relativistic electrons, as well as the density of the ambient medium. About half of the energy delivered by the jet is dumped into the mini-cocoon and the quasi-spherical shape of the bubble suggests that this energy may be transferred to a significantly larger volume of the interstellar medium than what would be accomplished by the well-collimated jet on its own. The pressure of the hot mini-cocoon also provides a natural explanation for the almost cylindrical jet profile seen in the 22\,GHz \emph{RadioAstron} image.
}

   \keywords{Galaxies: jets -- 
             Galaxies: active -- 
             Galaxies: individual: 3C\,84 -- 
             Techniques: interferometric -- 
             Techniques: high angular resolution
             }

   \maketitle
%

\section{Introduction}

Understanding how accreting supermassive black holes (SMBH) in active galactic nuclei (AGN) launch relativistic jets of magnetised plasma and how these jets affect their host galaxy environment continue to be focal questions for studies in both relativistic astrophysics and galaxy evolution \citep[e.g.][]{Blandford2019}. Rapid progress in the numerical simulations of astrophysical jets over the past couple of decades has shed light on these complex, dynamic systems, whose study is complicated by their non-linearity and the need to cover a range of scales greater than a factor of $10^{10}$\,\citep{Komissarov2021}. The three-dimensional (3D) general relativistic magnetohydrodynamic (GRMHD) simulations have, for example, now demonstrated how jets can be powered through the extraction of the rotational energy of the black hole by electromagnetic torques \citep{Tchekhovskoy2011}, a mechanism originally proposed by \citet{Blandford1977}. An alternative model posits that the jets are formed when accreting plasma is centrifugally flung out by poloidal magnetic fields co-rotating with the accretion disk \citep{Blandford1982}. There is observational evidence both for black hole spin-powered \citep[e.g.][]{Zamaninasab2014,EHT2019e}, and accretion disk-powered jets \citep[e.g.][]{Boccardi2016}. 

Comparing the simulation results to observations is challenging mainly due to two reasons. First, the magnetohydrodynamic (MHD) approach does not include the microphysics of the electron heating and cannot self-consistently describe the radiating particle population. This means that generating emission maps from the GRMHD output requires an ad hoc prescription for the electron energy distribution. Second, the extent of the 3D GRMHD jet formation simulations is typically $\lesssim 10^4$ gravitational radii ($r_\mathrm{g}$), which has made it difficult to resolve the relevant spatial scales in cm-wavelength ground-based very long baseline interferometry (VLBI) observations, with the notable exception of \object{M\,87} \citep[e.g.][]{Asada2012, Hada2016, Kim2018}. The development of VLBI at short millimetre wavelengths, especially with the advent of the Event Horizon Telescope (EHT) observing at 1.3\,mm, has recently made it possible to image horizon-scale structures in M\,87 \citep{EHT2019a,EHT2019d} and to probe the jet formation site down to $\sim 200$\,$r_\mathrm{g}$ in \object{Centaurus~A} \citep{Janssen2021}. Besides observing at shorter wavelengths, it is also possible to increase the angular resolution of the VLBI observations by increasing the baseline lengths beyond the limits set by the Earth's diameter. This provides resolutions similar to or even higher than the EHT, but at centimetre wavelengths, allowing us to probe a lower energy electron population compared to mm-VLBI.

The \emph{RadioAstron} mission carried out a ground-to-space VLBI array with baseline lengths up to 360\,000\,km \citep{Kardashev2013}. 
The Russian \emph{Spektr-R} satellite was launched in 2011 and placed in a highly elliptical, lunar-perturbed orbit around the Earth. The satellite deployed a 10-m Space Radio Telescope (SRT) after launch and operated successfully until 2019.
\emph{RadioAstron} had four receivers: 92\,cm, 18\,cm, 6\,cm, and a wide-band K band receiver covering 1.19$-$1.63\,cm. Thanks to concentric feed horns, the SRT could observe in two bands simultaneously. This combination of cm-wavelength receivers and unprecedented long ground-to-space baselines has made \emph{RadioAstron} the highest angular resolution instrument in the history of astronomy, with a 7\,$\mu$as nominal resolution in the K band \citep[e.g.][]{Kardashev2013,2020AdSpR..65..705K}. 

The \emph{RadioAstron} open science program included three key science programmes (KSP) concentrating on ultra-high-resolution imaging and polarimetry of the AGN jets, with experiments typically carried out at the times of the perigee passage of \emph{Spektr-R} and involving large networks of ground radio telescopes \citep{Bruni2020}. While observations at the longest baselines were made in visibility tracking mode only, observations carried out during perigee passes were suitable for imaging; and the ground-to-space baselines were changing quickly avoiding large gaps in the $(u,v)$ coverage. Nearby AGN KSP has been aimed at translating the exquisite angular resolution of \emph{RadioAstron} observations into ultra-high spatial resolution for the AGN jet studies by observing the nearest radio galaxies that exhibit bright and compact pc-scale emission structures. The main goal of the KSP is to resolve the target jets in or close to their formation site and to provide data for comparisons with the GRMHD simulations. The targets selected for the program were Cen~A at a distance of 3.8\,Mpc, M\,87 at a distance of 16.8\,Mpc, and 3C\,84 at a distance of 76\,Mpc.  

The radio source \object{3C\,84} (\object{Perseus~A}, \object{B0316+413}) is hosted by the peculiar elliptical galaxy \object{NGC\,1275} in the centre of the Perseus cluster. 3C\,84 has been noted as a source with an inverted spectrum (i.e. flux density increasing with frequency) by \citet{Dent1965}, which prompted \citet{Shklovsky1966} to predict both variable cm-wavelength radiation and strong X-ray emission produced by inverse Compton radiation from the source. Shortly afterwards, 3C\,84 was confirmed as one of the first variable radio sources, with an increasing flux density observed in the early 1960s \citep{Dent1966,PaulinyToth1966}. This outburst continued with peaks at cm wavelengths above 50\,Jy in the early 1970s and early 1980s. Moreover, following the launch of the {\it Uhuru} X-ray satellite in 1970, 3C\,84 was found to be a bright X-ray source \citep{Forman1972}.

As a bright radio source, 3C\,84 was one of the early targets studied on the parsec scale with VLBI at centimetre \citep{Clark1968} and millimetre \citep{Readhead1983} wavelengths. On the kiloparsec scale, \citet{Pedlar1990} imaged 3C\,84 with the Very Large Array (VLA), revealing a strongly core-dominated source with a bright component $\sim 4$\,arcseconds to the south-east, a second component $\sim 16$\,arcsec to the south-south-east of the core, and a faint component $\sim 12$\,arcsec to the north-west of the core. 

The outburst in the 1970s and 1980s was followed by almost two decades of decline in the radio flux density. Centimetre-wavelength VLBI observations made in the 1990s and the early 2000s showed slowly expanding (0.5\,$c$) small-scale radio lobes $\sim$10$-$15\,mas north and south of the core, with the northern lobe being strongly free-free absorbed at frequencies below 10\,GHz \citep{Walker2000, Asada2006}. By back-extrapolating, using the advance speed of 0.5\,$c$, \citet{Asada2006} estimated the ejection epoch of the lobes to be 1956$\pm$9, which coincides with the increasing radio flux density in the late 1950s and early 1960s \citep{Nesterov1995}. Lower frequency VLBI observations made at 2.3\,GHz showed fainter lobe-like structures also at about 30\,mas and 70\,mas south of the core \citep{Walker2000}. Combined with the VLA results of \citet{Pedlar1990}, this means that there are several lobe-like radio features in 3C\,84, from pc to kpc scales, which may be a sign of a recurrent jet activity.  

The source was also successfully imaged with mm-VLBI during the 1990s and these observations resolved the core region\footnote{We note that since the definition of the 'core' depends on the resolution of the observations and the observing frequency, we use the term 'core region' to refer to the central part of the source which is located between the northern mini-lobe and the first southern mini-lobe in cm-VLBI images, and has a size of $\lesssim 4$\,mas. The term "core" is from now on reserved for the bright, compact feature on the northern side of the 'core region' that is visible in the high-resolution VLBI images at frequencies $\gtrsim 15$\,GHz.}, revealing a complex structure within the innermost 3\,mas. This structure consists of the presumably stationary core at the northern end of the structure and a bending jet pointing towards the southwest \citep{Dhawan1998, Lister2001}. 

3C\,84 became active again in the 2000s with the emergence of a new component from the core \citep{Nagai2010}, strong increase in the radio flux density \citep[e.g.][]{Abdo2009, Nagai2010, Paraschos2022}, the discovery of GeV gamma-ray emission following the launch of the {\it Fermi} satellite in 2008 \citep{Abdo2009}, as well as detections of TeV emission in multiple instances with ground-based Cherenkov telescopes \citep{Aleksic2012,Aleksic2014,Benbow2015,MAGIC2018}. 

The VLBI monitoring carried out by \citet{Nagai2010} and \citet{Suzuki2012} revealed that a bright new emission feature was ejected from the core before November 2003 and moved towards the south on a curved trajectory at a sub-luminal speed \citep{Suzuki2012}. \citet{Nagai2010} called this new feature 'C3', whereas the other features visible in their 22\,GHz images taken with the VLBI Exploration for Radio Astrometry (VERA) array in 2006$-$2009 are at the core (which they call 'C1') located 0.6$-$1.2\,mas north of C3 and 'C2', which appears to be a remnant of the 1990s jet emission $\sim 1$\,mas southwest of the core. We use this nomenclature (C1, C2, C3) throughout the paper when discussing features in the core region. The single-dish radio flux density started to rise around 2005, first peaked in 2014 and a second time in 2016 \citep{Nagai2010,Paraschos2022}.  

A high-resolution 43\,GHz Very Long Baseline Array (VLBA) image by \citet{Nagai2014} revealed that C3 was connected to the core (C1) by a strongly limb-brightened jet. The limb-brightening can be explained either by velocity structure in a stratified jet or by a change of emissivity across the jet due to, for instance, shear acceleration of the particles in the jet boundary layer. As the direction of this jet between C1 and C3 differs significantly from the jet direction in the 1990s, it seems obvious that C3 marks a point where the new jet interacts with the ambient medium (i.e. the jet head). Furthermore, the significant increase in radio flux density, slow apparent speed of C3, and the detection of gamma-ray emission after 2008 -- while the Energetic Gamma-ray Experiment Telescope (EGRET) on board the Compton Gamma-ray Observatory did not detect gamma-rays in the 1990s \citep{Abdo2009} -- strongly suggest that we are seeing restarted jet activity. That is to say that the jet power has increased and also its direction has changed in 2003. In this paper, we refer to the feature of C3 and the jet connecting it to C1 as a 'restarted jet'. 

The complex interaction between the restarted jet and the ambient medium in 3C\,84 has been studied intensively over the last few years. There is a bright hot spot inside the jet head C3 and the movement of this hot spot is complex; it moves non-linearly, flips from one side of C3 to the other \citep{Kino2018}, and gets frustrated in 2016.7$-$2018.0 before finally breaking out after 2018.0 \citep{Kino2021}. After the breakout, the morphology of C3 appears to change from 'Fanaroff-Riley class II-like' to 'Fanaroff-Riley class I-like'. \citet{Kino2021} explained this behaviour as interaction between the jet and a dense cloud with an average density of $(4-6) \times 10^5$\,cm$^{-3}$. The high Faraday rotation measure observed in C3 supports the scenario of jet interaction with dense interstellar medium (ISM) clouds \citep{Nagai2017}.

Further information on the structure of this restarted jet in 3C\,84 has come from our \emph{RadioAstron} Nearby AGN KSP observations at 22\,GHz, which trace the nearly cylindrical jet profile previously seen on scales of thousands of gravitational radii down to a few hundred gravitational radii, implying that the bright outer jet layer has a rapid lateral expansion closer to the core or that the jet is launched from the accretion disk \citep{Giovannini2018}. Recently, based on measuring a core-shift between 15\,GHz, 43\,GHz, and 86\,GHz, \citet{Paraschos2021} suggested that the jet apex is 80\,$\mu$as north of the 86\,GHz core, which would alleviate the problem of a very wide jet base. However, this is difficult to reconcile with the limb-brightened counter-jet side structure visible in the \emph{RadioAstron} image, which (if it indeed corresponds to a counter-jet) limits the amount of core-shift at 22\,GHz to less than 30\,$\mu$as \citep{Giovannini2018}.

The quasi-cylindrical jet profile in the 22\,GHz \emph{RadioAstron} image requires a nearly flat density profile of the external medium, suggesting that the restarted jet may be surrounded by a cocoon of hot plasma. Until now, there have been no direct observations of such a structure around parsec-scale jets, although the existence of parsec-scale cocoons is hinted at by the mini-radio-lobes in, for example, compact symmetric objects \citep[e.g.][]{Lister2020} and 3C\,84 itself on the $\sim 10$\,pc scale \citep[e.g.][]{Asada2006}.

Here, we present a new, high-resolution 5\,GHz \emph{RadioAstron} image of 3C\,84 that reveals a cocoon-like emission around the restarted pc-scale jet. Furthermore, we describe in detail the data processing of \emph{RadioAstron} near-perigee imaging experiments of resolved sources that have low correlated flux density on space baselines. The CLEAN images made from the 22\,GHz data discussed in this paper have already been presented in \citet{Giovannini2018}; however, here we extend the scope to include also regularised maximum likelihood (RML) imaging of \textit{RadioAstron} data. In the two companion papers we plan to: a) discuss the unexpectedly high brightness temperature of the hot spot in the restarted pc-scale jet of 3C\,84 (T.~Savolainen et al., in prep.) and b) the implications of the small-scale sub-structure inside the ground-based VLBI beam, revealed by \emph{RadioAstron}, on the effective source position (L.~Petrov et al., in prep.). 

3C\,84 is at redshift of 0.0176 \citep{Strauss1992}, which means that one milliarcsecond corresponds to 0.354\,pc, assuming $\mathrm{H}_0 = 70.7$\,km\,s$^{-1}$\,Mpc$^{-1}$, $\Omega_\mathrm{M} = 0.27$ and $\Omega_\Lambda = 0.73$. We note here that the jet inclination angle, namely, the angle between the jet and line-of-sight directions, of 3C\,84 is rather uncertain, with reported values in the literature ranging from $11^\circ$ \citep{Lister2009b} to $65^\circ$ \citep{Fujita2017}. For the analyses in this paper, we adopted two different jet inclination values: $18^\circ$ and $45^\circ$. The reasoning behind the choice of these two values is discussed in \citet{Giovannini2018}. Here, $45^\circ$ represents the typical values derived from the jet-to-counter-jet length and brightness ratios \citep[e.g.][]{Walker1994,Asada2006}, while $18^\circ$ represents values derived from modelling the broadband spectral energy distribution \citep[e.g.][]{Tavecchio2014}.


\section{Observations and data reduction}

\subsection{\textit{RadioAstron} Space-VLBI observations at 5 and 22\,GHz}

The \textit{RadioAstron} space-VLBI imaging observations of 3C\,84 at 5 and 22\,GHz were carried out during a perigee passage of \textit{Spektr-R} on September 21-22, 2013. The whole observing session, including both the SRT and the ground array, lasted from 2013-09-21 15:00\,UT to 2013-09-22 13:00\,UT. During this time, the SRT sampled projected baselines from 0.2 to 10.4 Earth diameters ($D_\oplus$) in length. Since the drive motors that point the spacecraft's high-gain communications antenna to the ground station overheat in long operations, the SRT could not observe continuously through the whole imaging run. Instead, it was scheduled for 12 segments of 30$-$40\,min each with 70$-$90\,min cooling gaps in between. These gaps were used to make additional 15.4\,GHz and 43.2\,GHz observations with the frequency-agile ground array antennas, as we further discuss in Sect.~\ref{GRobs}. In addition to the target source, we observed two calibrators with the ground array only. The nearby compact source \object{B3\,0307+380} was observed frequently to track telescope gain changes and the bright blazar \object{TXS\,0355+508} (\object{4C\,+50.11}) was observed for a few times to allow an independent check on the amplitude calibration.

The SRT was operated in a mixed-band mode, whereby left circularly polarised (LCP) signals from both C band (4.836\,GHz) and K band (22.236\,GHz) receivers were observed simultaneously \citep{Kardashev2013}. Two sub-bands (IFs) of 16\,MHz apiece were recorded per polarisation, giving a total bandwidth of 32\,MHz for each observing band. One-bit sampling was used at the SRT, giving a recording bit rate of 128\,Mbps, while the ground telescopes employed two-bit sampling and, thus, a 256\,Mbps recording rate. The data from the SRT were downlinked in real time to two \textit{RadioAstron} science data acquisition stations, one in Puschino, Russia \citep{Kardashev2013} and one in Green Bank, USA \citep{Ford2014}. The Puschino station operated from 2013-09-21 15:00\,UT to 08:50\,UT on the following day, while the Green Bank station operated from 10:00\,UT to 12:40\,UT on 2013-09-22.

\begin{figure*}
\centering
\includegraphics[width=0.49\textwidth]{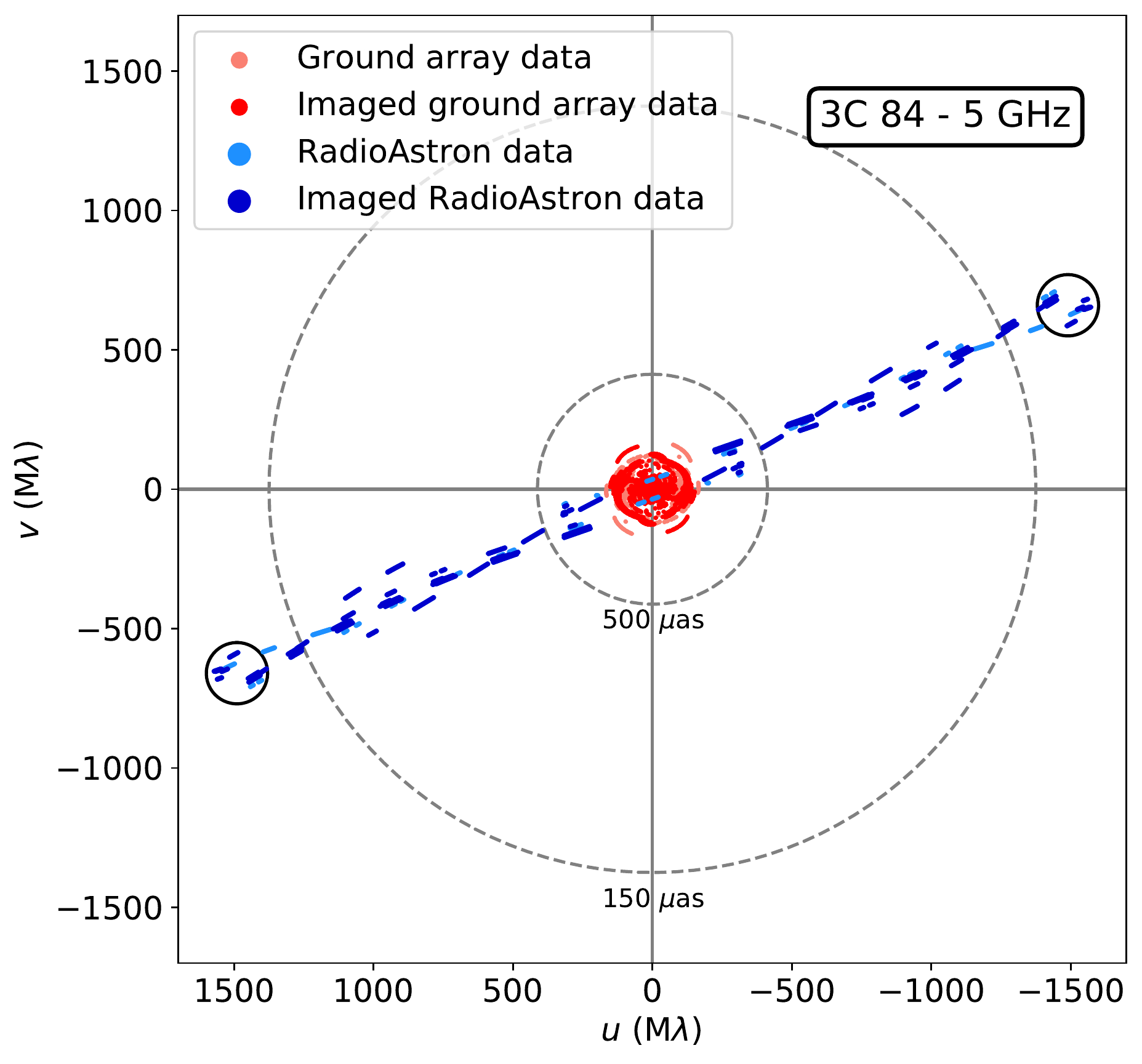}
\includegraphics[width=0.48\textwidth]{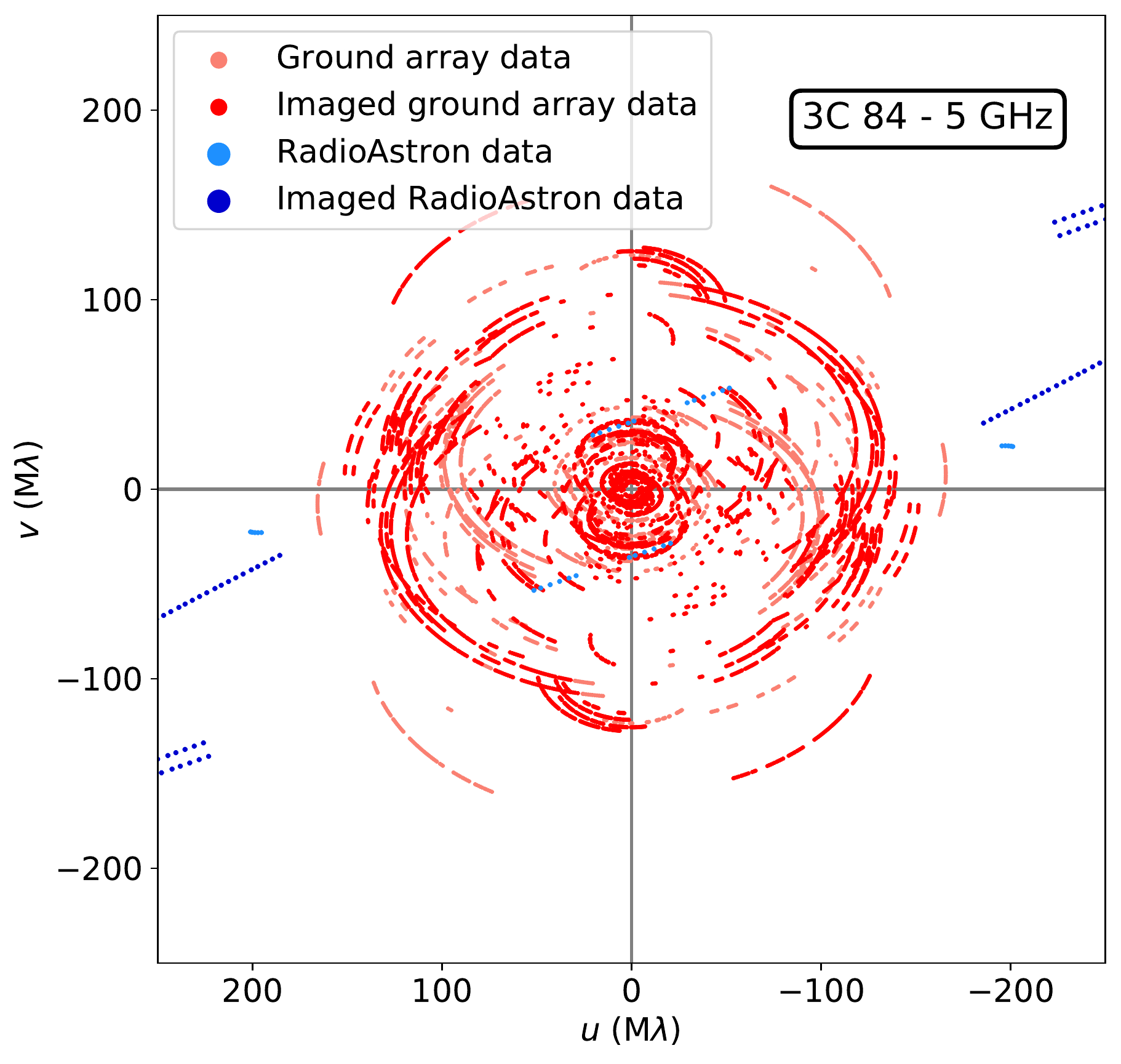} \\
\includegraphics[width=0.50\textwidth]{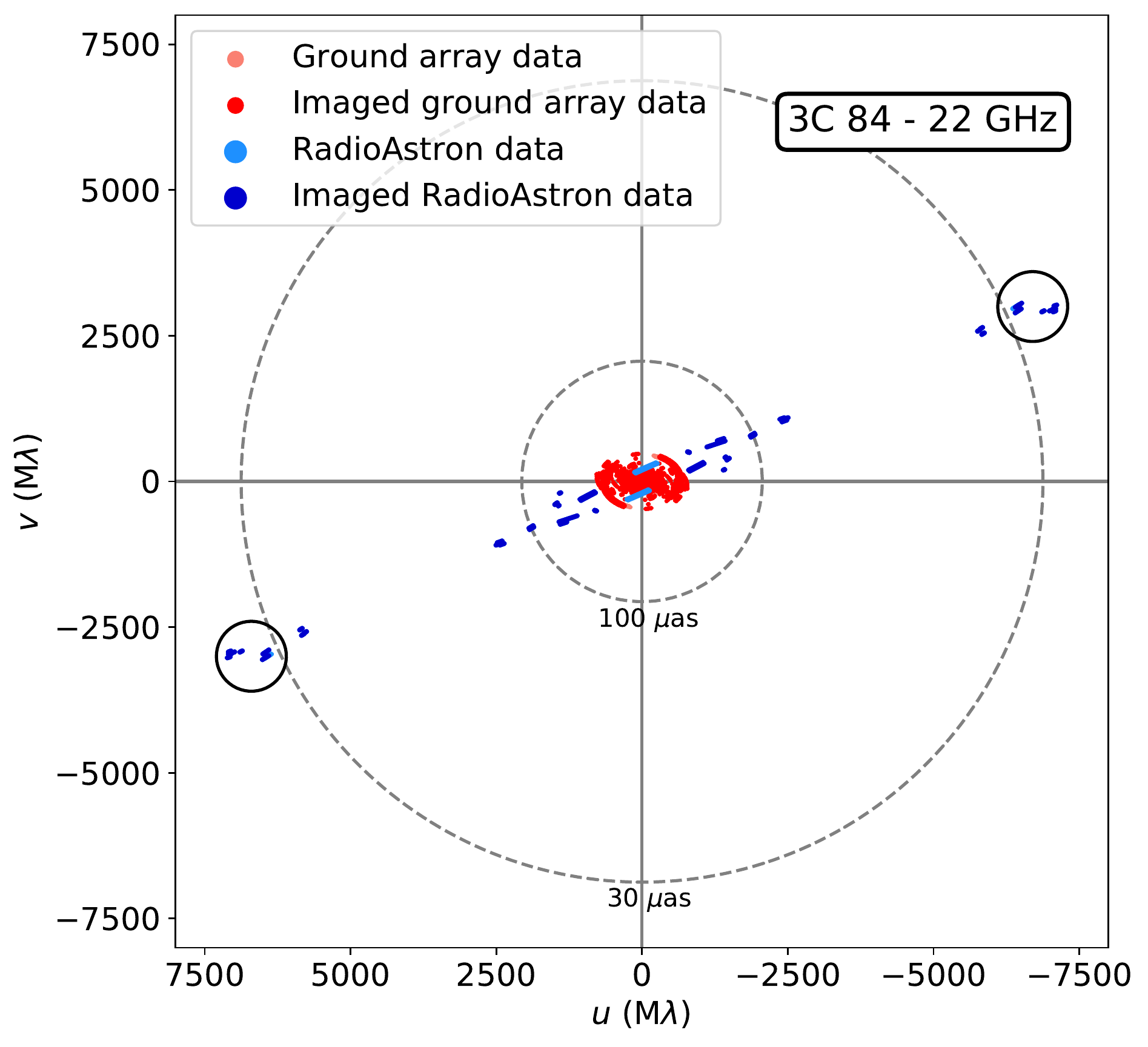}
\includegraphics[width=0.48\textwidth]{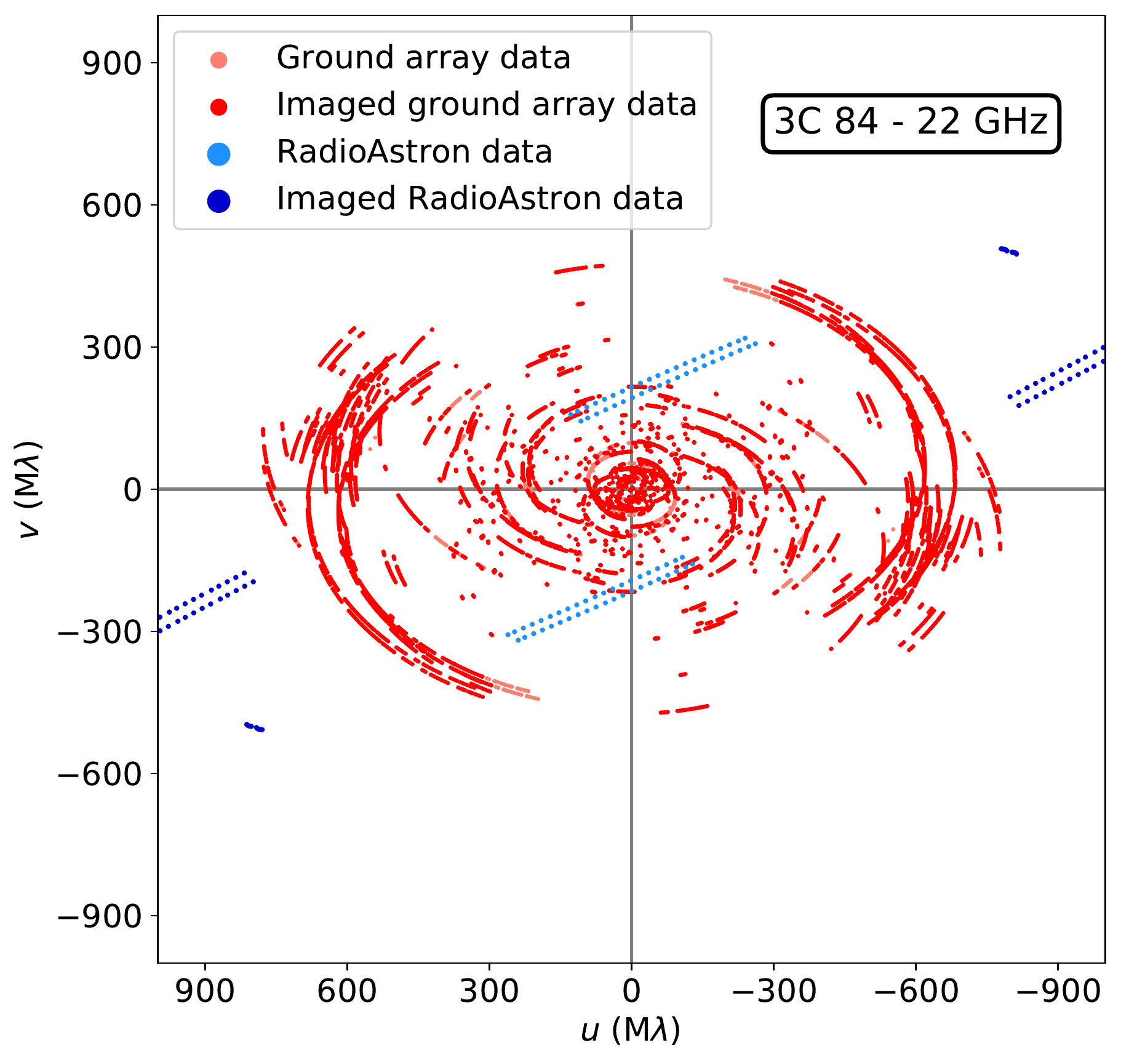}
\caption{$(u,v)$ coverages of the \textit{RadioAstron} imaging observations of 3C\,84. Red and blue points show the data used in the final images from the ground and space baselines, respectively. The visibility data that were detected in global fringe fitting with \textsc{AIPS} (see Fig.~\ref{aipsfringe}) but later flagged in the imaging stage are shown by light red for the ground baselines and by cyan for the space baselines. \textit{Top-left panel:} Full $(u,v)$ coverage in the C band. The dashed circles indicate the fringe spacings of 150\,$\mu$as and 500\,$\mu$as. \textit{Top-right panel:}  $(u,v)$ coverage for the baseline lengths $\lesssim 250$\,M$\lambda$ in the C band. \textit{Bottom left panel:} The full $(u,v)$ coverage in the K band. The dashed circles indicate the fringe spacings of 100\,$\mu$as and 30\,$\mu$as. \textit{Bottom-right panel:}  $(u,v)$ coverage for the baseline lengths $\lesssim 1000$\,M$\lambda$ in the K band. The black circles in the left panels highlight the scans between 06:10:00\,UT and 06:50:00\,UT on the second day of the observation. \emph{RadioAstron} baselines make almost straight lines in the $(u,v)$ plane while ground-only baselines are elliptical arcs.}
\label{uv}
\end{figure*}

Altogether 29 ground-based radio telescopes were originally scheduled for the observing run and 24 of them produced data that were successfully correlated. These are listed in Table~\ref{antennas} together with their sizes and sensitivities at the frequency bands they observed. Since the SRT recorded simultaneously in the C and K bands, the ground array was also split in two parts during the space-VLBI scans (see Table~\ref{antennas}). Effelsberg switched between 5\,GHz and 22\,GHz bands, spending half the time at each. All ground array telescopes observed in a dual-polarisation mode. The resulting $(u,v)$ coverages at 5\,GHz and 22\,GHz are shown in Fig.~\ref{uv}. As will be discussed in Sect.~\ref{stack} the longest (shortest) baselines with fringe detections are 1.7\,G$\lambda$ (3.6\,M$\lambda$) at 5\,GHz and 7.7\,G$\lambda$ (3.7\,M$\lambda$) at 22\,GHz.

\begin{table*}
\caption{Radio telescopes that successfully participated in the
  \textit{RadioAstron} imaging observations of 3C\,84}
\label{antennas}
\centering
\begin{tabular}{llccc}
\hline\hline
Telescope                  & Code &  Diameter  & \multicolumn{2}{c}{SEFD\tablefootmark{a}}  \\
                           &      &     &  C band & K band \\
                           &      & (m) &  (Jy)   &  (Jy) \\
\hline
\textit{Spektr-R} SRT (RU) & RA   & 10  &   11600\tablefootmark{b} & 46700\tablefootmark{b} \\
KVN-Tamna (KR)            & KT   & 21  &    --   & 1288 \\
KVN-Ulsan (KR)             & KU   & 21  &    --   & 1288 \\
VLBA-Brewster (USA)        & BR   & 25  &   210   & --   \\
VLBA-Fort~Davis (USA)      & FD   & 25  &   --    & 640  \\
VLBA-Hancock (USA)         & HN   & 25  &   --    & 640  \\
VLBA-Kitt~Peak (USA)       & KP   & 25  &   210   & --   \\
VLBA-Los~Alamos (USA)      & LA   & 25  &   --    & 640  \\
VLBA-Mauna~Kea (USA)       & MK   & 25  &   --    & 640  \\
VLBA-North~Liberty (USA)   & NL   & 25  &   210   & --   \\
VLBA-Owens~Valley (USA)    & OV   & 25  &   --    & 640  \\
VLBA-Pie~Town (USA)        & PT   & 25  &   210   & --   \\
VLBA-St.~Croix (USA)       & SC   & 25  &   210   & --   \\
Green~Bank~Telescope (USA) & GB   & 100 &   10    & --   \\
Very~Large~Array (USA)     & Y27  & 115\tablefootmark{c} & -- &  40 \\    
Shanghai (CN)              & SH   & 25  &   720   &  --  \\  
Kalyazin (RU)              & KL   & 46  &   150   &  --  \\
Onsala (SE)                & ON   & 25  &   600   &  --  \\
Medicina (IT)              & MC   & 32  &    --   &  700 \\
Noto (IT)                  & NT   & 32  &   260   &  --  \\
Jodrell~Bank\tablefootmark{d} (UK) & JB  & 25 & 320 & -- \\
Effelsberg (DE)            & EF   & 100 &   20    &  90  \\ 
Westerbork (NL)            & WB   & 93\tablefootmark{e} &  120 & -- \\
Yebes (ES)                 & YS   & 40  &   --    &  200 \\
Hartebeesthoek (ZA)        & HH   & 26  &   650   &  --  \\
\hline
\end{tabular}
\tablefoot{ 
\tablefoottext{a}{Typical system equivalent flux density.}
\tablefoottext{b}{Values for the left circular polarisation channel are shown \citep{Kovalev2014}.} 
\tablefoottext{c}{Equivalent diameter of the phased array of $27
  \times 25$\,m telescopes taking into account phasing losses in the CnB configuration.}
\tablefoottext{d}{Mk2 telescope.}
\tablefoottext{e}{Equivalent diameter of the phased array of $14
  \times 25$\,m telescopes without taking into account phasing
  losses.}
}
\end{table*}

In addition to five antennas that suffered from a complete data loss due to technical problems\footnote{These were KVN-Yonsei 21-m, Urumqi 25-m, Evpatoria 70-m, Torun 32-m and Robledo DSN 70-m.}, hardware problems of various degrees were also experienced at Kitt Peak, Kalyazin, and the Green Bank Telescope. Kitt Peak showed weak fringes, which was later found to have been caused by a poorly seated cable in the signal path. At Kalyazin, the low-noise amplifier (LNA) of one of the polarisation channels failed. Since Kalyazin has linearly polarised feeds and circular polarisation is formed after the LNA stage in a hybrid, the failed LNA resulted in recording of the same linearly polarised signal at both polarisation channels instead of two opposite circular polarisations. The C band receiver of the Green Bank Telescope exhibited strong cross-talk between the polarisation channels and its bandpass shape varied strongly with time. 
Kalyazin and the Green Bank Telescope were successfully used at the fringe detection stage, but they were dropped from the imaging due to the non-closing errors they introduced to the data.

\subsubsection{Correlation and baseline-based fringe search} \label{pimafring}

The data were correlated at the Max-Planck-Institut f\"ur Radioastronomie (MPIfR) using the DiFX software correlator \citep{Deller2011}, which had been modified to include a model for the path delay of an interferometer including an orbiting element by taking into account both special and general relativistic effects in a rigorous manner \citep{Bruni2016}. Post-correlation analysis was performed in two steps in order to deal with specific issues of space-VLBI data.

Typical accuracies (maximum deviation from the true value) in the position and velocity of the reconstructed SRT state vector are less than 150\,m and 0.01\,m/s, respectively \citep{2014CosRe..52..342Z,2020AdSpR..65..798Z}. These uncertainties lead to two complications. First, the fringe delay and delay rate may differ from the a priori values  typically by up to 0.5\,$\mu$s and $4 \times 10^{-11}$\,s/s, respectively. Secondly, the residual fringe phase may have a significant non-linear dependence on time over ten-minute-long scans. A non-linear term that is unaccounted for can lead to significant (a factor of three or more) coherence losses. 

To overcome the first problem, we used a wide fringe search window of 4096 spectral channels and 0.05\,s integration time ($\pm64$\,$\mu$s in delay and $\pm 2 \times 10^{-10}$\,s/s in rate) in the initial fringe search at the correlator. Since the time stamp information for the SRT is set by the tracking station at the moment it starts recording a scan (and since there were concerns that the tracking station clock offset could have random jumps between the scans), fringes to the SRT were searched separately for every scan at the correlator. However, it turned out that the SRT delays and rates changed relatively smoothly from scan to scan and therefore the interpolation or extrapolation of the clock offsets and rates were used for those scans that did not yield detections on baselines to the SRT in the initial fringe search at the correlator. The data were re-correlated applying the refined delay and delay rate model and averaged to have 128 spectral channels per sub-band (IF) and 0.5\,s integration time. As a result of re-correlation, the signal-to-noise ratio (S/N) is enhanced thanks to the centring of the search window near the true delay and the respective delay rate.

To overcome the second problem, namely, the residual fringe phase possibly having a large non-linear dependence on time, we modified the traditional fringe-fitting procedure and included evaluation of phase delay acceleration term in addition to estimation of phase delay rate and group delay. This procedure was implemented in the VLBI processing software \textsc{PIMA}\footnote{\url{http://astrogeo.org/pima/}} \citep{Petrov2011} by applying a sequence of trial acceleration terms in the specified range with the given step, performing a fringe search over phase delay rate and group delay for each trial acceleration term, and selecting the phase delay acceleration, phase delay rate, and group delay that maximise the time and frequency averaged fringe amplitude.

\begin{figure}
\centering
\includegraphics[angle=-90,width=0.99\columnwidth]{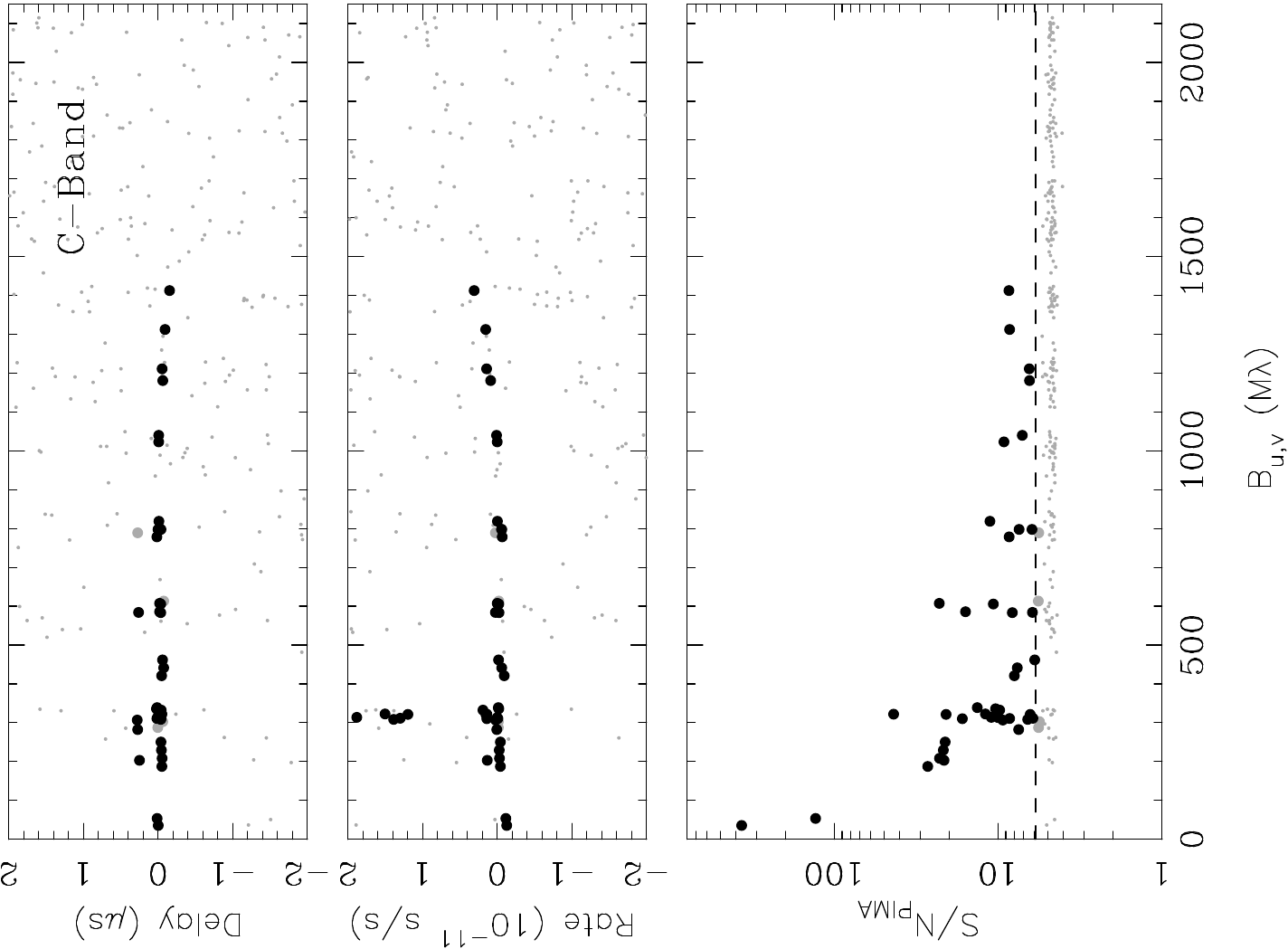}
\caption{Results of the initial C band fringe search on \textit{RadioAstron} baselines using \textsc{PIMA}. Delay (\textit{top panel}), rate (\textit{middle panel}), and S/N (\textit{bottom panel}) of the \textsc{PIMA} fringe-fit solutions are plotted against the projected baseline length (in M$\lambda$). Large black dots correspond to solutions that have $\mathrm{S/N}_\mathrm{PIMA}$ higher than the detection threshold of 5.92, which is indicated by the dashed line in the bottom panel. Large grey dots are solutions with 5.52\,$ < \mathrm{S/N}_\mathrm{PIMA} \leq$\,5.92 (potential detections) and small grey dots are solutions with $\mathrm{S/N}_\mathrm{PIMA} \leq$\,5.52 (non-detections).}
\label{pimaC}
\end{figure}

\begin{figure}
\centering
\includegraphics[angle=-90,width=\columnwidth]{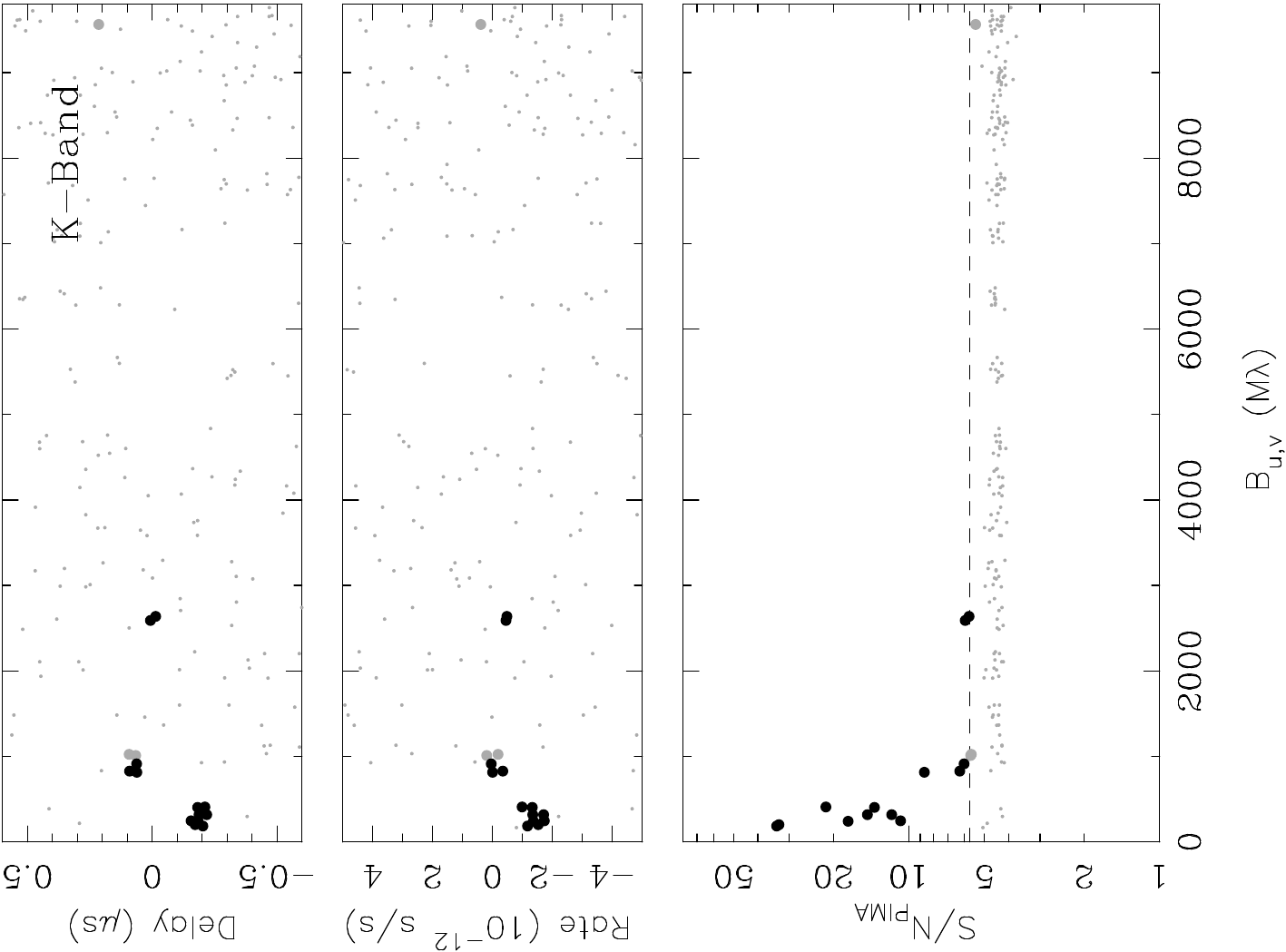}
\caption{Results of the initial K band fringe search on \textit{RadioAstron} baselines using \textsc{PIMA}. Delay (\textit{top panel}), rate (\textit{middle panel}), and S/N (\textit{bottom panel}) of the \textsc{PIMA} fringe-fit solutions are plotted against the projected baseline length (in M$\lambda$). The threshold for a detection is $\mathrm{S/N}_\mathrm{PIMA} > 5.72$ (large black dots).  Solutions with $5.30 < \mathrm{S/N}_\mathrm{PIMA} \leq 5.72$ are marked by large grey dots and solutions with $\mathrm{S/N}_\mathrm{PIMA} \leq 5.30$ are marked with small grey dots.}
\label{pimaK}
\end{figure}

We started the fringe-search with PIMA in the C band with a search space of $\pm2 \mathrm{\mu s}$ in delay, $\pm2\times10^{-11} \mathrm{s/s}$ in rate and $\pm1 \times 10^{-13}$\,s/s$^2$ in acceleration. The detection threshold was chosen to be $\mathrm{S/N}_\mathrm{PIMA} >$\,5.92, which corresponds to a false detection probability of $P_e < 0.1$\% for this search space (see Appendix~\ref{false_det})\footnote{The S/N in \textsc{PIMA}, $\mathrm{S/N}_\mathrm{PIMA}$, is defined as the ratio of fringe amplitude at the maximum to the \emph{average} fringe amplitude of the randomly picked visibilities in the absence of signal \citep{Petrov2011}. This definition differs from the S/N definition used in the AIPS task \textsc{fring}, where the S/N is the fringe amplitude at the maximum divided by an approximation of the standard deviation of the fringe amplitude of the noise.}. The results of the initial C band fringe search with PIMA are shown in Fig.~\ref{pimaC}. \textit{RadioAstron} fringes were detected from projected baseline lengths of 35\,M$\lambda$ (0.2\,$D_\oplus$) to 1412\,M$\lambda$ (6.9\,$D_\oplus$) at 5\,GHz. The longest baseline detection was with Effelsberg and had an $\mathrm{S/N}_\mathrm{PIMA}$ of 8.6. The fringes are generally centred around the delay and rate values used at the correlator with the exception of two scans at the beginning of the experiment that had a large residual acceleration term and an offset in rate $\big($(1.2--1.9)$\times 10^{-11}$\,s/s$\big)$. The results of the C band fringe search were then used to limit the search space in the K band to $\pm0.6\,\mathrm{\mu s}$ in delay and $\pm5\times10^{-12}\,\mathrm{s/s}$ in rate. The corresponding detection threshold giving a false detection rate of 0.1\% was $\mathrm{S/N}_\mathrm{PIMA} >$\,5.72. The results are shown in Fig.~\ref{pimaK}. At 22\,GHz, \textit{RadioAstron} fringes were detected from projected baseline lengths of 186\,M$\lambda$ (0.2\,$D_\oplus$) to 2638\,M$\lambda$ (2.8\,$D_\oplus$). As we show in Sect.~\ref{stack}, more fringes can be recovered if we combine baselines to several ground radio telescopes.

\begin{figure}
\centering
\resizebox{\hsize}{!}{\includegraphics[angle=-90]{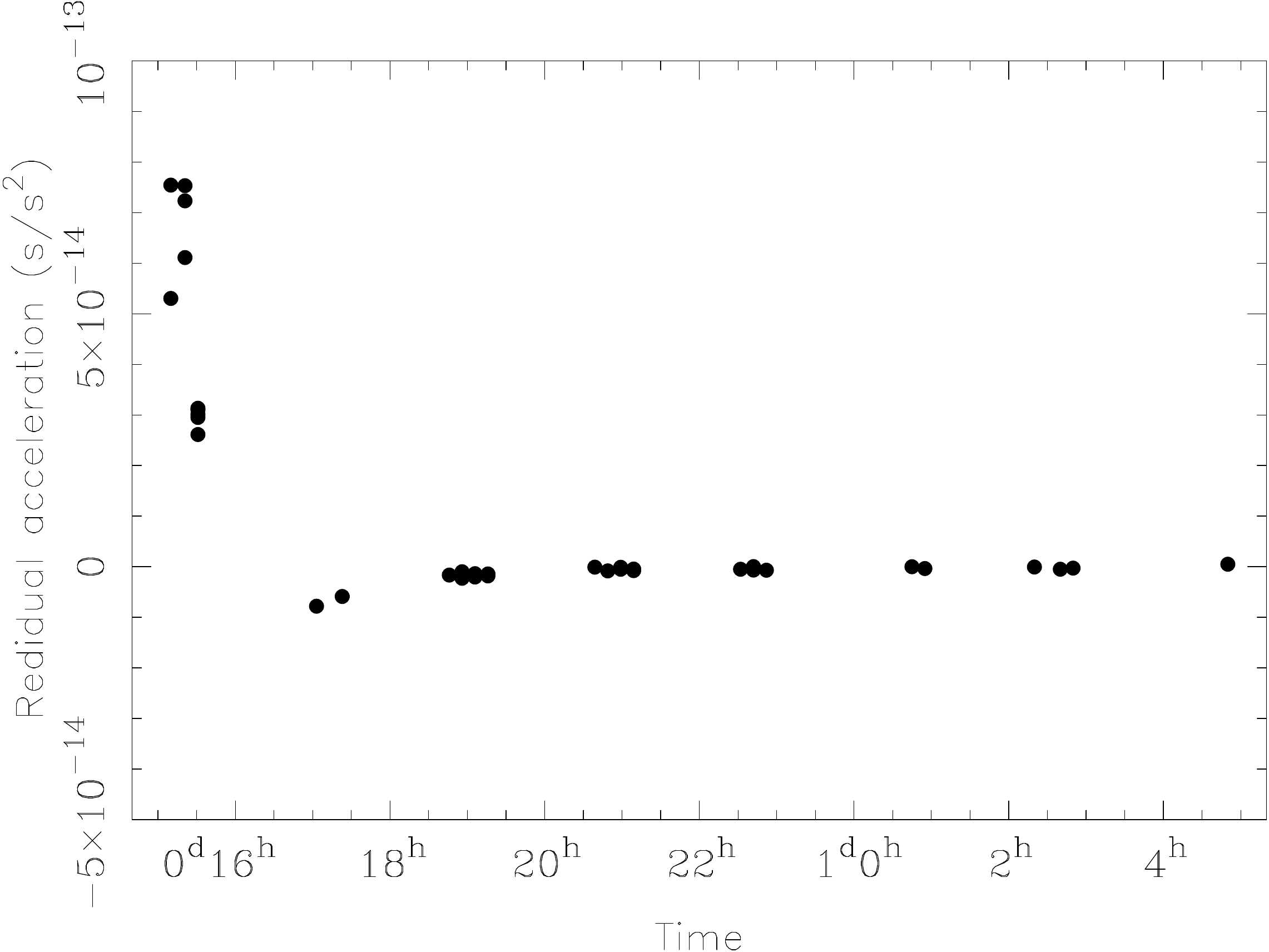}}
\caption{Residual acceleration term of the Space Radio Telescope measured at 5\,GHz after correlation. The SRT passed through perigee at the beginning of the observing run.}
\label{acceleration}
\end{figure}

The initial fringe-fitting with \textsc{PIMA} revealed large residual accelerations close to the orbital perigee, $(2-8) \times 10^{-14}$\,s/s$^2$, corresponding to $0.1-0.4$\,mHz/s at 4.836\,GHz and $0.4-1.8$\,mHz/s at 22.236\,GHz (see Fig.~\ref{acceleration}). It is clear that (especially in the K band), such high acceleration values significantly limit the coherence time (to some tens of seconds), unless they are corrected for. As the SRT moves away from perigee, the residual acceleration term decreases and approaches zero at large distances from the Earth. The derived residual acceleration terms were used to correct the data in \textsc{AIPS} (see Sect.~\ref{stack}). We note that it would have been desirable to include the acceleration correction already in the correlator model and re-correlate the data. Unfortunately, this feature was not yet implemented in the correlator software at the time of our experiment \citep{Bruni2016}.

\subsubsection{Global fringe-fitting and phasing of the ground array to calibrate the space baselines} \label{stack}

The success of a space-VLBI observation crucially depends on our ability to detect weak fringes on noisy space baselines. The standard method to lower the detection threshold in fringe search is to employ a global fringe-fitting scheme that uses data from all the baselines to estimate the residual phases, delays and rates \citep{Schwab1983}. For $N_a$ identical antennas, the reduction factor compared to a single baseline case is $\approx \sqrt{2/N_a}$ for a point-like source \citep{Thompson2007}. In the case of space-VLBI, the source can be usually easily detected on the ground-ground baselines and, therefore, we could phase up the ground array and correlate its coherently averaged signal with the space antenna signal. For $N_g$ identical ground array antennas, this would reduce the detection threshold by a factor of $\sqrt{1/N_g}$ compared to a single baseline case.\footnote{If the ground array is heterogeneous, the reduction is obviously not as good as $\sqrt{1/N_g}$, since the sensitivity is mostly set by the largest dishes. Furthermore, a non-point-like source structure also results in a smaller improvement in the sensitivity than this theoretical value. Therefore, using a source model is important for sources that are partially resolved on the ground-ground baselines.} Interestingly, \citet{Kogan1996} has shown that, in fact, the same reduction in the detection threshold for ground-space baselines can be achieved by global fringe-fitting, if the ground-ground baselines are much more sensitive than the space baselines. Therefore, after determining the residual acceleration terms with baseline-based fringe fitting in \textsc{PIMA}, we have carried out further rounds of global fringe fitting using the \textsc{AIPS} task \textsc{fring}.

First we calibrated, imaged, and self-calibrated the ground-array data in a standard manner using \textsc{AIPS} and \textsc{Difmap} \citep[see, e.g.][]{Lister2009a}. This provided us with models (images) of the target source and the calibrators on ground baselines. Then we ran another round of fringe-fitting in \textsc{AIPS} using these source models and first solving only for the ground array antennas. Next, we corrected the SRT data for the large residual acceleration terms obtained with PIMA (see Sect.~\ref{pimafring}). For this, we wrote a ParselTongue \citep{Kettenis2006} script that calculates the phase corrections for every SRT scan from the input acceleration terms and then writes these in \textsc{AIPS} SN tables that can be applied by the \textsc{AIPS} task \textsc{clcal}. We also used a single scan with a strong signal on the space baselines (typically, the shortest projected space baseline in the observation) to solve the single-band delays and the phase offsets between the IFs for the SRT.

After these steps, fringes were iteratively searched on the space baselines by running the AIPS task \textsc{fring} with the input model of 3C\,84 (i.e. the ground-array image) and solving only for the SRT. The weighted average of up to three baseline combinations was used in the coarse fringe search stage using simple fast Fourier transform (FFT) to estimate the phase on a given baseline to the SRT (parameter \textsc{dparm(1)=3} in \textsc{fring}). Since, at this point, the ground array is already calibrated, the data are divided by the source model, and the residual acceleration of the SRT is corrected, the coherence time on the baselines to the SRT is long, about 8\,min even at 22\,GHz (see Appendix~\ref{apxcoh}), thus allowing for long coherent averaging times. We used a solution interval of 10\,min in \textsc{fring} and ran three rounds of fringe search: the first one with the FFT stage $\mathrm{S/N}_\mathrm{AIPS,FFT}$ cutoff set to 5, then the second round using a much tighter search window around the position of the neighbouring high-S/N solutions and with a lower $\mathrm{S/N}_\mathrm{AIPS,FFT}$ cutoff, and then finally an exploratory search for weak fringes on the longest baseline scans for which the clock offsets and rates had been extrapolated at the correlator (scans after 05:00:00\,UT on the second day of observations).

Setting the S/N cutoff is always a trade-off between the sensitivity and the risk of false detection. It is known that \textsc{fring} underestimates the $\mathrm{S/N}_\mathrm{AIPS,FFT}$ at the low-S/N limit \citep{Desai1998}. Therefore, we evaluated the false-detection probability by examining the distribution of the maximum $\mathrm{S/N}_\mathrm{AIPS,FFT}$ from fringe searches limited to the part of the parameter space where no signal was expected (see Appendix~\ref{false_det}). The $\mathrm{S/N}_\mathrm{AIPS,FFT}$ cutoff in the initial FFT step of the second round of \textsc{fring} was set to 3.1, while the search windows were constrained to $\pm100$\,ns in delay and $\pm25$\,mHz (C band) or $\pm50$\,mHz (K band) in rate. The corresponding false detection rates are 0.08\% and 0.2\% for the C and K bands, respectively. In the K band, we searched for weak fringes in 26 scans, which gives a 5\% probability of at least one false detection with $\mathrm{S/N}_\mathrm{AIPS,FFT} \ge 3.1$ in our experiment. In the C band, all the detections have a higher $\mathrm{S/N}_\mathrm{AIPS,FFT}$ (minimum is 3.3) and weak fringes ($\mathrm{S/N}_\mathrm{AIPS,FFT} < 5$) needed to be searched only for eight scans. Therefore, the probability of at least one false detection in the C band from the second round of fringe fitting is only 0.2\%.

Finally, we searched for weak fringes for scans after 05:00:00\,UT on the second day of observations. Since we did not have any strong detections for these longest space baseline scans (baseline lengths longer than 1.5\,G$\lambda$ and 7.4\,G$\lambda$ in the C and K bands, respectively), we could not use small windows to constrain the search. Instead, we looked for cases where several $\mathrm{S/N_{AIPS,FFT}} > 3$ solutions group in the same area of the parameter space. For four scans between 06:10:00\,UT and 06:50:00\,UT, we find such a grouping around a delay value of $-500$\,ns and a rate value of $0.5 \times 10^{-11}$\,s/s with the highest $\mathrm{S/N_{AIPS,FFT}}$ being 4.5. We consider these solutions as detections in imaging, but since their false detection probability cannot be easily estimated, we carried out the brightness temperature analysis in the companion paper (T.~Savolainen et al., in prep), also without these data.

\begin{figure}
\centering
\includegraphics[angle=-90,width=\columnwidth]{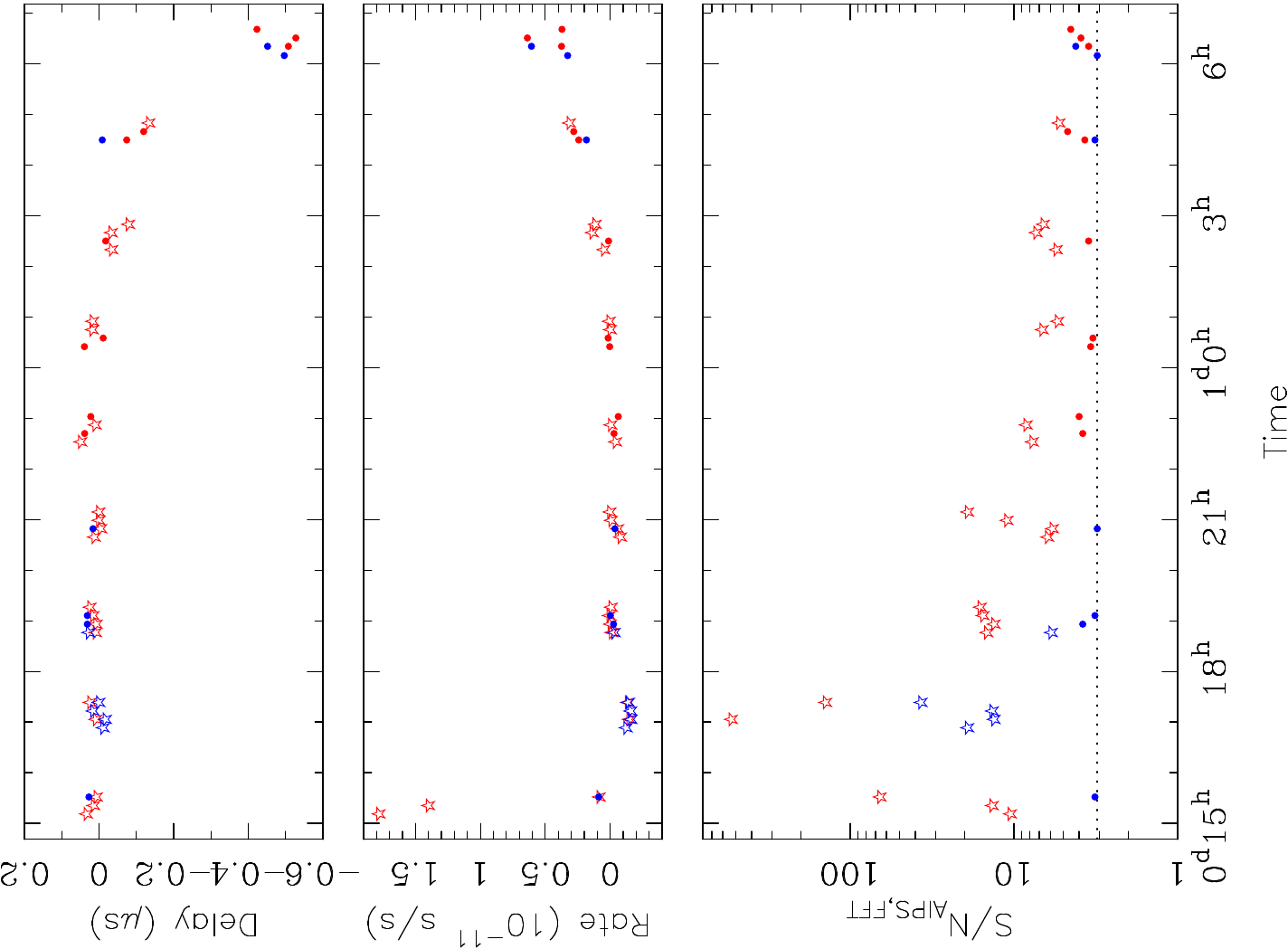}
\caption{Detected fringes from the global fringe search in \textsc{AIPS} for baselines to the \textit{RadioAstron} satellite. The top panel shows the residual delay of the SRT, the middle panel shows the residual rate of the SRT, and the bottom panel shows the S/N of the initial FFT stage. C band solutions are marked in red and K band solutions are marked in blue. Solutions with $\mathrm{S/N_{AIPS,FFT}} \ge 5.0$ are marked with stars and solutions with $\mathrm{S/N_{AIPS,FFT}} < 5.0$ are marked with dots. The dotted line shows the cutoff-value of $ \mathrm{S/N_{AIPS,FFT}} \ge 3.1$ adopted for the second round of fringe search using narrow windows.}
\label{aipsfringe}
\end{figure}

The results of the global fringe search in \textsc{AIPS} are presented in Fig.~\ref{aipsfringe}. While fringes were found for 22 SRT scans with \textsc{PIMA} in the C band on individual baselines, the global search in \textsc{fring} resulted in detections for 33 SRT scans with the longest baselines being 1656\,M$\lambda$ (8.1\,$D_\oplus$) to the GBT and 1585\,M$\lambda$ (7.7\,$D_\oplus$) to Effelsberg. In the K band the number of detections increased from 7 to 12 scans with the longest baselines being 7640\,M$\lambda$ (8.1\,$D_\oplus$) to the VLA and 7136\,M$\lambda$ (7.6\,$D_\oplus$) to Effelsberg. The fringe spacing of the longest baseline with a detection at 22\,GHz corresponds to 27\,$\mu$as. We describe how we used these data from the AIPS global fringe search in the imaging in Sect.~\ref{sec:imaging}.

We note that the source structure causes an additional complication when searching for fringes on the long space baselines. If the structure is not modelled accurately, the residual phase slopes on the long baselines can be large and a fast-moving spacecraft can sample multiple $2\pi$ rotations of the structural phase during a single scan. Fringe-fitting can absorb these slopes to residual delays and rates (especially when the triangles to the SRT are long and narrow) and this can increase scatter of the solutions significantly. It is difficult to judge whether the increased scatter in the delay and rate solutions on the longest baselines at the end of the experiment in Fig.~\ref{aipsfringe} is due to this effect or solely due to the noise.

\subsubsection{Amplitude calibration}

The gain amplitude calibration of the SRT and the ground radio telescopes was carried out in a standard manner using measured system temperatures and a priori gain curves. Ground-telescope K band data were also corrected for the atmospheric opacity in the \textsc{AIPS} task \textsc{apcal}. This a priori calibration was improved by determining an overall gain correction factor for individual antennas from the average amplitude self-calibration solutions of the target source and the observed calibrators. If this amplitude correction for a given ground array antenna was larger than 15\%, we applied it to the data using the \textsc{AIPS} task \textsc{clcor}.

\subsubsection{Imaging and self-calibration} \label{sec:imaging}

\begin{figure*}
\centering
\includegraphics[angle=-90,width=0.4\textwidth]{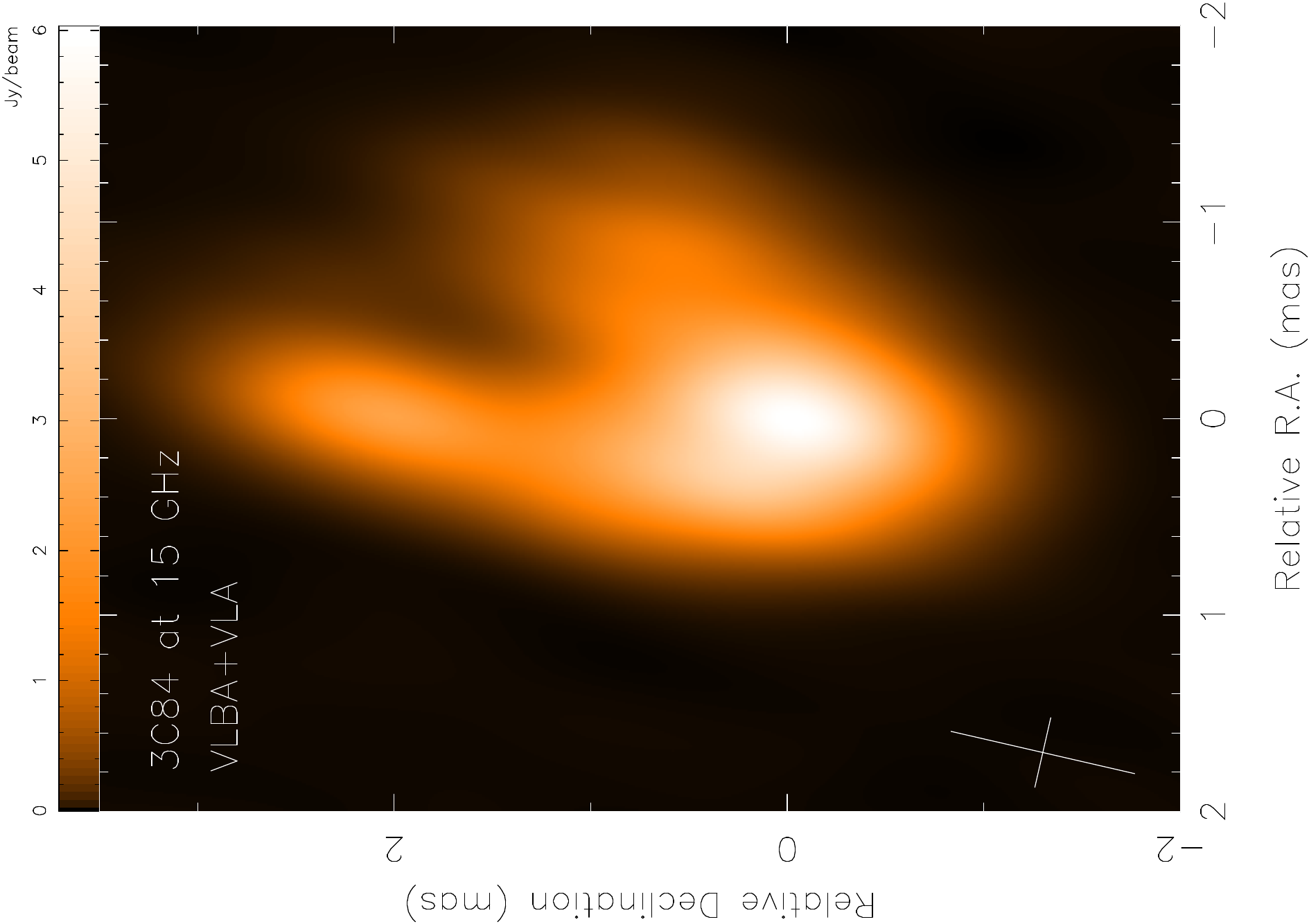}
\includegraphics[angle=-90,width=0.4\textwidth]{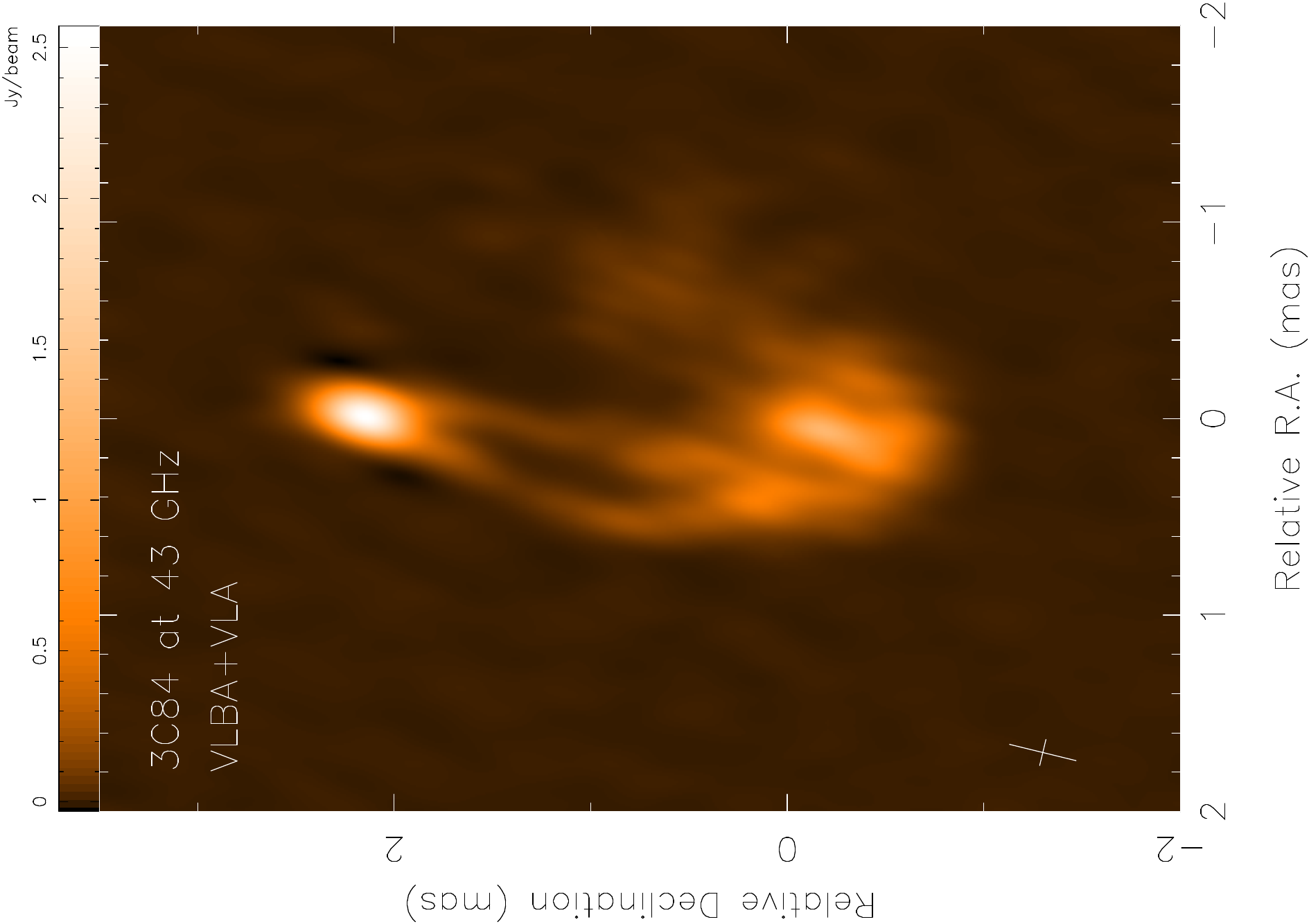}
\caption{Uniformly weighted VLBA+VLA CLEAN images of the core region of 3C\,84 at 15\,GHz (\textit{left panel}) and at 43\,GHz (\textit{right panel}). The restoring beam sizes are $1.0 \times 0.4$\,mas at PA=$-$12$^\circ$ (15\,GHz) and $0.36 \times 0.134$\,mas at PA=$-$13$^\circ$ (43\,GHz).}
\label{GRTcolor}
\end{figure*}

We imaged the data using the standard non-linear deconvolution algorithm CLEAN \citep{Hogbom1974,Clark1980} implemented in the \textsc{Difmap} software package \citep{Shepherd1997}. The editing, imaging, and self-calibration of the data were all carried out in \textsc{Difmap}. The procedure used for this process differed slightly from the standard one. In order to achieve the high angular resolution that is the whole aim of space-VLBI observations, it is necessary to give more weight to the space baselines than in the usual natural weighting scheme \citep{Murphy2000}. However, we have to balance this weighting against adding excessive amounts of noise to the image. We find that a combination of uniform weighting with a bin size of $5 \times 5$ pixels (i.e. grid elements) in the $(u,v)$ grid and additional weighting by the amplitude errors to the power of $-1$ in \textsc{Difmap} gives a good balance (i.e. setting \textsc{uvweight} 5,$-$1). We note that not using any weighting at all by the amplitude errors is dangerous, as this causes an over-fitting of the CLEAN model on the noisy space baselines. When self-calibrating the ground array data, we did use natural weighting in order to ensure we did not miss any large scale emission before running amplitude self-calibration. A low loop gain (a percentage of the map amplitude removed by each CLEAN component) of 0.01$-$0.02 was used to obtain a better CLEAN performance for extended emission structure even though it meant increased computation time.

Another important step is the self-calibration of the SRT phases. For this, we used a 2\,min solution interval, while the ground array was self-calibrated with a 10\,s solution interval. This ensures a high-enough S/N of the space baseline solutions and prevents generation of spurious flux from noise that could otherwise happen when the SRT is at end of long, narrow baseline triangles that are noisy. Such a spurious flux generation is a special case of the more general phenomenon described by \citet{MartiVidal2008}. For the ground array, we carried out amplitude self-calibration down to 0.5\,min solution intervals, while for the SRT, only one round of amplitude self-calibration with the solution interval equal to the observation length was employed. The amplitude corrections for the SRT were $15 - 20$\,\%.

The sparse $(u,v)$ coverage on the space baselines makes the imaging of the \emph{RadioAstron} data challenging. As described in Sect.~\ref{sec:images}, the CLEAN algorithm tends to create artificial `clumpiness' in the final images due to the significant gaps in the $(u,v)$ coverage of space baselines. Therefore, in addition to inverse modelling with CLEAN, we also carried out forward modelling with the RML method \citep[see e.g.][]{EHT2019d} implemented in the \textsc{eht-imaging} library \citep[\textsc{ehtim};][]{Chael2016, Chael2018}. The general approach in RML imaging is to find the image that minimises an objective function $\sum \alpha_D \chi_D^2 - \sum \beta_R F_R$, where $\chi_D^2$ is a goodness-of-fit function of the data term $D$ and $F_R$ is a regulariser function $R$. Hyperparameters $\alpha$ and $\beta$ give the relative weights between different data terms and regularisers. Since one can use regularisers that emphasise, for instance, smoothness, RML imaging can avoid some of the problematic issues of CLEAN that are related to the CLEAN's inherent assumption of the sky brightness distribution being composed of point sources. Also, the output image from the RML imaging does not require convolving with a Gaussian beam, since the image itself fits the visibility data -- unlike in CLEAN, where it is the collection of point sources that fits the data.

When imaging with the \textsc{ehtim}, we started with an initial image from ground-only CLEAN imaging to speed up the convergence. At 5\,GHz, we chose a field of view (FoV) of 40.96\,mas, divided in a grid of $1024 \times 1024$ pixels, and further blurred the initial image with a Gaussian kernel of 0.9\,mas FWHM. At 22\,GHz, the FoV was 10.24\,mas, the grid was $1024 \times 1024$ pixels and the initial blurring was done with a Gaussian kernel of 0.25\,mas FWHM. The number of pixels in the model is large due to the wide range of angular scales present in the data and having the ground-only image as the initial model is essential for the convergence. This is not a major issue in terms of the independence of our analysis, however, since we are interested in the small-scale structure in the core region, which is not present in the ground-only images. In the first round of imaging, the data terms included closure phases, log closure amplitudes and, with a small weight, visibility amplitudes. After convergence, visibility phases were self-calibrated and imaging was started over with data terms including both closures and complex visibilities. Before the third and final round, amplitudes were also self-calibrated. We used the same self-calibration intervals as in \textsc{Difmap}.

The choice of regularisers and their weights has an impact on the final images from RML and, therefore, we carried out a small parameter survey of $\beta_R$ for four regulariser terms: relative entropy (MEM), total variation (TV), total squared variation (TV2), and the $l_1$ norm. A description of the parameter survey and a representative set of image reconstructions from different hyperparameter combinations both at 5\,GHz and 22\,GHz are  presented in Appendix~\ref{appendix:eht}. 

\subsection{Ground-array observations at 15\,GHz and 43\,GHz} \label{GRobs}

The SRT cooling gaps were used to make additional 15 and 43\,GHz observations of 3C\,84 with the frequency-agile VLBA and phased VLA antennas. The source was observed for eight scans of four minutes each at both frequencies using all eleven stations. The data were calibrated in a standard manner (including an opacity correction) using \textsc{AIPS}, after which they were imaged and self-calibrated in \textsc{Difmap}. The errors, $\sigma$, of the visibilities on baselines to VLA are much smaller than for the rest of the array and they do not seem to correctly represent the systematical uncertainties. Therefore, we weighted the visibilities by $\sigma^{-1}$ instead of the usual $\sigma^{-2}$. Figure~\ref{GRTcolor} shows the resulting images of the core region of 3C\,84 and image parameters are given in Table~\ref{tab:images}.


\section{Space-VLBI images of the restarted jet} \label{sec:images}

\begin{figure*}
\centering
\includegraphics[width=1.0\textwidth]{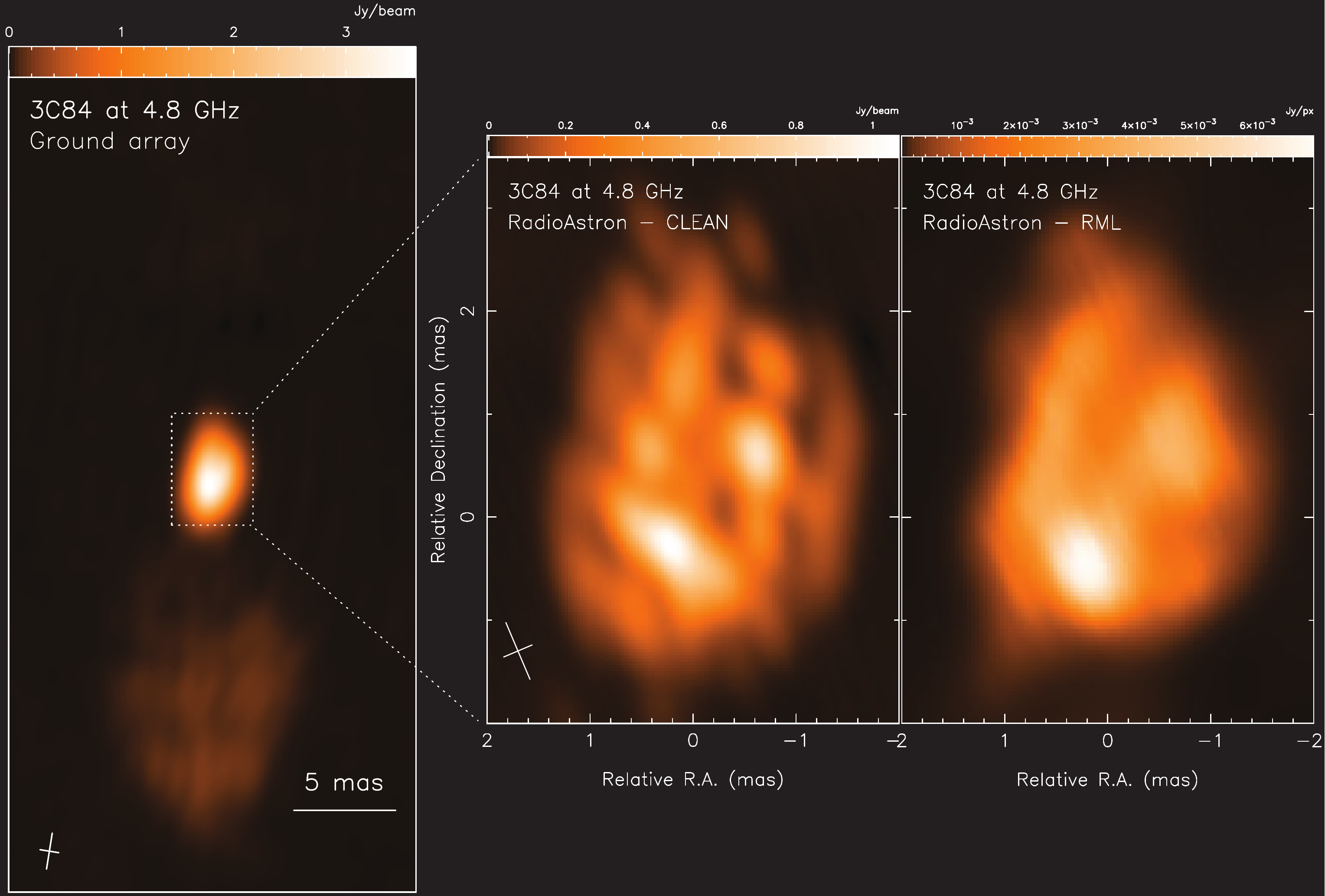}
\caption{Ground-based and space-based VLBI images of 3C\,84 at 5\,GHz. \textit{Left panel:} Naturally weighted global-VLBI image. The restoring beam size is $1.8 \times 0.9$\,mas at PA=$-9^\circ$ and the rms noise is 0.8\,mJy/beam. We note that the low-intensity emission south of the core region is the well-studied "mini-lobe" related to the 1959 outburst \citep[e.g.][]{Walker2000,Asada2006}. \textit{Middle panel:} Super-uniformly weighted space-VLBI CLEAN image of the central part of 3C\,84. The restoring beam size is $0.60 \times 0.30$\,mas at PA=23$^\circ$ and the rms noise is 0.9\,mJy/beam. \textit{Right panel:} Regularised maximum likelihood image of the central part of 3C\,84 made with \textsc{ehtim}. The used hyperparameters are MEM=0, TV1=1, TV2=1 and L1=1. The colour scale in all images has a lightness  proportional to the square-root of the pixel intensity.}
\label{RAcombined}
\end{figure*}

\begin{figure*}
\centering
\includegraphics[angle=-90,width=0.32\textwidth]{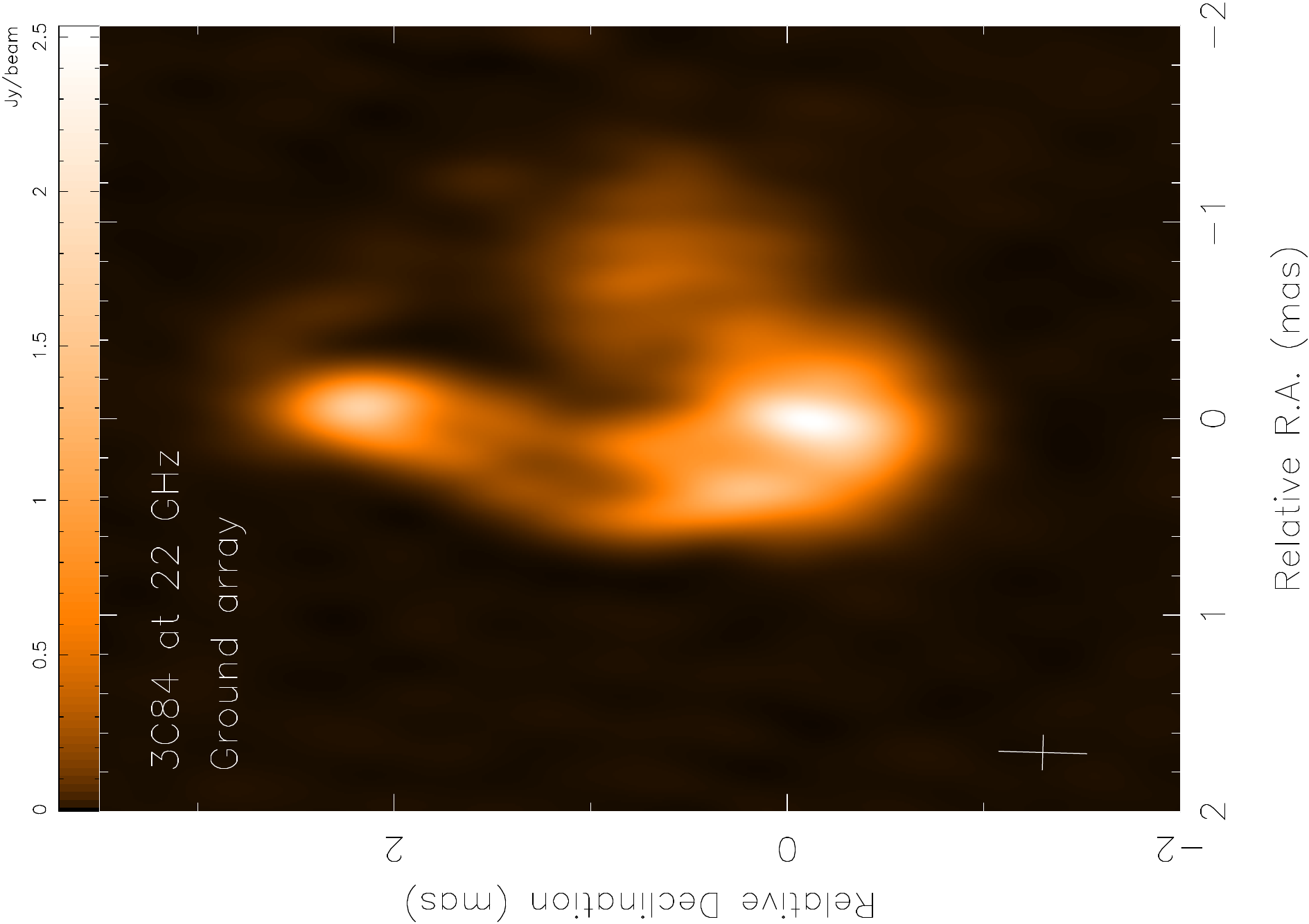}
\includegraphics[angle=-90,width=0.32\textwidth]{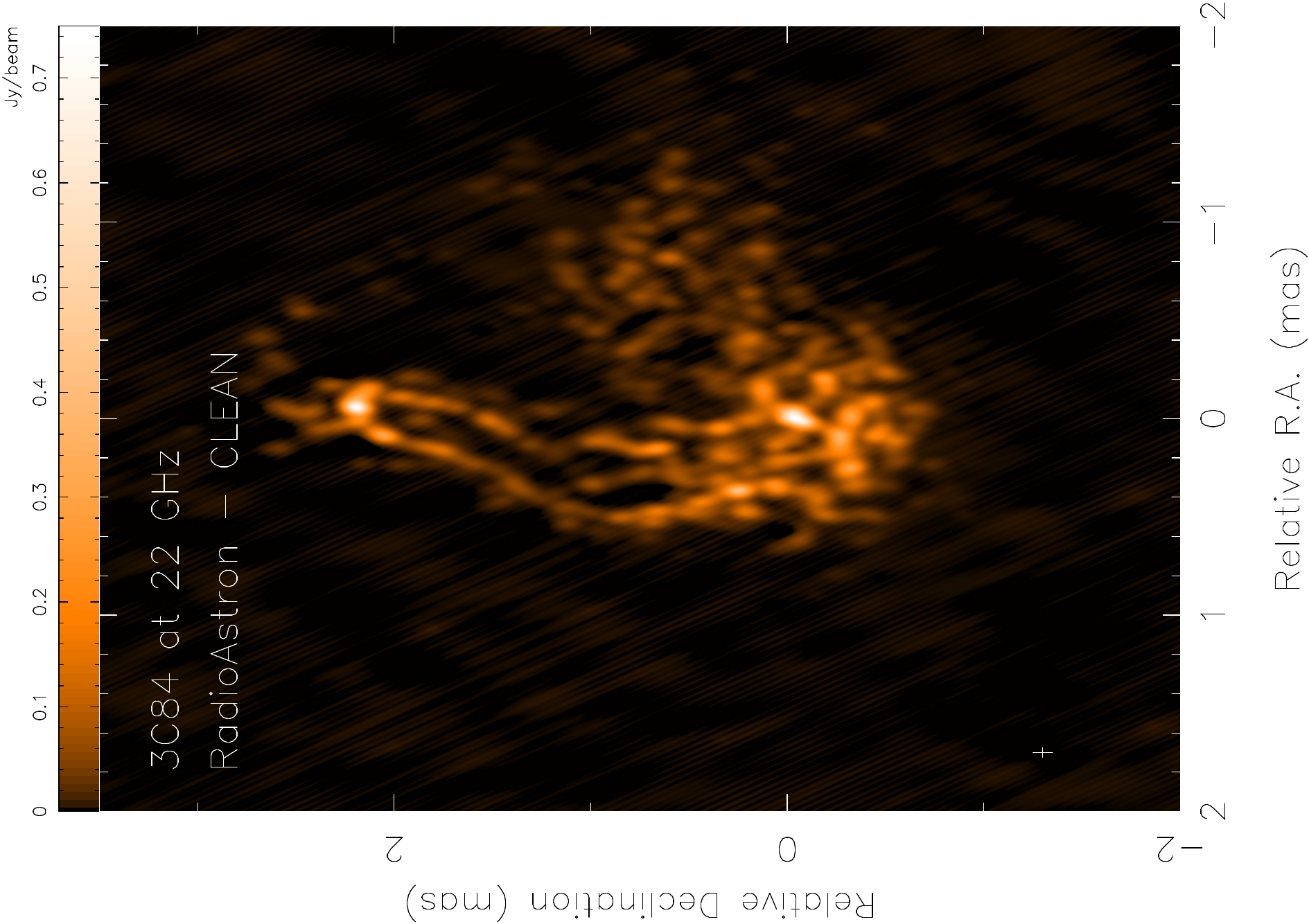}
\includegraphics[angle=-90,width=0.32\textwidth]{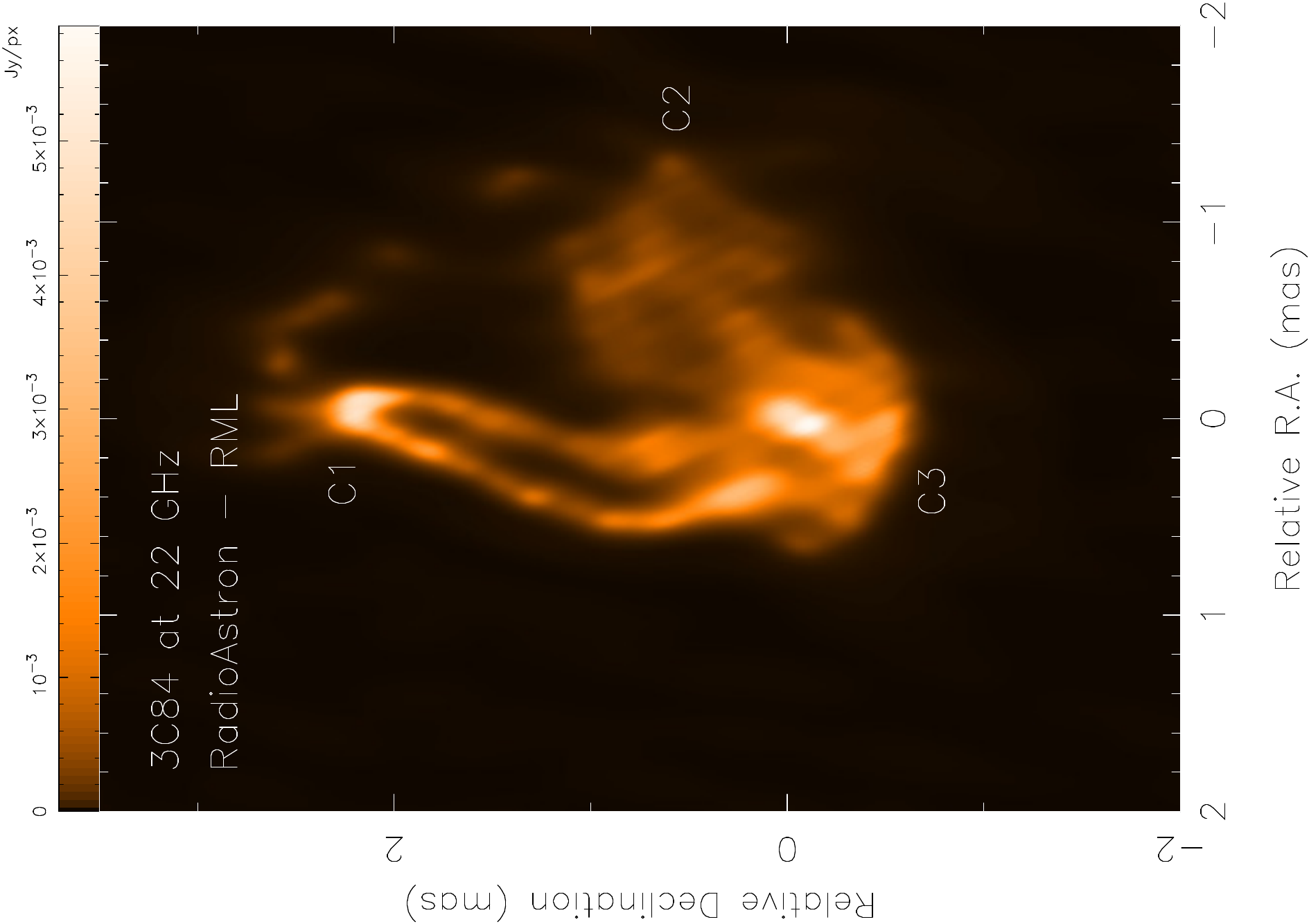}
\caption{Ground-based and space-based VLBI images of 3C\,84 at 22\,GHz. \textit{Left panel:} Uniformly weighted global-VLBI image. The restoring beam size is $0.45 \times 0.18$\,mas at PA=$-2^\circ$ and the rms noise is 1.5\,mJy/beam. \textit{Middle panel:} Super-uniformly weighted space-VLBI CLEAN image. The restoring beam size is $0.10 \times 0.05$\,mas at PA=0$^\circ$ and the rms noise is 1.4\,mJy/beam. This image was originally published in \citet{Giovannini2018}. The colour scale is restricted to positive intensity values in order to help comparison with the regularised maximum likelihood image on the right. An image with the full intensity range in the colour scale can be found in Appendix~\ref{appendix:clean}. \textit{Right panel:} Regularised maximum likelihood space-VLBI image made with \textsc{ehtim}. The used hyperparameters are MEM=10, TV1=1, TV2=0 and L1=1. The colour scale in all images has a lightness  proportional to the square-root of the pixel intensity.}
\label{RAcombined_K}
\end{figure*}

\begin{figure}
\centering
\includegraphics[trim={3.5cm, 5cm, 2cm, 5cm},clip,width=\columnwidth]{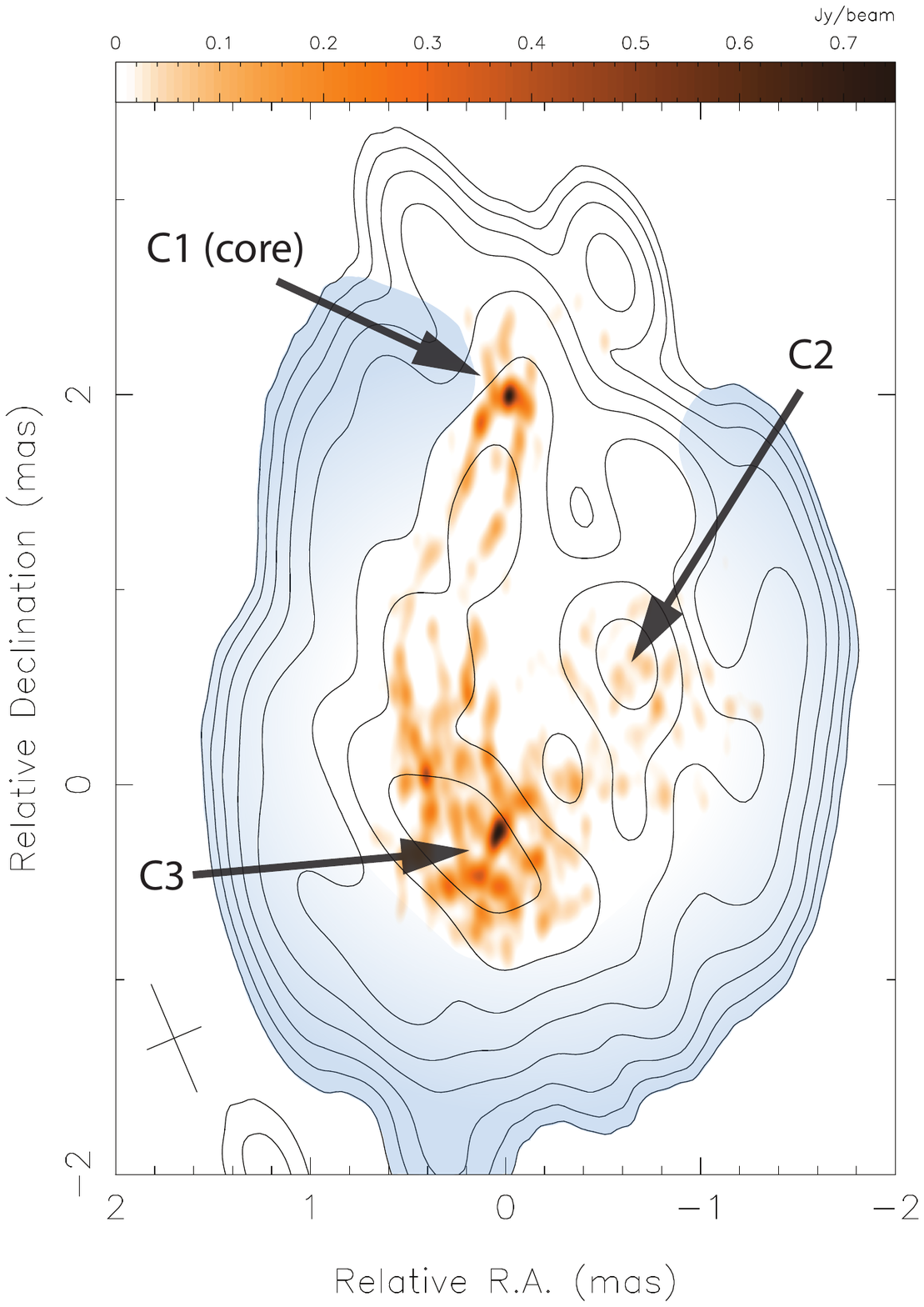}
\caption{Combined 5\,GHz (\emph{contours}) and 22\,GHz (\emph{colour scale}) \emph{RadioAstron} CLEAN images that show the location of the mini-cocoon (marked in \emph{blue}) and the features C1, C2, and C3. Note: C3 refers to the whole extended jet head, not just the bright 22\,GHz hot spot inside it. The restoring beam sizes are the same as in Figs.~\ref{RAcombined} and \ref{RAcombined_K}. The contours increase in steps of two from 5\,mJy/beam.}
\label{RAcombined_labels}
\end{figure}

Figure~\ref{RAcombined} shows the ground-only image of 3C\,84 at 5\,GHz as well as the CLEAN and RML space-VLBI images at 5\,GHz. The space baselines significantly increase the resolution at 5\,GHz, which allows us to clearly resolve the bright core region that has a size of $4.1 \times 2.9$\,mas or $1.4 \times 1.0$\,pc (projected). It is clear that the core region is much more complex than what can be resolved from the ground at 5\,GHz. This has consequences to the accuracy of VLBI astrometry, which will be discussed in another work (L.~Petrov et al., in prep). In Fig.~\ref{RAcombined_K}, we show the 22\,GHz ground-only image of 3C\,84 together with two space-VLBI images: a CLEAN image previously published in \citet{Giovannini2018} and a new RML image. It is clear that the space-VLBI images allow one to discern significantly smaller details than the ground-only images, but at the same time, the sparse $(u,v)$ coverage of the space baselines limits the image fidelity and generates visible imaging artefacts. Using different imaging methods that are based on different underlying assumptions about the source structure is essential for judging the robustness of different features in the images.

\begin{table*}
\centering
\caption{CLEAN image parameters}
\label{tab:images}
\begin{tabular}{lcccccllrccc}
\hline\hline
Frequency & Figure & $r_{uv, \mathrm{min}}$ & $r_{uv, \mathrm{max}}$ & Pol. & Weights\tablefootmark{a} & B$_\mathrm{maj}$ & B$_\mathrm{min}$ & B$_\mathrm{PA}$ & $I_\mathrm{tot}$\tablefootmark{b} & $\sigma_\mathrm{I}$ & $I_\mathrm{peak}$ \\
 (GHz)                     &    & (M$\lambda$) & (M$\lambda$) &         &      & (mas)  & (mas)  & ($^\circ$) & (Jy)         & (mJy/beam) & (mJy/beam) \\
\hline
 5\tablefootmark{c}        & 7  &  3.6         &  164         &  I  & 0,$-2$   &  1.8   & 0.9    & $-9$      & $15.6\pm0.8$ &  0.8       & 3620       \\
 5\tablefootmark{d}        & 7  &  3.6         &  1710        &  LL & 5,$-1$   &  0.6   & 0.3    &  23       & $15.2\pm0.8$ &  0.9       & 1060       \\
 15.4\tablefootmark{c}     & 6  &  2.6         &  440         &  I  & 2,$-1$   &  1.0   & 0.4    & $-12$     & $26.4\pm1.3$ &  1.4       & 6350       \\
 22.2\tablefootmark{c} & 8  &  3.7 &  789 &  I  & 2,$-$1 & 0.45 & 0.18 & $-$2 & $22.4 \pm 1.2$ & 1.5 & 2540    \\
 22.2\tablefootmark{d}     & 8  &  3.7         &  7740        &  LL & 5,$-1$   &  0.10  & 0.05   &  0        & $23.0\pm1.2$ &  1.4       &  746       \\
 43.2\tablefootmark{c}     & 6  &  7.1         &  1230        &  I  & 2,$-1$   &  0.36  & 0.13   & $-13$     & $15.0\pm0.8$ &  2.6       & 2560       \\
\hline
\end{tabular}
\tablefoot{ 
\tablefoottext{a}{Parameters of \textsc{uvweight} command in \textsc{Difmap}. The first number gives the size of the grid used in determining weights in the uniform weighting scheme; "5" = $5\times5$. "2" = $2\times2$, and "0" means natural weighting. The second number gives the exponent of the weighting based on the visibility errors, i.e, data are weighted either by $\sigma^{-1}$ or $\sigma^{-2}$}.
\tablefoottext{b}{Error estimate includes random noise and 5\% systematic uncertainty.}
\tablefoottext{c}{Ground radio telescopes only.}
\tablefoottext{d}{Super-resolved restoring beam.}
}
\end{table*}

The $(u,v)$ coverage of space-baselines mostly covers a narrow range of position angles around $\sim 110^\circ$, which leads to a large axial ratio of the interferometer's beam. The standard practice in radio interferometric imaging using CLEAN is to convolve the point sources found by CLEAN with a so-called 'nominal restoring beam' which is a 2D Gaussian fit to the central peak of the interferometer's beam. This is done to represent the finite resolution of the interferometer and to blur out artefacts that result from CLEAN algorithm's often poor extrapolation of the visibilities to the unsampled high spatial frequencies. In our case, the \emph{RadioAstron} CLEAN images have nominal beam sizes of $1.22 \times 0.17$\,mas (5\,GHz) and $0.32 \times 0.05$\,mas (22\,GHz) and the images convolved with these very elongated restoring beams are shown in Appendix~\ref{appendix:clean}. The very elongated restoring beam makes it difficult to inspect those images by eye and especially to compare the structures in the core region with the structures visible in the RML images, which are not convolved with a Gaussian.

Therefore, we have chosen to show in Figs.~\ref{RAcombined} and \ref{RAcombined_K}, CLEAN images, which are restored with Gaussian beams that are more circular than the interferometer beam, that is, we have super-resolved in the direction perpendicular to the space-baselines. While this increases the risk of creating small scale imaging artefacts (and indeed creates them, especially at 22\,GHz), we find the  super-resolution in the direction transverse to the space-baselines necessary for a meaningful comparison with the RML images. Moreover, for high-S/N visibility data the effective resolution of the interferometer is better than the nominal beam size \citep{Lobanov2005} and a modest amount of super-resolution is indeed often achievable with the RML methods and (to a lesser degree) also with CLEAN \citep{Akiyama2017,Chael2016}.  

The 5\,GHz space-VLBI CLEAN image in Figure~\ref{RAcombined} has a restoring beam of $0.60 \times 0.30$\,mas, namely, the image is super-resolved by a  factor of two in one direction and under-resolved by a factor of 1.8 in the other direction. The 22\,GHz restoring beam used in \citet{Giovannini2018} was $0.10 \times 0.05$\,mas, which corresponds to a factor of three super-resolution in the direction perpendicular to the space baselines. In \citet{Giovannini2018} this relatively large amount of super-resolution was used to study the high-S/N structures in the core region of 3C\,84. Other examples of CLEAN images with different restoring beam sizes are presented in Appendix~\ref{appendix:clean}.

A crucial question when using super-resolved images to draw conclusions about the source structure is related to the features that are likely to be real and which are imaging artefacts. We base our judgement both to a common understanding of typical imaging artefacts generated by different algorithms and to a comparison between the images generated by the different imaging algorithms. The latter is a good diagnostic, since CLEAN and RML typically produce different kinds of artefacts \citep[e.g.][]{EHT2019d}.

It is clear that the super-resolved CLEAN space-VLBI images in Figs.~\ref{RAcombined} and \ref{RAcombined_K} suffer from artificial `clumpiness' that is due to the CLEAN algorithm's inherent difficulty in presenting smooth structures combined with the sparse $(u,v)$ coverage of the ground-to-space baselines and the chosen super-resolved restoring beam especially at 22\,GHz. Regularised maximum likelihood images made with the \textsc{ehtim} package indeed have significantly smoother features. The RML images, on the other hand, show low-level striping, especially at 22\,GHz, which is also an artefact created by the gaps in the $(u,v)$ coverage. With these caveats in mind, we can now turn to the core region structures that are common in the CLEAN and RML maps and which we consider robust.

\subsection{Structure of the core region} \label{Sec:structure}

In Fig.~\ref{RAcombined_labels}, the 5\,GHz (in contours) and the 22\,GHz (in colour) \emph{RadioAstron} CLEAN images are overlaid in order to better show how the structures in the core region seen at different frequencies relate to each other. The three main features in the core region are labeled as C1, C2, and C3 following the naming convention of \citet{Suzuki2012} and \citet{Nagai2010}. The compact northern component C1 has a strongly inverted spectrum and is identified as the core. C1 is clearly visible both in the CLEAN and the RML images at 22\,GHz (Fig.~\ref{RAcombined_K}), but it appears more centre-brightened in the CLEAN image, which may be due to the CLEAN algorithm's point source assumption. The bright, resolved southern feature C3 is the head of the restarted jet and its morphology suggests a strong interaction between the jet and the surrounding medium. In particular, its overall size, the presence and the location of the compact hot spot inside it, and the brightened edge south-south-east of the hot spot are consistent between the 22\,GHz RML and CLEAN images. The small scale spottiness inside C3 in the 22\,GHz CLEAN image reflects CLEAN algorithm's difficulties in reproducing extended diffuse emission.

In the both 22\,GHz images, there is also a strongly edge-brightened, almost cylindrical jet connecting C1 and C3 that was previously detected at 43\,GHz \citep{Nagai2014} and discussed in detail in \citet{Giovannini2018}. The edge-brightening continues all the way down to the core in the both CLEAN and RML images, but the edges are smoother in the RML image. The western component C2 is diffuse and again its overall size is consistent between the 22\,GHz RML and CLEAN images, but the spotty sub-structure in the CLEAN image is a clear artefact. The component C2 presents remains of the jet of the 1990s, when the jet direction was significantly different from its current direction \citep{Dhawan1998,Lister2001}.

\begin{figure}
\centering
\includegraphics[angle=0,width=1.0\columnwidth]{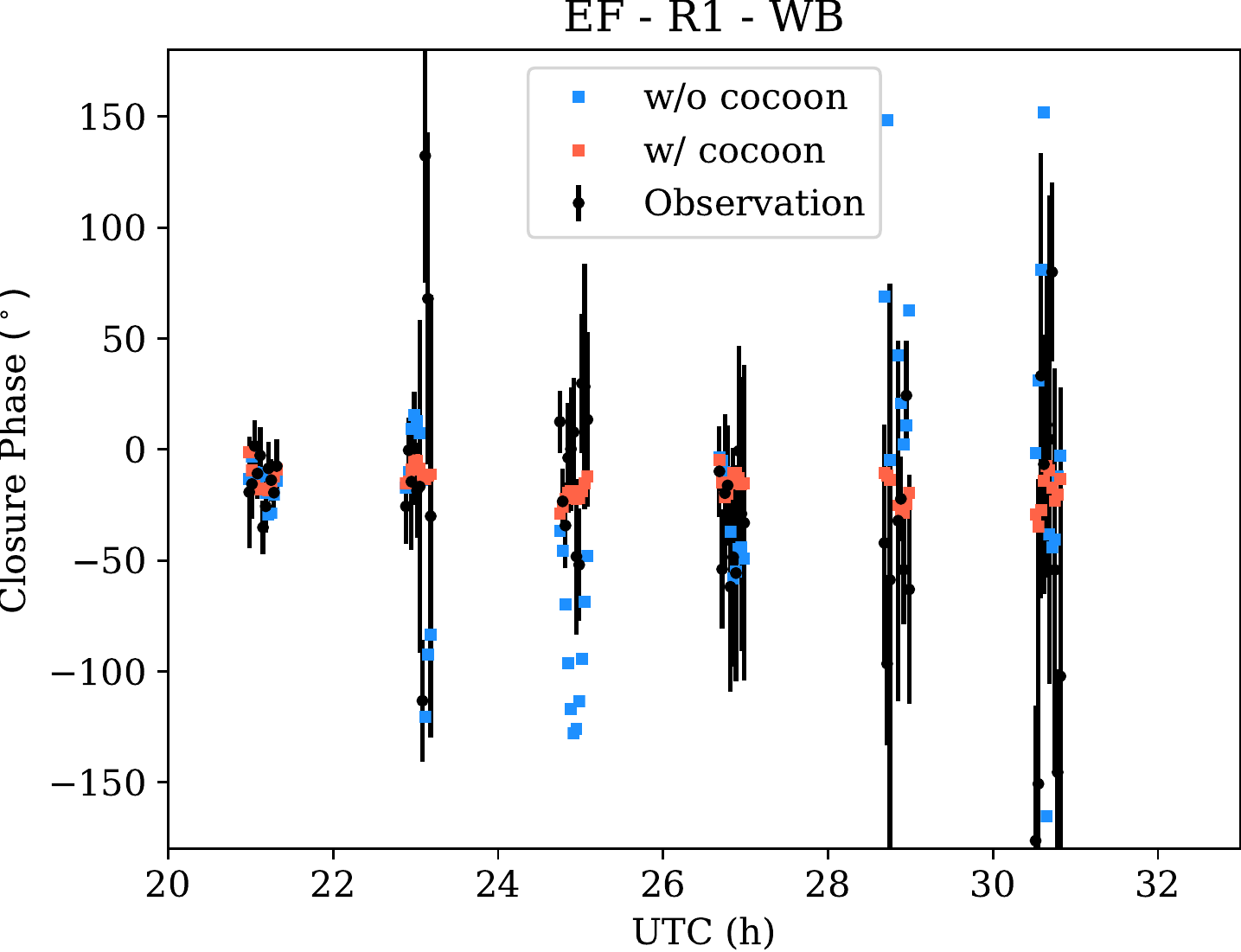} 
\includegraphics[angle=0,width=1.0\columnwidth]{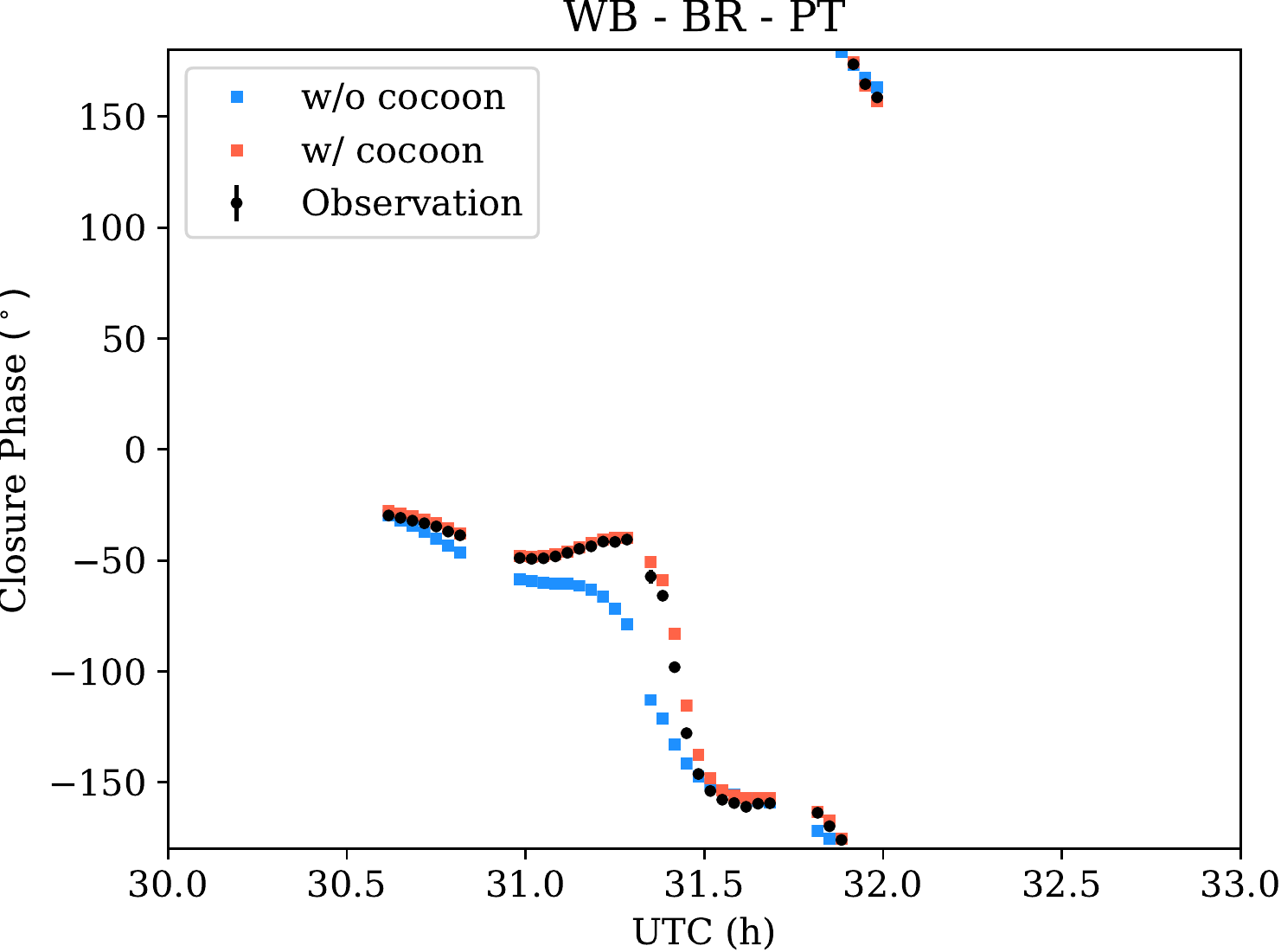}
\caption{Observed 5\,GHz closure phases (black data points) on the triangles Effelsberg $-$ Westerbork $-$ RadioAstron (\emph{top}) and Westerbork $-$ Brewster $-$ Pie Town (\emph{bottom}). Blue points show the closure phases of the best-fitting model without a cocoon structure, while red points correspond to the closure phases of a model including the cocoon emission.}
\label{CocoonCP}
\end{figure}

The components C2 and C3 as well as the restarted jet are visible also in the both CLEAN and RML images at 5\,GHz. The component C1 (core) appears heavily absorbed at 5\,GHz. Surrounding C2, C3, and the jet, there is a large, low-intensity emission region visible in the 5\,GHz images. It is present both in the RML and CLEAN images, although it appears to taper off more quickly towards the edges in the RML image. In the following, we refer to this emission the `mini-cocoon' (marked in the blue colour in Fig.~\ref{RAcombined_labels}), since it resembles diffuse emission structures surrounding many kpc-scale jets and radio lobes. As we can see in Fig.~\ref{RAcombined_labels}, the mini-cocoon envelops the restarted jet including the jet head C3, as well as the old emission feature C2, making the kpc scale cocoon analogy fitting. 

As a fidelity test for the robustness of the mini-cocoon structure, we attempted imaging at 5\,GHz with CLEAN windows that excluded the cocoon area, but it was not possible to obtain a high dynamic range image this way: without including the cocoon emission, the rms noise in the final image increased from 0.9\,mJy/beam to 3.2\,mJy/beam and the dynamic range decreased from 1200 to 320. In addition, the fit to the closure phases became significantly worse when excluding the cocoon area: the reduced $\chi^2_\mathrm{CP}$ increased from 1.4 to 12.8. Therefore, we conclude that the cocoon structure is real.  To illustrate the differences between the two models, Figure~\ref{CocoonCP} shows an example of closure phases of two triangles together with model values from images with and without the cocoon structure. We note that the long space-baselines provided by \textit{RadioAstron} are essential for a robust identification of the mini-cocoon. The cocoon structure has a steep spectrum, as shown in Sect.~\ref{Sect:Spix}, and as a result, it is not visible in the images taken at 22\,GHz or higher frequencies. On the other hand, at 5\,GHz the ground-based VLBI imaging does not have high enough resolution\footnote{The east-west resolution of the ground-only image at 5\,GHz is 0.9\,mas, while the cocoon structure has an east-west thickness of $\sim 0.3$\,mas.} to robustly resolve the cocoon structure -- even though the overall size of the core region can be inferred from the ground-only data. 

Finally, there is low-intensity emission on the counter-jet side in the both 22\,GHz images and this emission is also likely to be real. The counter-jet side emission persisted in the CLEAN imaging even when we deliberately tried to self-calibrate it away. Also, the fit to closure phases becomes worse if the counter-jet side emission is excluded from the model --- the reduced $\chi^2_\mathrm{CP}$ increases from 1.3 to 5.2. There is some weak counter-jet-side emission visible $\lesssim 1$\,mas from the core also at 5\,GHz (see Appendix~\ref{appendix:eht} and Appendix~\ref{appendix:clean}), and this may be a part of the counter-jet-side mini-cocoon. However, beyond $\sim 1$\,mas north of the core, the cocoon in the counter-jet-side is invisible, which is likely due to the free-free absorbing screen that hides low-frequency emission from the counter-jet \citep{Walker2000}.

As mentioned earlier in this paper, the feature C3 likely presents a slowly moving head of the re-started jet that strongly interacts with the ambient medium. The earlier study of the motion and brightness evolution of the `hot spot' inside C3 has shown that it even became temporarily frustrated in 2016$-$2018 due to a collision with a dense cloud, after which it broke out \citep{Kino2021}. The hot spot in our 22\,GHz \emph{RadioAstron} image from 2013 has unusually high brightness temperature and this, together with its self-absorbed synchrotron spectrum, will be analysed in a companion paper (T.~Savolainen et al. in prep). We have already analysed the structure of the C1 and the collimation profile of the edge-brightened jet in the 22\,GHz image in \citet{Giovannini2018}. In the rest of this paper, we concentrate on the low-intensity cocoon-like emission around the jet that is visible in the 5\,GHz image, since this is (to our knowledge) the first time such a structure has been seen around a re-started (sub)parsec-scale jet. 

\subsection{Properties of the mini-cocoon}

\subsubsection{Size and surface brightness} \label{sect:cocoon_size}

\begin{figure*}
\centering
\includegraphics[width=1.0\textwidth]{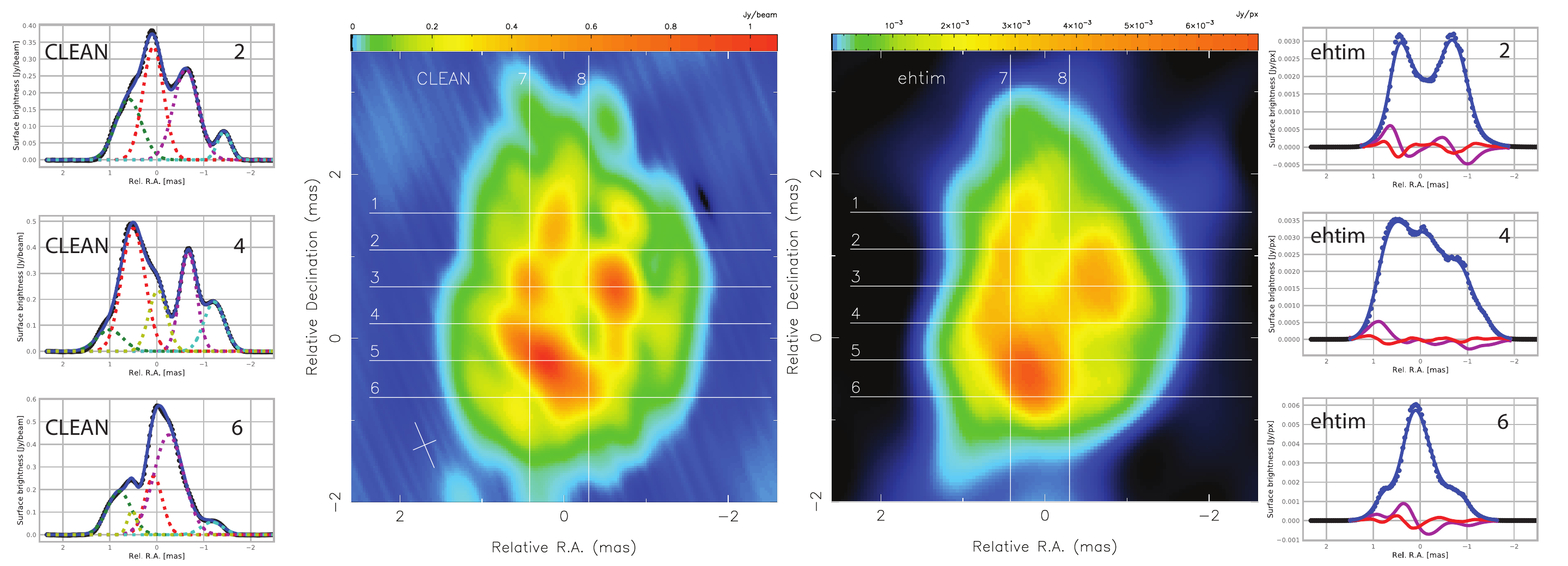}
\caption{Intensity cross-sections across the core region of 3C\,84 at 5\,GHz. The second and third panels from the left show the CLEAN image ($0.6 \times 0.3$\,mas at 23$^\circ$) and the \textsc{ehtim} (MEM=0, TV=1, TV2=1, L1=1; see Appendix~\ref{appendix:eht} for an explanation of the parameters) image, both in logarithmic colour scale with the analysed cross-sections overlaid. The left panel shows three examples of CLEAN image cross-sections with Gaussian fits. The right panel again shows three examples of \textsc{ehtim} cross-sections, where the black dots are the pixel intensities, the blue curve is the Gaussian-smoothed intensity cross-section, the magenta curve is the first derivative of the smoothed cross-section, and the red curve is the second derivative of the smoothed cross-section.}
\label{RA_slices_combined}
\end{figure*}

\begin{table}
\caption{Deconvolved FWHM width of the mini-cocoon using the 5\,GHz CLEAN images.}
\label{table:cocoon_size}
\centering
\begin{tabular}{cccc}
\hline\hline
\multicolumn{4}{c}{Slices along R.A.} \\ 
\hline
Slice\tablefootmark{a}   &  Rel. Dec. & Eastern shell    & Western shell  \\
        &  (mas)     & (mas)            & (mas) \\
\hline
1   & $+1.53$ & $0.39\pm0.02$  &  $0.14\pm0.05$ \\
2   & $+1.08$ & $0.37\pm0.10$  &  $0.11\pm0.03$ \\
3   & $+0.63$ & $0.29\pm0.08$  &  $0.16\pm0.05$ \\
4   & $+0.18$ & $0.20\pm0.03$  &  $0.35\pm0.04$ \\
5   & $-0.27$ & $0.37\pm0.09$  &  $0.20\pm0.05$ \\
6   & $-0.72$ & $0.64\pm0.08$\tablefootmark{b}  &  $0.62\pm0.06$\tablefootmark{b} \\
\hline
\multicolumn{4}{c}{Slices along Dec} \\ 
\hline
        &  Rel. R.A. & \multicolumn{2}{c}{Southern shell}  \\
        &  (mas)     & \multicolumn{2}{c}{(mas)}           \\
\hline
7   &  $+0.30$ & \multicolumn{2}{c}{$0.98\pm0.03$}  \\  
8   &  $-0.42$ & \multicolumn{2}{c}{$0.70\pm0.23$}  \\
\hline
\end{tabular}
\tablefoot{
\tablefoottext{a}{See Fig.~\ref{RA_slices_combined} for the slice positions.}
\tablefoottext{b}{Modelled with two Gaussians. Size refers to the separation of the peaks plus FWHM/2 of each Gaussian.}
}
\end{table}

\begin{table*}
\caption{Width of the mini-cocoon using the 5\,GHz \textsc{ehtim} images.}
\label{table:cocoon_size_ehtim}
\centering
\begin{tabular}{ccccc}
\hline\hline
\multicolumn{5}{c}{Slices along R.A.} \\ 
\hline
Slice\tablefootmark{a}   &  \multicolumn{2}{c}{Eastern shell}    & \multicolumn{2}{c}{Western shell}  \\
        &  $W_\mathrm{10}$ & $W_\mathrm{50}$ & $W_\mathrm{10}$ & $W_\mathrm{50}$ \\ 
        &  (mas) & (mas) & (mas) & (mas) \\
\hline
1   & $0.59\pm0.04$  & $0.23\pm0.04$  & $0.50\pm0.07$ & $0.17\pm0.04$ \\
2   & $0.41\pm0.04$  & $0.14\pm0.04$  & $0.47\pm0.07$ & $0.16\pm0.04$ \\
3   & $0.41\pm0.07$  & $0.16\pm0.06$  & $0.52\pm0.05$ & $0.21\pm0.05$ \\
4   & $0.34\pm0.04$  & $0.13\pm0.04$  & $0.55\pm0.04$ & $0.25\pm0.04$ \\
5   & $0.64\pm0.06$  & $0.34\pm0.05$  & $1.07\pm0.04$ & $0.62\pm0.04$ \\
6   & $0.69\pm0.04$  & $0.37\pm0.04$  & $0.97\pm0.08$ & $0.63\pm0.12$ \\
\hline
\multicolumn{5}{c}{Slices along Dec} \\ 
\hline
        &  \multicolumn{4}{c}{Southern shell}  \\
        &  \multicolumn{2}{c}{$W_\mathrm{10}$}  & \multicolumn{2}{c}{$W_\mathrm{50}$} \\
        &  \multicolumn{2}{c}{(mas)}  & \multicolumn{2}{c}{(mas)} \\
\hline
7   & \multicolumn{2}{c}{$0.83\pm0.11$} & \multicolumn{2}{c}{$0.51\pm0.11$} \\  
8   & \multicolumn{2}{c}{$1.11\pm0.07$} & \multicolumn{2}{c}{$0.41\pm0.06$} \\
\hline
\end{tabular}
\tablefoot{
\tablefoottext{a}{See Fig.~\ref{RA_slices_combined} for slice positions.}
}
\end{table*}

We interpret the low-intensity emission surrounding the core region in the 5\,GHz \emph{RadioAstron} image as a mini-cocoon that has developed around the new jet already in the first ten years after the re-start of the activity. In this interpretation, the cocoon radio emission is due to synchrotron radiation from particles (shock-)accelerated by the interaction between the jet and the ISM. As discussed by \citet{Nagai2017} and \citet{Kino2021}, the re-started jet in \object{3C\,84} is making its way through a dense and clumpy ISM present at the centre of the galaxy. The jet-ISM interaction transfers energy from the jet into the ISM and this happens not only via ram pressure acceleration of the gas clouds but also via heating of the gas swept-up by the shock front. This heated gas can form a cocoon-like structure around the jet and provide confining pressure affecting the jet geometry.

In order to confirm that this interpretation is correct and to investigate the physical properties of the cocoon and the surrounding medium, we need to quantify its size and energy content. To measure the cocoon width (i.e. thickness), we analysed eight intensity cross-sections across the central emission region, as shown in Fig.~\ref{RA_slices_combined}. We have carried out the analysis using different methods for CLEAN images and for images from \textsc{ehtim}, since the former are convolved with a Gaussian restoring beam, while the latter are not. In the case of CLEAN images, we have fitted the cross-sections with between four and six Gaussian components, using the routines from \textsc{astropy} Python library \citep{Astropy2013}, and deconvolved their FWHM width with the restoring beam size. The results for the eastern, western, and southern parts of the cocoon shell are given in Table~\ref{table:cocoon_size}. We carried out the analysis for three different restoring beams ($0.6 \times 0.3$\,mas at 23$^\circ$, $0.6 \times 0.3$\,mas at 0$^\circ$, and a circular 0.4\,mas) and we report the mean values in Table~\ref{table:cocoon_size} . The errors include the average uncertainty of the individual fits from the respective covariance matrices (which are clearly underestimated) and the standard error of the mean added in quadrature. The left side of Fig.~\ref{RA_slices_combined} shows three examples of these Gaussian fits.

The same cross-sections of the four \textsc{ehtim} images shown in the Appendix~\ref{appendix:eht} were analysed in the following way. First, the intensity cross-sections were smoothed with a Gaussian filter of 6 pixels in width (1 pixel in \textsc{ehtim} images has a size of 40\,$\mu$as) in order to remove spurious local extrema created by small-scale pixel-to-pixel variations\footnote{The selected size of the filter is a compromise between the resolution and the number of spurious local extrema left in the intensity curve. For filter sizes around 6\,px, the final results on the cocoon width depend only very weakly on the exact size of the filter.}. Then we calculated the second derivative of the smoothed curve and found its local maxima. These maxima can be used to identify the point where the steep intensity slope of the bright central part becomes flatter towards the edge of the slice due to the cocoon emission. We took this point as the inner edge of the cocoon. As the outer edge of the cocoon, we used the points at which the cocoon intensity drops to either 50\% or 10\% of its inner edge value. The corresponding widths, $W_\mathrm{50}$ and $W_\mathrm{10}$, are given in Table~\ref{table:cocoon_size_ehtim}. The values are averages over the four analysed images and the errors represent the standard error of the mean with the pixel size added in quadrature. 

The three measured cocoon widths (Gaussian FWHM, $W_\mathrm{50}$, and $W_\mathrm{10}$)  are differently defined and thus they differ (on average). The ratio between $W_\mathrm{50}$ and the Gaussian FWHM for the eastern shell is about 0.62, which is reasonably close to half as one would expect. On the other hand, the western shell is much thicker in the \textsc{ehtim} images than in the CLEAN images for cross-sections 5 and 6. The average ratio between $W_\mathrm{10}$ and the Gaussian FWHM for the eastern shell is 1.49, namely, the widths for the eastern shell are $W_\mathrm{50} <$ deconvolved single Gaussian FWHM $< W_\mathrm{10}$. In the following discussion, we mainly use the (deconvolved) Gaussian FWHM as the shell thickness.

The average thickness is 0.38\,mas or 0.13\,pc for the eastern shell and 0.26\,mas or 0.09\,pc for the western shell. The total (deconvolved) width of the core region estimated from the CLEAN image cross-sections ranges from 2.3\,mas (slice 1) to 2.7\,mas (slice 4), corresponding to 0.8$-$0.9\,pc. We use the distance between the two outermost Gaussians in the fit plus their deconvolved FWHM/2 values as the total width of the core region. Now, we can assume that the cocoon is axisymmetric and calculate its volume assuming that the overall thickness for every cross-section corresponds to the average thickness of the eastern and western shells. The distance along the jet depends on the assumed viewing angle, $\theta$, hence, we give the volume as a function of $\theta$. Summing over the volume elements that correspond to the individual slices gives a total volume of the shell, $V_\mathrm{c,FWHM} = \frac{0.35\pm0.03}{\sin{\theta}}$\,pc$^3$. Carrying out the same exercise for $W_\mathrm{10}$ widths from the \textsc{ehtim} images results in $V_\mathrm{c, W_\mathrm{10}} = \frac{0.40\pm0.03}{\sin{\theta}}$\,pc$^3$. Here we see that the estimate of the total volume of the cocoon depends only weakly on the imaging algorithm used to create the \emph{RadioAstron} 5\,GHz images. We take $V_\mathrm{c,min} = \frac{0.4}{\sin{\theta}}$\,pc$^3$ as a lower limit to the cocoon volume, since the above measurements exclude, for example, the volume between the jet and the feature C2. An upper limit to the cocoon volume can be obtained by considering the total (deconvolved) width of the slices and summing over these volume elements. The resulting upper limit is $V_\mathrm{c,max} = \frac{0.6}{\sin{\theta}}$\,pc$^3$.

We also want to calculate the line-of-sight thickness of the cocoon. At the centre of the core region it can be roughly estimated as the average thickness of the eastern and western shells: $\Delta l_\mathrm{c,centre} \approx \frac{0.11}{\sin \theta}$\,pc. In the middle of the eastern or western shell, the line-of-sight thickness can be calculated as:
\begin{equation}
    \Delta l_\mathrm{c,limb} = 2 \frac{ r + \Delta r }{\sin \theta} \sin \bigg( \cos ^{-1} \bigg( \frac{r + \Delta r / 2}{r + \Delta r} \bigg) \bigg),
\end{equation}
where $r$ is the inner radius of the cocoon cylinder and $\Delta r$ is the thickness of the cylinder. Taking $\Delta r \approx 0.11$\,pc and the average $\langle r \rangle \approx 0.32$\,pc, we get $\langle \Delta l_\mathrm{c,limb} \rangle \approx \frac{0.42}{\sin \theta}$\,pc. 

The average surface brightness of the cocoon was measured from the CLEAN image to be $114\pm7$\,mJy/beam and the corresponding average brightness temperature is $\sim 3 \times 10^{10}$\,K. The \textsc{ehtim} images give similar average values of $\sim$440\,mJy/mas$^2$ and $T_\mathrm{c,B} \sim 2 \times 10^{10}$\,K. These brightness temperatures as well as the steep spectrum of the cocoon (see Sect.~\ref{Sect:Spix}) indicate that the cocoon emission is of synchrotron origin. The total measured flux of the cocoon shell, $F_\mathrm{c,limbs} = 2.5\pm0.2$\,Jy at 5\,GHz, which can be considered as the lower limit, since it does not take into account the cocoon emission from the central part of the core region where it is superimposed on the emission from the jet and the hot spots. However, we can make a crude estimation of this emission by using the ratio of line-of-sight path lengths through the centre of the cocoon and through the limbs together with the ratio of the corresponding surface areas. The resulting $F_\mathrm{c,centre} \approx 2.5\mathrm{Jy} \cdot (0.11\mathrm{pc}/0.42\mathrm{pc}) \cdot (0.62\mathrm{pc^2}/0.43\mathrm{pc^2}) = 0.9$\,Jy and the total flux from the cocoon would therefore be roughly 3.4\,Jy at 5\,GHz. 

\subsubsection{Spectral index} \label{Sect:Spix}

\begin{figure}
\centering 
\includegraphics[angle=-90,width=0.49\textwidth]{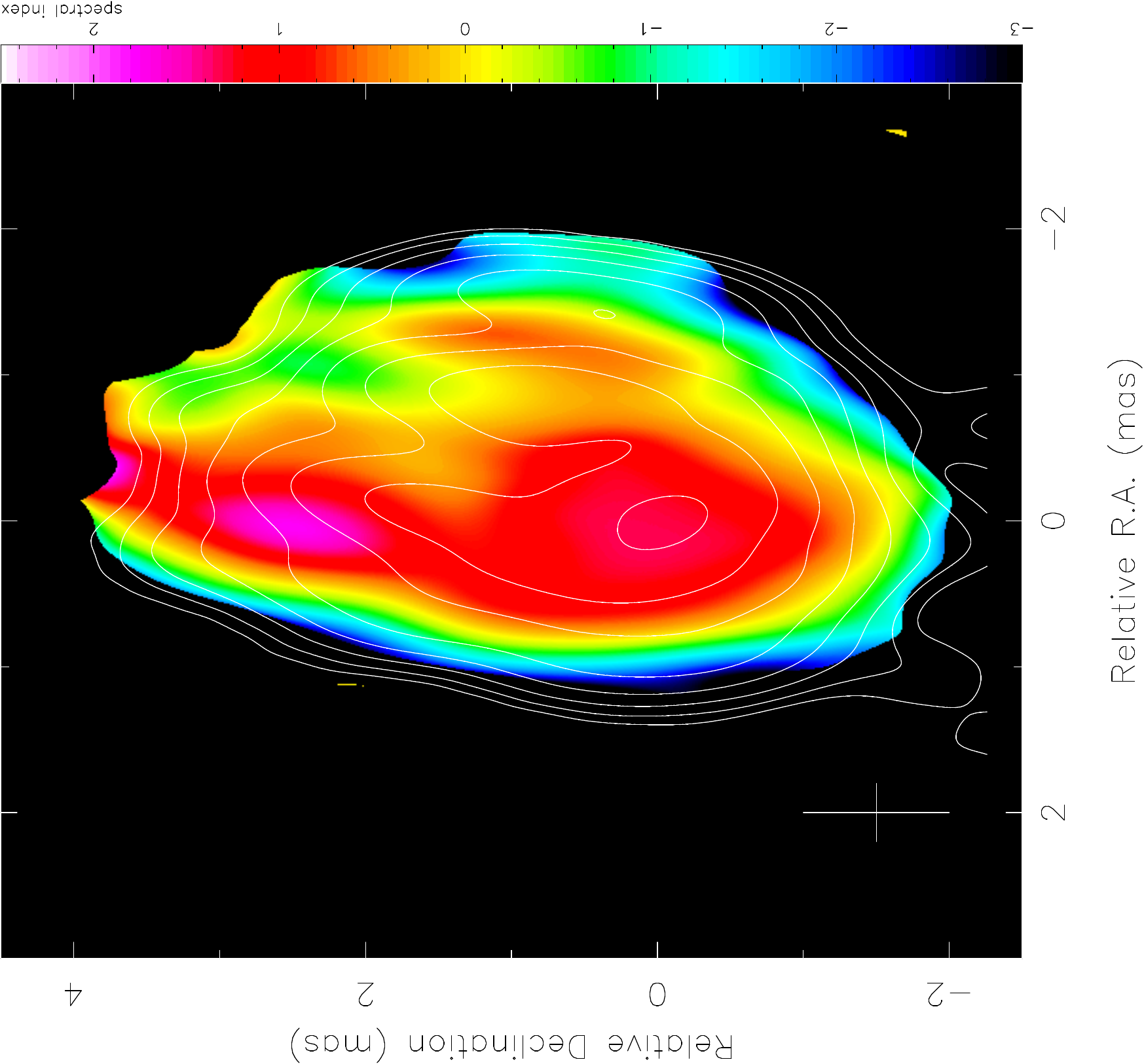}
\caption{Spectral index ($\alpha$ defined as $I \propto \nu^{+\alpha}$) image of 3C\,84 between 15 and 5\,GHz. Contours show the total intensity at 5\,GHz. Restoring beam size is $1.0 \times 0.4$\,mas at PA=0$^\circ$.}
\label{spectraCU}
\end{figure}

In Sect.~\ref{sec:discuss}, we use the  measured volume and surface brightness of the cocoon to estimate the energy content of cocoon in the minimum energy conditions. Yet before we can do so, we need an estimate of the cocoon spectrum between 5 and 15\,GHz. In order to analyse the spectrum, we aligned the images taken at 5 and 15\,GHz by using a 2D cross-correlation of the jet features excluding C1 \citep[e.g.][]{Pushkarev2012}. This extra alignment step is necessary, since the phase self-calibration removes the absolute position information. The image shifts from the cross-correlation are $x_{15} - x_5 = -0.14$\,mas in R.A. and $y_{15} - y_5 = 0.27$\,mas in Dec.  We considered also an alternative means for alignment by matching the peak of C3 using both Gaussian fitting and the pixel of maximum brightness. The differences in the alignment between different methods are less than 0.03\,mas. The fact that it is not only the core, but also C2, C3, and the jet that are all optically thick at these frequencies may create biases for all the above alignment methods that rely on the assumption of downstream jet features having frequency-independent positions. Luckily, the spectrum of the cocoon mainly depends on the alignment in the direction transverse to the jet which is typically less affected by spectral index gradients than the alignment along the jet.

Figure~\ref{spectraCU} shows the 15$-$5\,GHz spectral index image of the inner parsec formed after the alignment. In order to match the spatial frequencies sampled at different observing frequencies, data at short ($u,v$) spacings were clipped to a matching spatial frequency and tapering was used on the long ($u,v$) spacings. The image pairs were also convolved with a common restoring beam of $1.0 \times 0.4$\,mas at a PA of 0$^\circ$. The spectral index image shows that C1, C2, and C3 are indeed all optically thick between 5 and 15\,GHz. The mini-cocoon, on the other hand, has a very steep spectrum. The average spectral index $\alpha_\mathrm{5-15} \approx -1.0$ for the western shell and $\alpha_\mathrm{5-15} < -1.0$ for the eastern shell. The full spectra of features C1, C2, and C3 covering 5$-$43\,GHz are discussed elsewhere (T.~Savolainen et al., in prep).


\section{Discussion} \label{sec:discuss}

As pointed out previously by a number of authors, C3 appears to be the head of a radio mini-lobe, namely, a working surface between the new jet ejected in 2003 and the ambient medium \citep[e.g.][]{Nagai2010, Nagai2014, Fujita2017, Hiura2018, Giovannini2018, Kino2018, Kino2021}. Interestingly, the ambient medium that C3 is expanding towards is the funnel (or cocoon) that was created by the earlier radio mini-lobes that are currently located at $\lesssim 20$\,mas --- or $\lesssim 7$\,pc (projected) from the core and ejected by the previous outburst that started in 1959 \citep{Nesterov1995, Walker2000, Asada2006, Fujita2016}. Such a funnel could be filled with very hot, low-density gas heated by the termination shock of the jet related to the 1959 outburst. 

The jet--external medium interaction taking place at C3 is complex; for example, the hot spot in C3 is moving non-linearly and possibly even jumping around inside the jet head \citep{Kino2018,Kino2021}. It has been suggested that this behaviour, together with the high Faraday rotation measure observed in C3, is due to the jet interacting with clouds having high electron density of $(2-60) \times 10^4$\,cm$^{-3}$ \citep{Nagai2017,Kino2021}. This would imply a clumpy medium inside the funnel and we are potentially witnessing a restarted jet interacting with a non-homogeneous, multi-phase medium. 

Below, we estimate the physical conditions in the new mini-cocoon and discuss the power requirements to inflate it, as well as the densities of the different media inside the central $\sim 5$\,pc (de-projected) of 3C\,84.  

\subsection{Energy content of the mini-cocoon}

Assuming the minimum energy condition, which corresponds to a near-equipartion between magnetic and particle energies, we can use the synchrotron theory to derive a lower limit to the energy stored in the mini-cocoon \citep{Pacholcyk1970}:
\begin{equation}
    U_\mathrm{c,min} = 0.541 \, c_{13} (1+k)^{4/7} V_\mathrm{c}^{3/7} L_\mathrm{c}^{4/7} \quad [\mathrm{erg\,s^{-1}}],
\end{equation}
where $V_\mathrm{c}$ is the volume of the mini-cocoon in cm$^3$, $L_\mathrm{c}$ is the total synchrotron luminosity of the mini-cocoon in erg\,s$^{-1}$, emitted between frequencies $\nu_1$ and $\nu_2$, and $k$ is the energy ratio between heavy particles and relativistic electrons. The factor $c_{13}$ depends on $\alpha$, $\nu_1$, and $\nu_2$ and it is given in \citet{Pacholcyk1970}. Since the cocoon spectrum is steep, both $L_\mathrm{c}$ and $U_\mathrm{c,min}$ depend only weakly on $\nu_2$ and we can safely assume here that $\nu_2 = 10^{11}$\,Hz. The low-frequency cut-off has a more significant impact, therefore we calculate the minimum energy for different values of $\nu_1$. Using $F_\mathrm{c} = 3.4$\,Jy and $\alpha = -1$, we get $L_\mathrm{c} = 8 \times 10^{41}$\,erg\,s$^{-1}$ and $U_\mathrm{c,min} = (3 - 4) \times 10^{51} (1+k)^{4/7} \sin^{-3/7}\theta$\;erg for $\nu_1 = 10^8$\,Hz. The range corresponds to the minimum and maximum cocoon volume (see Sect.~\ref{sect:cocoon_size}). The corresponding minimum energy density is $u_\mathrm{c,min} = (2 - 3) \times 10^{-4} (1+k)^{4/7} \sin^{4/7}\theta$\;erg\,cm$^{-3}$ and the equipartition magnetic flux density is $B_\mathrm{c,eq} = 6 \times 10^{-2} (1+k)^{2/7} \sin^{2/7}\theta$\;G. Table~\ref{tab:cocoon_phys} lists $L_\mathrm{c}$, $B_\mathrm{c,eq}$, $u_\mathrm{c,min}$, and $U_\mathrm{c,min}$ for other values of $\nu_1$.

Given the brightness temperature of the mini-cocoon in excess of $10^{10}$\,K, it is reasonable to assume an ultra-relativistic equation-of-state for the cocoon gas, namely, the ratio of specific heats $\Gamma_\mathrm{c} = 4/3$, which results in the following pressure: 
\begin{equation}
    p_\mathrm{c} = (\Gamma_\mathrm{c} - 1) u_\mathrm{c} = \frac{1}{3} u_\mathrm{c} \quad \mathrm{[dyn\,cm^{-2}]}.
\end{equation}
The minimum pressure of the cocoon is therefore $p_\mathrm{c,min} \approx 1 \times 10^{-4} (1+k)^{4/7} \sin^{4/7}\theta$\;dyn\,cm$^{-2}$ for $\nu_1 = 10^8$\,Hz.

\begin{table*}
\caption{Energy content of the mini-cocoon in the minimum energy condition}
\label{tab:cocoon_phys}
\centering
\begin{tabular}{ccccc}
\hline\hline
$\nu_1$ &   $L_\mathrm{c}$   &   $B_\mathrm{c,eq}$  &  $u_\mathrm{c,min}$ &  $U_\mathrm{c,min}$ \\
  (Hz)  &  (erg s$^{-1}$) & $(1+k)^{2/7} \sin^{2/7}\theta$ (G)  & $(1+k)^{4/7} \sin^{4/7}\theta$ (erg cm$^{-3}$) &  $(1+k)^{4/7} \sin^{-3/7}\theta$ (erg) \\
\hline
$10^7$  & $1 \times 10^{42}$ & $(8 - 9) \times 10^{-2}$ & $(4 - 5) \times 10^{-4}$ & $(6 - 8) \times 10^{51}$ \\
$10^8$  & $8 \times 10^{41}$ & $6 \times 10^{-2}$ & $(2 - 3) \times 10^{-4}$ & $(3 - 4) \times 10^{51}$ \\
$10^9$  & $6 \times 10^{41}$ & $(4 - 5) \times 10^{-2}$ & $(1 - 2) \times 10^{-4}$ & $2 \times 10^{51}$ \\
\hline
\end{tabular}
\end{table*}

\subsubsection{Expansion speed and densities inside and outside of the cocoon} \label{sec:densities}

Since the advance of the jet head C3 has been followed with the high-resolution ground-based VLBI through several studies, there is a good estimate of its speed and we also know that this new jet was born before November 2003 \citep{Suzuki2012}. We can use the age of C3, $\Delta t \approx 10$\,yr to estimate the transverse expansion speed of the cocoon if we assume that the cocoon is filled by gas heated by the interaction taking place at C3. Estimation of the expansion speed is complicated by the fact that the distance from the jet to the western shell is greater than to the eastern shell. Hence, the cocoon appears to be expanding faster in the western side, which indicates a potentially denser environment in the eastern side than in the western side. This is understandable, since the fading component, C2, in the western side indicates that the previous jet activity may have evacuated the region west of the current jet, which would allow the cocoon to expand more freely in that direction. A lower limit to the average transverse expansion speed can be calculated from the thickness of the eastern shell, $v_\mathrm{c,\perp} \gtrsim 0.13\,\mathrm{pc} / 10\,\mathrm{yr} = 0.04$\,$c$, and the upper limit from the total radius of the cocoon, $v_\mathrm{c,\perp} \lesssim 0.45\,\mathrm{pc} / 10\,\mathrm{yr} = 0.15$\,$c$. Hence, the overpressured gas in the cocoon expands sideways roughly at a speed of 0.1\,$c$. 

The average apparent speed of the jet head C3 was reported to be $0.23\pm0.06$\,$c$ in 2003$-$2007 from 43\,GHz VLBA data \citep{Suzuki2012} and $0.27\pm0.02$\,$c$ in 2007$-$2013 from 22\,GHz VERA data \citep{Hiura2018}. Furthermore, \citet{Suzuki2012}  found the C3 feature to accelerate from an apparent speed of 0.10\,$c$ to 0.47\,$c$ between 2003 and 2008. This acceleration seems to be at odds with the slower speed reported by \citet{Hiura2018}, but it is possible that the discrepancy is due to different angular resolutions between the VLBA at 43\,GHz and the VERA at 22\,GHz --- the VLBA kinematics likely correspond to the hotspot inside C3 while VERA follows the overall movement of C3. Here, we adopt an average apparent speed of 0.25\,$c$, which corresponds to the physical speed of the jet head $v_\mathrm{h} = 0.46$\,$c$ for $\theta = 18^\circ$, and $v_\mathrm{h} = 0.28$\,$c$ for $\theta = 45^\circ$. Hence, the ratio $v_\mathrm{h} / v_\mathrm{c,\perp}$ is between 2 and 12, namely, $v_\mathrm{h} > v_\mathrm{c,\perp}$ as required by self-consistency. 

Equating the cocoon pressure with the ram pressure of the ambient medium, 
\begin{equation}
    p_\mathrm{c} = \rho_\mathrm{a} v_\mathrm{c,\perp}^2, \label{eq:ram_pres_cocoon}
\end{equation}
where $\rho_\mathrm{a}$ is the density of the ambient medium, and using $p_\mathrm{c} \gtrsim p_\mathrm{c,min}$ and $v_\mathrm{c,\perp} \lesssim 0.15$\,$c$, we get an estimate of the minimum $\rho_\mathrm{a} \gtrsim 5 \times 10^{-24} \sin^{4/7} \theta \, (1+k)^{4/7}$\,g\,cm$^{-3}$. For $\theta = 18^\circ$ and assuming normal plasma, this translates to a number density of $n_\mathrm{a} \gtrsim 1.5\,(1+k)^{4/7}$\,cm$^{-3}$. In the previous calculation, we ignored the thermal mass-energy of the ambient medium. \citet{Fujita2017} estimated the ambient gas density in the direction of the motion of C3 to be $\sim 8$\,cm$^{-3}$, if ($2\times$) the jet power equals the average jet power measured from the X-ray cavities. Their density estimate is consistent with our lower limit, if $k \lesssim 20$. Our lower limit for $n_\mathrm{a}$ is also larger than the upper limit to the electron number density in the jet, $n_{e,\mathrm{j}} \leq 0.6$\,cm$^{-3}$, estimated from the Faraday rotation in mm-wavelength core \citep{Kim2019}. Thus, the jet is lighter than the ambient medium. 

If we indeed assume that $\rho_\mathrm{a}$ is same in the direction of the jet head's motion as it is in the direction of the sideways expansion of the mini-cocoon, we can use Eq.~\ref{eq:ram_pres_cocoon} together with the equation for the pressure balance at the jet head,
\begin{equation}
    \rho_\mathrm{a} v_\mathrm{h}^2 = \frac{P_\mathrm{j}}{v_\mathrm{j} A_\mathrm{h}},
\end{equation}
where $P_\mathrm{j}$ is the jet power, $v_\mathrm{j} \approx c$ is the jet speed, and $A_\mathrm{h}$ is the cross-sectional area of the jet head, to obtain the following estimate of the jet power:
\begin{equation} \label{eq:pressure_balance}
    P_\mathrm{j} = p_\mathrm{c} A_\mathrm{h} c \bigg( \frac{v_\mathrm{h}}{v_\mathrm{c,\perp}} \bigg)^2.
\end{equation}
From the space-VLBI images we estimate the total width of the jet head to be about $1.2$\,mas, which results in $A_\mathrm{h} = 0.13$\,pc$^2$. Using again $p_\mathrm{c} \gtrsim p_\mathrm{c,min}$ and assuming $\theta = 18^\circ$, we get a lower limit to the jet power, $P_\mathrm{j} \gtrsim 2 \times 10^{43} \, (1+k)^{4/7}$\,erg\,s$^{-1}$. This lower limit can be compared with the long-term average jet power of 3C\,84 measured from the X-ray cavities in Perseus~A, 
$P_\mathrm{cav} = 1.5^{+1.0}_{-0.3} \times 10^{44}$\,erg\,s$^{-1}$ \citep{Rafferty2006}. Taking $2P_\mathrm{j} = P_\mathrm{cav}$, our lower limit is compatible with the upper limit of the long-term jet power if $k \lesssim 30$. We note that Eq.~\ref{eq:pressure_balance} assumes that all the jet power is used to advance the hot spot. We relax this requirement in the calculations presented in Sect.~\ref{sec:power_req}. 

If we adopt $k = (m_\mathrm{+} \langle \gamma_\mathrm{+} \rangle ) / (m_e \langle \gamma \rangle ) \lesssim 30$, where $m_\mathrm{+} = 1836\,m_e$ and $\langle \gamma_\mathrm{+} \rangle \approx 1$ for the electron-proton plasma with cold protons, we can place a lower limit to the mean Lorentz factor of the relativistic electrons in the case of normal plasma, $\langle \gamma \rangle = \gamma_\mathrm{min} \ln{(\gamma_\mathrm{max}/\gamma_\mathrm{min})} \gtrsim 60$. This suggests a minimum electron Lorentz factor $\gamma_\mathrm{min} \gtrsim 10$ in the cocoon. Using $B_\mathrm{c,eq}$ from Table~\ref{tab:cocoon_phys}, this gives a lower frequency cut-of of $\nu_1 = 2.8 \times 10^6 B \gamma_\mathrm{min}^2 \gtrsim 3 \times 10^7$\,Hz, thus confirming that the values of $\nu_1$ used in our minimum energy condition estimates are reasonable. Smaller values of $k$ would indicate larger $\langle \gamma \rangle$ and consequently larger $\gamma_\mathrm{min}$. It is, of course, possible that the current jet power exceeds the long-term average, which would allow larger values of $k$. However, as we see in Sect.~\ref{sec:synchro_lifetime}, also other arguments point to $k$ being on the order of a few tens.

What about the density of the cocoon? \citet{Nagai2017} observed a rotation measure of $6 \times 10^5$\,rad\,m$^{-2}$ in a hot spot located at the edge of C3 in 2015$-$2016 using the VLBA. Rotation measure depends on the number density of thermal electrons, $n_e$, the line-of-sight component of the magnetic field, $B_\parallel$, and the path length through the Faraday rotating plasma from the source to observer:
\begin{equation}
    \mathrm{RM} = 8.1 \times 10^5 \int n_e B_\parallel dl \quad \mathrm{[rad\,m^{-2}]},
\end{equation}
where $n_e$ is in cm$^{-3}$, $B_\parallel$ is in G, and $dl$ is in pc. We do not know how much of the Faraday rotation takes place in the cocoon, but we can use the measured RM, $B_\parallel \approx B_\mathrm{c,eq}/\sqrt{3}$ and $dl \approx \Delta l_\mathrm{c,centre}$ to place an upper limit to the number density of thermal electrons in the cocoon:
\begin{equation}
    n_{e,\mathrm{c}} \lesssim 2.1 \times 10^{-6} \frac{\mathrm{RM}}{B_\mathrm{c,eq} \Delta l_\mathrm{c,centre}} \quad \mathrm{[cm^{-3}]}.
\end{equation}
We get an upper limit of $n_{e,\mathrm{c}} \lesssim 1.9 \times 10^2 \sin ^{5/7} \theta (1+k)^{-2/7}$\,cm$^{-3}$, or for $\theta = 18^\circ$, $n_{e,\mathrm{c}} \lesssim 80 (1+k)^{-2/7}$\,cm$^{-3}$. 

\subsection{Considering whether the mini-cocoon may be inflated by the recently re-started jet activity}

\subsubsection{Power requirements} \label{sec:power_req}

A simple sanity test for our interpretation of the low-surface-brightness emission around the jet in the 5\,GHz \emph{RadioAstron} image as a mini-cocoon inflated by the increased jet activity since 2003 is to estimate the jet power required to inflate the cocoon. Using the energy conservation for the cocoon and the jet head and assuming constant pressure,
\begin{equation}
    dU_\mathrm{c} \approx P_\mathrm{j} dt - dU_\mathrm{h} - p_\mathrm{h} dV_\mathrm{h} - p_\mathrm{c} dV_\mathrm{c} - dW_\mathrm{h},
\end{equation}
where $P_\mathrm{j}$ is the jet power, $dU_\mathrm{h}$ is the change in the internal energy stored in the jet head, $p_\mathrm{h}$ is the pressure in the jet head, $dV_\mathrm{h}$ is the change in the volume of the jet head, and $W_\mathrm{h}$ is the work done by the jet to advance the jet head. Here, we neglect  radiation losses.  Applying again the minimum energy conditions, we get a rough estimate for the required jet power:
\begin{equation} \label{eq:power_req}
    P_\mathrm{j} \gtrsim \frac{U_\mathrm{c} + p_\mathrm{c} V_\mathrm{c} + U_\mathrm{h} + p_\mathrm{h} V_\mathrm{h} + W_\mathrm{h}}{\Delta t} = \frac{4/3 (U_\mathrm{c} + U_\mathrm{h}) + W_\mathrm{h}}{\Delta t},
\end{equation}
where $\Delta t \approx 10$\,yr is the time since the ejection of C3. The work $W_\mathrm{h} = F_\mathrm{j} r_\mathrm{h}$, where $F_\mathrm{j}$ is the thrust that can be estimated from the pressure balance (assuming constant ambient density)
\begin{equation}
    \frac{F_\mathrm{j}}{A_\mathrm{h}} \simeq \rho_\mathrm{a} v_\mathrm{h}^2 \simeq p_\mathrm{c} \bigg( \frac{v_\mathrm{h}}{v_\mathrm{c,\perp}} \bigg) ^2    
\end{equation}
and $r_\mathrm{h}$ is distance of the hot spot from the core, $r_\mathrm{h} = 0.76  \sin ^{-1} \theta$\,pc. We take $V_\mathrm{h} = 0.035 \sin ^{-1} \theta$\,pc$^{3}$. Using the 5$-$43\,GHz images, we estimate the integrated synchrotron luminosity $L_\mathrm{h} \sim 6 \times 10^{42}$\,erg\,s$^{-1}$ between $10^8$\,Hz and $2 \times 10^{11}$\,Hz. This gives a minimum energy stored in the jet head $U_\mathrm{h} \sim 2 \times 10^{51} (1+k)^{4/7} \sin ^{-3/7} \theta$\,erg. Assuming $\theta = 18^\circ$, this is $U_\mathrm{h} \sim 3 \times 10^{51} (1+k)^{4/7}$\,erg. For $\theta = 18^\circ$, $v_\mathrm{h}/v_\mathrm{c,\perp} > 0.46c / 0.15c = 3$, and the lower limit to the work done to advance the jet head is $W_\mathrm{h} \gtrsim 4 \times 10^{51} (1+k)^{4/7}$\,erg. The resulting requirement for the average jet power is $P_\mathrm{j} \gtrsim 5 \times 10^{43} (1+k)^{4/7}$\,erg\,s$^{-1}$. If the jet viewing angle is $\theta = 45^\circ$, the same calculations give $P_\mathrm{j} \gtrsim 3 \times 10^{43} (1+k)^{4/7}$\,erg\,s$^{-1}$.   

The above lower limit can be again compared with the long-term average jet power from the cavities. We find that the power required to inflate the mini-cocoon in ten years is compatible with the upper limit of the long-term average jet power if $k \lesssim 5$ $(\theta = 18^\circ)$ or $k \lesssim 15$ $(\theta = 45^\circ)$. It is interesting to note that in these very early stages of the jet evolution, the amount of energy dumped in the cocoon, that is, its enthalpy $\Gamma_\mathrm{c}/(\Gamma_\mathrm{c}-1) p_\mathrm{c} V_\mathrm{c} = 4/3 U_\mathrm{c}$, is about half of the energy delivered by the jet in Eq.~\ref{eq:power_req} and the fraction that goes to driving the shock in to the ambient medium is about 0.11$-$0.14.

\subsubsection{An alternative jet power estimate}

\citet{Marti1997} showed that the 1D advance speed of the jet head can be related to the jet speed by the following expression:
  \begin{equation}
   v_\mathrm{h,1D} = \frac{\sqrt{\eta}}{1+\sqrt{\eta}}\,v_\mathrm{j}, \label{eq:v_1D}
\end{equation}
where $\eta$ is the relativistic enthalpy ratio $\rho_\mathrm{j} h_\mathrm{j} / \rho_\mathrm{a} h_\mathrm{a}$ between the jet and the ambient medium. If we take the reasonable assumptions $h_\mathrm{a} \simeq c^2$ (i.e. the internal energy of the ambient medium is negligible as compared to its rest-mass energy) and $v_\mathrm{j} \simeq c$, considering the case of $\theta = 18^\circ$ and $v_\mathrm{h} \simeq 0.46\,c$, taking into account that $v_\mathrm{h} \leq v_\mathrm{h,1D}$, we find that $\eta \geq 0.85$. Thus, we can safely consider that $\eta \sim 1$, namely, $\rho_\mathrm{j} h_\mathrm{j} \simeq \rho_\mathrm{a} c^2$. 
 
The power of a relativistic jet can be expressed as:
\begin{equation}
P_\mathrm{j} = \left( \rho_\mathrm{j} \gamma_\mathrm{j} (h_\mathrm{j} \gamma_\mathrm{j} - 1)  + \frac{B^2}{4\pi} \right)\,v_\mathrm{j}\,\pi\,R_\mathrm{j}^2, 
\end{equation}
where $\gamma_\mathrm{j}$ is the jet Lorentz factor, $B$ is its magnetic field, and $R_\mathrm{j}$ is the jet radius. Using $\rho_\mathrm{j} h_\mathrm{j} \simeq \rho_\mathrm{a} c^2$, $v_\mathrm{j} \simeq c$, assuming that the jet is relativistic ($h_\mathrm{j} \gamma_\mathrm{j} >>1$), and adding the estimated value for the equipartition magnetic field and ambient density, we find $P_\mathrm{j} \simeq (6.75 \times 10^7 \gamma_\mathrm{j}^2 + 4.4 \times 10^6) (1+k)^{4/7} \pi R_\mathrm{j}^2$. Taking into account that $\gamma_\mathrm{j} > 1$, we find that the jet is probably particle-dominated because the first term in the bracket (standing for the particle kinetic plus internal energy) evidently becomes  larger than the second (Poynting flux). By neglecting the Poynting flux term in the equation, and taking $R_\mathrm{j} \simeq 0.2\,{\rm mas} \simeq 0.07 \,{\rm pc}$, we can compare the total jet energy flux with that estimated in the previous section, to find $9.35 \times 10^{42} \gamma_\mathrm{j}^2 \geq 4 \times 10^{43}$, which gives $\gamma_\mathrm{j} \geq 2.2$. This result is consistent with previous assumptions, but the exact value of the jet power remains uncertain, as a function of the Lorentz factor. 

It is worth discussing the implications of the enthalpy ratio being close to 1. If $\rho_\mathrm{j} h_\mathrm{j} \simeq \rho_\mathrm{a} c^2$, $h_\mathrm{j} \simeq \rho_\mathrm{a} / \rho_\mathrm{j} c^2$, which might be $\gg c^2$ -- considering that results in Sect.~\ref{sec:densities} indicate  $n_\mathrm{a}/n_{e,\mathrm{j}} > 2.5 (1+k)^{4/7}$ and the jet plasma can in principle include a significant electron-positron component further increasing $\rho_\mathrm{a} / \rho_\mathrm{j}$. Therefore, the jet is probably formed by hot plasma.

\subsubsection{Synchrotron-loss time scale} \label{sec:synchro_lifetime}

Assuming the magnetic flux density that minimises the total energy in the mini-cocoon, we can calculate the synchrotron half-lifetime of the cocoon electrons radiating at the different observing frequencies $\nu$: 
\begin{equation}
    t_\mathrm{1/2} = 1.04 \, B^{-3/2} \nu^{-1/2} \quad \mathrm{[yr]}, 
\end{equation}
where $B$ is the cocoon magnetic field in gauss and $\nu$ is in GHz. Now, assuming $B \sim B_\mathrm{c,eq}$, $\nu_1 = 10^8$\,Hz and $\theta = 18^\circ$, we have $t_\mathrm{1/2} \sim 50\,(1+k)^{-3/7}$\,yr at 5\,GHz, $t_\mathrm{1/2} \sim 30\,(1+k)^{-3/7}$\,yr at 15\,GHz, and $t_\mathrm{1/2} \sim 25\,(1+k)^{-3/7}$\,yr at 22\,GHz. If the viewing angle is $\theta = 45^\circ$, all the half-lifetimes are 30\% shorter. Again, if $k$ is a few tens, one can easily understand why the mini-cocoon is visible at 5\,GHz, but not at 22\,GHz -- the synchrotron lifetime is too short for the higher frequency emitting electrons to fill the whole cocoon.

\subsubsection{Mini-cocoon and the cylindrical shape of the jet}

The edge-brightened jet in the 22\,GHz image has an almost cylindrical shape \citep{Giovannini2018}, which  significantly differs from the parabolic one seen in M\,87 \citep{Asada2012}. This collimation profile requires a very shallow pressure profile of the ambient medium, $p_\mathrm{a} \propto z^{-b}$ with $b \leq 0.7$, and the observed mini-cocoon can provide a nearly constant collimating pressure for the jet. This is a well-known effect in large scale jets, where the cocoon pressure can re-collimate the jet before it enters the termination hot spot \citep{Komissarov1998}. Although \citet{Komissarov1998} analyse a kpc scale structure, we use their Eq.~47 to make a crude estimate of the distance at which the re-collimation shock reaches the jet axis, $z_\mathrm{r} \lesssim 2.5 (1+k)^{-2/7}$\,pc. Here we have assumed that the jet power is equal to the long-term average, $p_\mathrm{c} \gtrsim p_\mathrm{c,min}$, and $\theta = 18^\circ$. If $k=20$, $z_\mathrm{r} \lesssim 1$\,pc and an initially conical jet can expand only up to about half this distance. With $\theta = 18^\circ$, this corresponds to $\lesssim 0.4$\,mas in the images. This is consistent with the rapid collimation at $z \sim 0.1$\,mas reported by \citet{Giovannini2018}. However, we do not see clear signs of a re-collimation shock in the 22\,GHz image.

An alternative explanation, suggested, for example, by \citet{Blandford2019}, could be that the sub-parsec-scale jet in 3C\,84 is collimated by a magnetised sheath. We, however, consider the cocoon pressure to be a more natural explanation, since a) there \emph{is} evidence for hot plasma in the cocoon, and b) there is also cocoon emission in front of the C3 hot spot -- not only around the jet.

\subsection{Numerical simulations of jet propagation in an inhomogeneous medium}

Since the interaction between the jet and the external medium in 3C\,84 is clearly quite complex, it is interesting to turn to numerical simulations in the hope of gaining more insight in the physical processes involved. We have run a numerical simulation based on a new version of the code \emph{Ratpenat} \citep{Perucho2010}, which includes hydrogen ionisation and recombination physics \citep[based on][]{Vaidya2015}. The first results obtained with this code have been presented in \citet{Perucho2021}. This simulation was run during $\sim 36~{\rm hours}$ in 1024 cores, at \emph{Tirant}, the local supercomputer at the University of Val\`encia. The jet is injected in the grid with a radius of 0.07~pc, velocity $v_\mathrm{j}\,=\,0.98\,c$, and temperature $2.4\times10^{10}\,{\rm K}$ (jet specific enthalpy is $h_\mathrm{j}\,=\,12\,c^2$), which results in a kinetic power $L_\mathrm{j}\,=\,3\times10^{44}\,{\rm erg\, s^{-1}}$. The grid is set up with an inhomogeneous ambient medium with mean density of $15\,m_\mathrm{p}\,{\rm cm}^{-3}$, with a density distribution ranging from $3000\,m_\mathrm{p}\,{\rm cm}^{-3}$ to $0.01\,m_\mathrm{p}\,{\rm cm}^{-3}$, and temperatures ranging from $1000\,\mathrm{K}$ to $4\times10^8\,\mathrm{K}$. To define the initial density distribution, we used the publicly available PyFC code \citep{Wagner2011}, a package which is useful to represent a dense, inhomogeneous component embedded in a smooth background. The distribution of clouds is set up by an iterative process following the work on terrestrial clouds in \citet{Lewis2002}. The 3D scalar field follows a log-normal single-point statistics in cartesian space and the fractal distribution is established by Fourier transforming and multiplying the cube by a Kolmogorov power-law with spectral index $\beta=-5/3$. The mean ($\mu$) of the parent distribution is 1.0 and the variance $\sigma^2=5.0$ \citep{Wagner2011}. The dimensions of the numerical box are 512$^3$ cells, with a resolution of 5.8\,mpc/cell (resulting in a resolution of 12 cells/$R_\mathrm{j}$), so the box covers the inner three parsecs of evolution of the relativistic, electron-positron jet propagation. Based on the computational box dimensions, we set a minimum sampling wave-number $k_\mathrm{min}=17$, such that the scale of the largest fractal structure in the cube (i.e. the cloud size) is approximately 0.1\,pc.

\begin{figure*}
\centering
\includegraphics[width=0.57\textwidth]{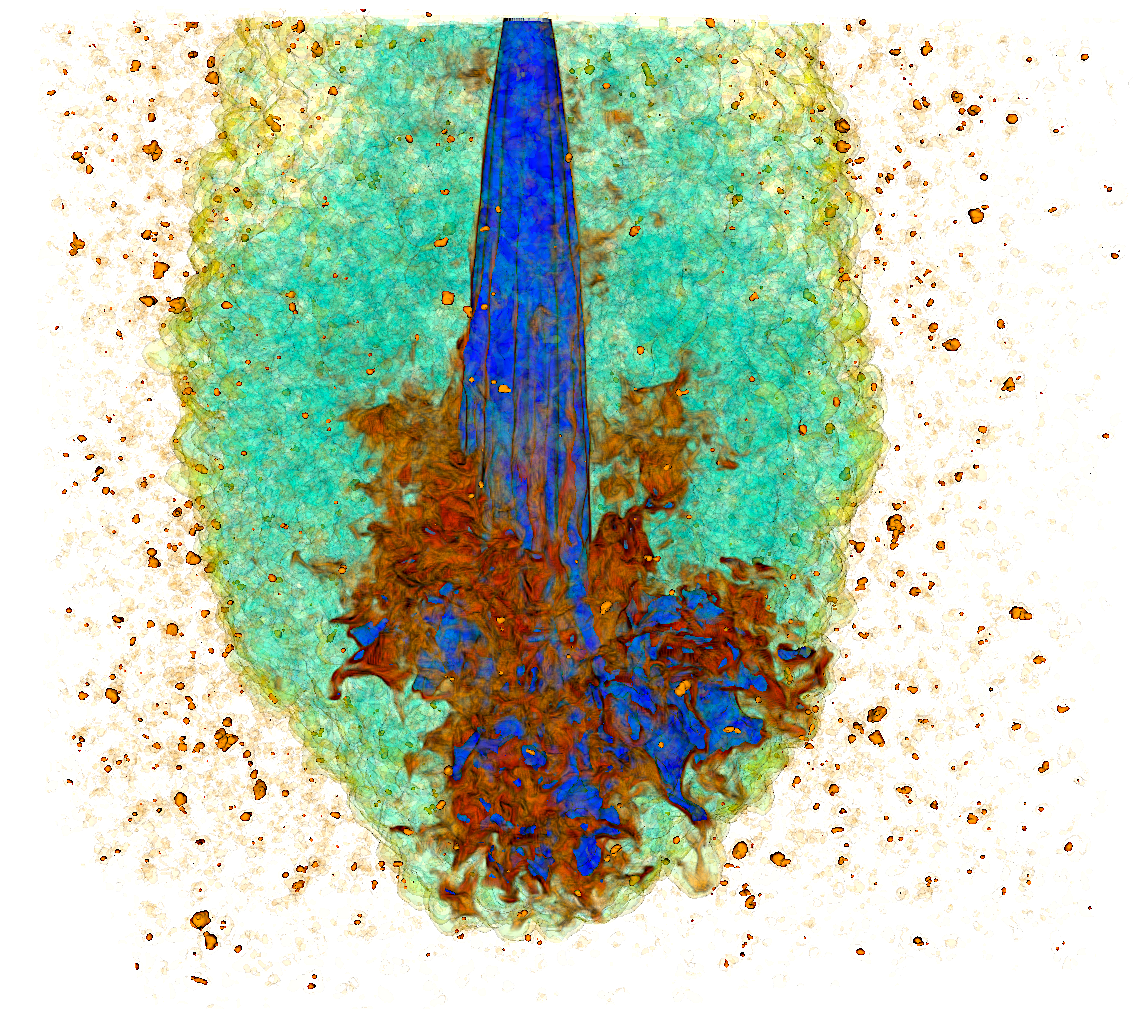}
\includegraphics[width=0.42\textwidth]{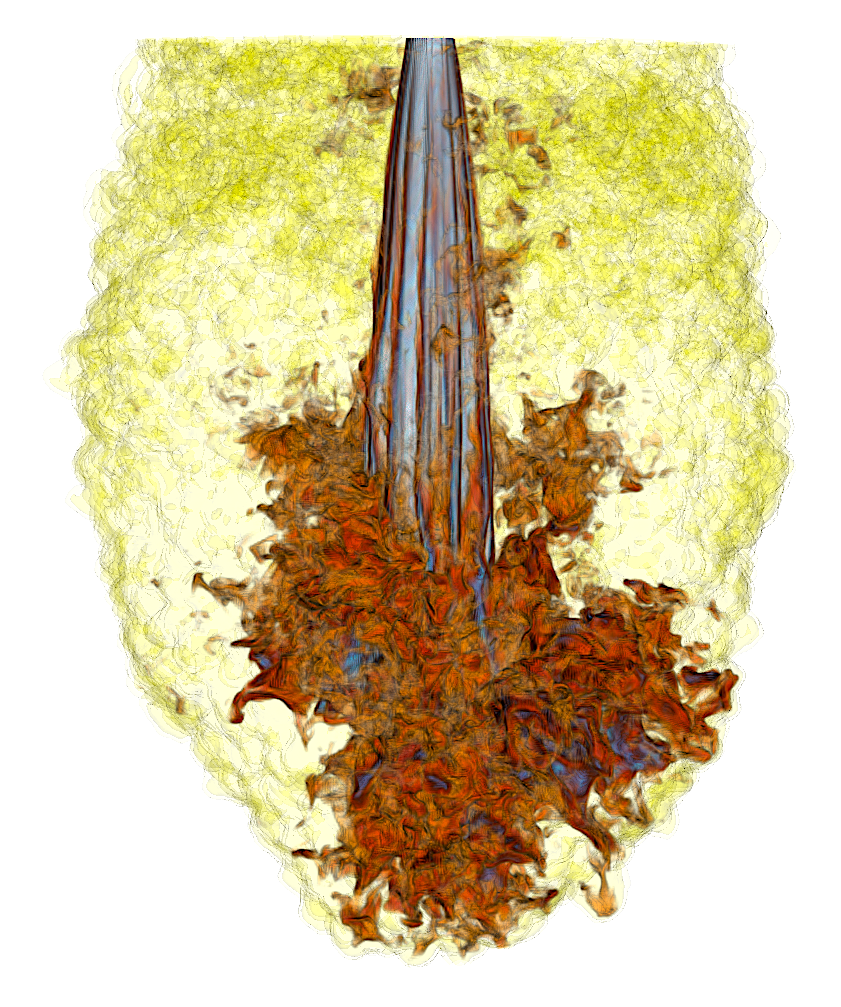}
\caption{Snapshots of a simulation of a relativistic, electron-positron jet propagating through a dilute ambient medium, as expected to be the case for the re-started jet in 3C\,84. The left panel shows a 3D render of the leptonic mass fraction (which identifies the jet), three temperature contours ($2.5\times 10^8$\,K in faint yellow, $2.1\times 10^{10}$\,K in faint blue and $1.8\times 10^{12}$\,K in dark blue), and atomic hydrogen density (brown clumps). The right panel shows a 3D render of the leptonic mass fraction and a pressure contour (faint yellow), to show the position of the jet bow shock. The plots were produced with LLNL VisIt software \citep{Childs2012}. 
}
\label{fig:simulations}
\end{figure*}

Figure~\ref{fig:simulations} shows the results of the simulation, for which we have used the approximate parameters derived for the jet in 3C\,84 in this paper. The images show the initial jet expansion and a transition to nearly cylindrical geometry caused by the jet's cocoon pressure. Because these represent the initial stages of evolution, the jet head is close enough to the injection point that the jet has not covered enough distance to recollimate in a conical shock. Following the jet evolution for longer distances would allow us to see how, as the jet head evolves to larger distances, a recollimation shock would be formed due to the collimating effect --- already observed in the simulations --- of the cocoon \citep[this has been observed in previous simulations of jet evolution, see, e.g.][]{Perucho2007b}. This is precisely the structure observed in the jet of 3C\,84, which shows no bright features from initial expansion until the hotspot.    

Interestingly, the plots also show, not only the aforementioned qualitative morphological resemblance, but also that the hottest regions in the jet are the shear layer, where dissipation takes place, and the hotspot itself. The dissipation of the kinetic energy due to the friction between different velocity layers, or the development of instabilities that play this same role can create a layer of hot gas where particles are accelerated \citep[e.g.][and references therein]{Rieger2019}. In the case of 3C\,84, energy dissipation at the jet boundaries could well be produced by the shear, as it is the case in the simulation\footnote{The generation and effects of shear-layers have been recently reported for the case of the radio galaxy 3C\,111 \citep{Beuchert2018, Schulz2020}}. The bow shock (external contours) shows a granular structure produced by the interaction of the jet with the colder, denser clouds.  

The main difference between the simulation and the observed jet is the bend observed in 3C\,84 close to the hot-spot. The jet shows no signs of instability development before this bend, which would correspond to an already non-linear amplitude. Another option would be jet precession, which, considering that jet propagation is basically ballistic, could also be excluded as the cause of the bend because there is no continuous change in direction either. Finally, the presence of a local, relatively large-scale inhomogeneity in the density distribution could trigger a differential cocoon pressure on both sides of the jet, forcing it to change direction. The fact that the lobe seems to develop in the direction of the bend, too, favours this interpretation. The reason is that the lobe plasma will preferentially move towards the lower pressure regions. Interestingly, \citet{Kino2021} reached the same conclusion based on 43\,GHz VLBI monitoring during 2016$-$2020 which shows $\sim 1$\,yr long frustration of the hot spot motion in 2017 and a subsequent breakout.

\subsection{Jet feedback}

It is well established that fully developed, large-scale jets of massive radio galaxies can heat the intergalactic or intracluster medium suppressing the cooling flows and controlling the growth of the massive galaxies \citep{MacNamara2007,McNamara2012}. There is good evidence that this `radio mode' or `maintenance mode' feedback mechanism is at work also in the Perseus Cluster and generations of outbursts from 3C\,84 have inflated multiple X-ray cavities in this cluster \citep{Boehringer1993,Fabian2000,Fabian2003}. A more recent idea is that the jet, especially a young, confined jet, could be an important feedback mechanism also inside the host galaxy heating gas and driving molecular outflows \citep[e.g.][]{Morganti2005,Morganti2013,Wagner2012}. One of the central problems of the latter type of feedback has been how well-collimated jets can affect the entire volume of the bulge, that is, how the jet couples to the ISM and how it can deposit energy and momentum relatively isotropically. There are several mechanisms that can play a role here, which will be discussed next. 

First of all, jets are light and overpressured. They can inflate a highly overpressured cocoon that expands quasi-isotropically. We can now see evidence for this in sub-pc scales. The mini-cocoon expands at a transverse velocity of $\sim 0.1$\,$c$ and its aspect ratio is about 1.4, so the cocoon can process a significantly larger volume of the ISM than the collimated jet alone. However, the transverse expansion speed of the cocoon will decrease as the cocoon pressure decreases over time and this will increase the aspect ratio -- for the southern mini-lobe, created by the previous activity period, the aspect ratio is about 2. The results in Sect.~\ref{sec:power_req} indicate that the jet can dump about half of its energy in the mini-cocoon.

In a series of numerical simulations, first 2D axisymmetric \citep{Perucho2011,Perucho2014}, and later 3D simulations \citep[][]{Perucho2019, Perucho2021}, have shown that collimated, relativistic jets can deposit a large fraction (up to $80-90\%$) of their energy into heating and displacing the ambient medium both in the galactic halos and intergalactic medium. In \citet{Perucho2017} the authors gave an explanation to this efficiency, on the basis of the aforementioned jet collimation and relativistic nature of AGN jets. In this work, they studied the ratio of the cavity (cocoon) pressure to the maximum cavity pressure defined as:
\begin{equation}
    p_\mathrm{c,max} = \frac{P_\mathrm{j}\,\Delta t}{V_\mathrm{c}}, \label{eq:power_pres_cocoon}
\end{equation}
where it is implicitly assumed that all the injected energy is transferred to the cocoon to drive the shock \citep{Begelman1989}. \citet{Perucho2017} found that in the relativistic simulations $p_\mathrm{c} / p_\mathrm{c,max} \simeq 0.4$ throughout the whole propagation phase. Taking this ratio and assuming $\theta = 18^\circ$, $P_\mathrm{j} \simeq p_\mathrm{c} V_\mathrm{c} / (0.4 \Delta t) \gtrsim 2 \times 10^{43} (1+k)^{4/7}$\,erg\,s$^{-1}$. This lower limit for $P_\mathrm{j}$ matches the lower limit calculated in Sect.~\ref{sec:densities} and is a factor of two smaller than the lower limit calculated in Sect.~\ref{sec:power_req}. This is consistent with a large proportion of the energy transfer from the jet taking place through a strong shock with the ambient medium, which, according to the cited works, could be assigned to its collimation and relativistic nature.

Numerical simulations of the jet propagation using a multi-phase description of the ambient medium have also shown that the jets couple strongly to a 'clumpy' interstellar medium \cite[e.g.][and the work described in the previous section]{Wagner2011, Wagner2012, Mukherjee2018}. When a jet in these simulations struggles through a dense cluster of gas clouds of various sizes, it is deflected and the flow is channelled to many different directions in the porous ISM. According to the simulations by \cite{Wagner2012} these channel flows provide an efficient means of energy transfer from the jet to the ISM with up to 40\% of the jet energy being deposited to the cold and warm gas. The earlier works by \citet{Nagai2017},  \citet{Kino2018}, and \citet{Kino2021} indicate that the restarted jet in 3C\,84 is likely moving through a clumpy, multi-phase medium and this mechanism can also be (at least partly) responsible for creating the cocoon bubble. The interaction between the jet and a dense gas clump is also the most likely explanation for the molecular outflow found by \citet{Nagai2019}. 

Finally, yet another effect that can increase the volume of the ISM that a jet can have an impact on, is the wobbling or precession of the jet. If the jet direction changes significantly with time, it naturally interacts with a larger volume of ambient gas. The 3C\,84 has clearly ejected sub-pc scale jets in different directions over the past 30 years as can be seen by comparing VLBI observations made in the 1990s to the jet direction visible in the 22\,GHz \textit{RadioAstron} image. In the 43\,GHz VLBA images of \citet{Dhawan1998} made in 1995$-$1996, the sub-pc jet points to southwest, to the direction of the current feature C2; in 2013, the jet was pointing in the south-southeasterly direction. 

It is quite possible that all three mechanisms -- an overpressured cocoon, interaction with a clumpy ISM, and changes in the jet direction --- work at the same time in 3C\,84 helping to transfer energy and momentum from the jet to the ISM.


\section{Conclusions}

We present a new 5\,GHz space-VLBI image of the radio galaxy 3C\,84 obtained with a global VLBI array and the 10-m Space Radio Telescope of the \emph{RadioAstron} mission, as well as 15 and 43\,GHz VLBA+VLA images taken quasi-simultaneously with the space-VLBI image. We offer an extensive  discussion of the methods for finding weak space-VLBI fringes as well as issues in imaging that are specific to the \emph{RadioAstron} data. Fringes were detected at baseline lengths up to about 8\,$D_\oplus$, which allowed us to resolve structures in the innermost $\sim 1$\,pc that are not visible in the ground-based images. Here, we report  the discovery of a cocoon-like, low-intensity, emission structure surrounding the jet in the 5\,GHz \emph{RadioAstron} image. This is, to our knowledge, the first time such a mini-cocoon has been seen in the (sub)parsec scale. Following our analysis of the cocoon structure in detail, our main findings are:
\begin{itemize}
    \item The minimum energy stored in the cocoon is $(3 - 4) \times 10^{51} (1+k)^{4/7} \sin^{-3/7}\theta$\;erg, where $k$ is the energy ratio between heavy particles and relativistic electrons, $\theta$ is the viewing angle of the jet, and we have assumed that the low-frequency cutoff of the synchrotron spectrum is at $10^8$\,Hz. We find that the long-term average jet power measured from the X-ray cavities is enough to inflate the mini-cocoon in 10 years if $k$ is less than a few tens. Also the calculated synchrotron-loss time scales are compatible with our observations if the mini-cocoon is less than ten years old.  If the minimum energy assumption holds and the jet power is not significantly different from the long-term average, then about half of the energy delivered by the jet is dumped in the cocoon and about 11\%$-$14\% goes to driving the shock to the ambient medium.
    \item We estimated the ambient density assuming minimum energy condition in the cocoon, $\rho_\mathrm{a} \gtrsim 5 \times 10^{-24} (1+k)^{4/7} \, \sin^{4/7} \theta$\,g\,cm$^{-3}$. This is compatible with the $\sim 8$\,cm$^{-3}$ number density estimate by \citet{Fujita2017} -- if we assume normal plasma and $k \lesssim 20$. Our lower limit is larger than the upper limit to the sub-pc jet electron number density estimated by \citet{Kim2019} from the Faraday rotation measurements. Therefore, the jet in 3C\,84 is light compared to the ambient medium.
    \item The average transverse expansion speed of the mini-cocoon is about 0.1\,$c$ while the average propagation speed of the jet head, C3, is about 0.25\,$c$. The aspect ratio of the mini-cocoon is 1.4 and it can process a significantly larger volume of the ISM compared to a narrow jet alone, which enhances the energy transfer from the jet to the ISM gas. 
    \item The pressure from a hot mini-cocoon around the restarted sub-pc scale jet can explain the almost cylindrical shape of the jet seen in the 22\,GHz \emph{RadioAstron} image \citep{Giovannini2018} and in the 43\,GHz VLBA image \citep{Nagai2014}.
\end{itemize}

The sub-structure in the innermost $\sim 1$\,pc of the 3C\,84 revealed by the \emph{RadioAstron} at 5\,GHz illustrates how valuable the long space baselines are for studying AGN jets. While a similar angular resolution can be obtained with the ground-based VLBI at mm-wavelengths, steep-spectrum structures such as the mini-cocoon are visible only at longer wavelengths.   

\begin{acknowledgements}
We thank Eduardo Ros for a careful reading of the manuscript and valuable comments. We also thank the anonymous referee for constructive criticism that helped us to improve the presentation of the results.  
The \emph{RadioAstron} project is led by the Astro Space Center of the Lebedev Physical Institute of the Russian Academy of Sciences and the Lavochkin Scientific and Production Association under a contract with the State Space Corporation ROSCOSMOS, in collaboration with partner organisations in Russia and other countries.
The National Radio Astronomy Observatory is a facility of the National Science Foundation operated under cooperative agreement by Associated Universities, Inc. The European VLBI Network is a joint facility of independent European, African, Asian, and North American radio astronomy institutes. This research is based on observations correlated at the Bonn Correlator, jointly operated by the Max Planck Institute for Radio Astronomy (MPIfR), and the Federal Agency for Cartography and Geodesy (BKG). 
TS was supported by the Academy of Finland projects 274477, 284495, 312496, and 315721. 
YYK, AVP, and PAV were supported by the Russian Science Foundation grant 21-12-00241.
Computer simulations have been carried out in \emph{Tirant}, the local supercomputer and at the \emph{Servei d'Inform\`atica de la Universitat de Val\`encia}. MP acknowledges support by the Spanish Ministry of Science through Grants PID2019-105510GB-C31, PID2019-107427GB-C33, and from the Generalitat Valenciana through grant PROMETEU/2019/071.
JLM acknowledges the support of a fellowship from ‘La Caixa’ Foundation (ID 100010434). The fellowship code is LCF/BQ/DR19/11740030.
BWS is grateful for the support by the National Research Foundation of Korea (NRF) funded by the Ministry of Science and ICT (MSIT) of Korea (NRF-2020K1A3A1A78114060).

\end{acknowledgements}

\bibliographystyle{aa} 
\bibliography{radioastron} 

\begin{thebibliography}{108}
\expandafter\ifx\csname natexlab\endcsname\relax\def\natexlab#1{#1}\fi

\bibitem[{{Abdo} {et~al.}(2009){Abdo}, {Ackermann}, {Ajello}, {Asano},
  {Baldini}, {Ballet}, {Barbiellini}, {Bastieri}, {Baughman}, {Bechtol},
  {Bellazzini}, {Blandford}, {Bloom}, {Bonamente}, {Borgland}, {Bregeon},
  {Brez}, {Brigida}, {Bruel}, {Burnett}, {Caliand ro}, {Cameron}, {Caraveo},
  {Casand jian}, {Cavazzuti}, {Cecchi}, {Celotti}, {Chekhtman}, {Cheung},
  {Chiang}, {Ciprini}, {Claus}, {Cohen-Tanugi}, {Colafrancesco}, {Cominsky},
  {Conrad}, {Costamante}, {Dermer}, {de Angelis}, {de Palma}, {Digel},
  {Donato}, {do Couto e Silva}, {Drell}, {Dubois}, {Dumora}, {Farnier},
  {Favuzzi}, {Finke}, {Focke}, {Frailis}, {Fukazawa}, {Funk}, {Fusco},
  {Gargano}, {Georganopoulos}, {Germani}, {Giebels}, {Giglietto}, {Giordano},
  {Glanzman}, {Grenier}, {Grondin}, {Grove}, {Guillemot}, {Guiriec},
  {Hanabata}, {Harding}, {Hartman}, {Hayashida}, {Hays}, {Hughes},
  {J{\'o}hannesson}, {Johnson}, {Johnson}, {Johnson}, {Kadler}, {Kamae},
  {Kanai}, {Katagiri}, {Kataoka}, {Kawai}, {Kerr}, {Kn{\"o}dlseder}, {Kuehn},
  {Kuss}, {Latronico}, {Lemoine-Goumard}, {Longo}, {Loparco}, {Lott},
  {Lovellette}, {Lubrano}, {Madejski}, {Makeev}, {Mazziotta}, {McEnery},
  {Meurer}, {Michelson}, {Mitthumsiri}, {Mizuno}, {Moiseev}, {Monte},
  {Monzani}, {Morselli}, {Moskalenko}, {Murgia}, {Nakamori}, {Nolan}, {Norris},
  {Nuss}, {Ohsugi}, {Omodei}, {Orlando}, {Ormes}, {Paneque}, {Panetta},
  {Parent}, {Pepe}, {Pesce-Rollins}, {Piron}, {Porter}, {Rain{\`o}}, {Razzano},
  {Reimer}, {Reimer}, {Reposeur}, {Ritz}, {Rodriguez}, {Romani}, {Ryde},
  {Sadrozinski}, {Sambruna}, {Sanchez}, {Sander}, {Sato}, {Parkinson},
  {Sgr{\`o}}, {Smith}, {Smith}, {Spandre}, {Spinelli}, {Starck}, {Strickman},
  {Strong}, {Suson}, {Tajima}, {Takahashi}, {Takahashi}, {Tanaka}, {Taylor},
  {Thayer}, {Thompson}, {Torres}, {Tosti}, {Uchiyama}, {Usher}, {Vilchez},
  {Vitale}, {Waite}, {Wood}, {Ylinen}, {Ziegler}, {Aller}, {Aller},
  {Kellermann}, {Kovalev}, {Kovalev}, {Lister}, \& {Pushkarev}}]{Abdo2009}
{Abdo}, A.~A., {Ackermann}, M., {Ajello}, M., {et~al.} 2009, \apj, 699, 31

\bibitem[{{Akiyama} {et~al.}(2017){Akiyama}, {Kuramochi}, {Ikeda}, {Fish},
  {Tazaki}, {Honma}, {Doeleman}, {Broderick}, {Dexter}, {Mo{\'s}cibrodzka},
  {Bouman}, {Chael}, \& {Zaizen}}]{Akiyama2017}
{Akiyama}, K., {Kuramochi}, K., {Ikeda}, S., {et~al.} 2017, \apj, 838, 1

\bibitem[{{Aleksi{\'c}} {et~al.}(2012){Aleksi{\'c}}, {Alvarez}, {Antonelli},
  {Antoranz}, {Asensio}, {Backes}, {Barres de Almeida}, {Barrio}, {Bastieri},
  {Becerra Gonz{\'a}lez}, {Bednarek}, {Berger}, {Bernardini}, {Biland},
  {Blanch}, {Bock}, {Boller}, {Bonnoli}, {Borla Tridon}, {Bretz},
  {Ca{\~n}ellas}, {Carmona}, {Carosi}, {Colin}, {Colombo}, {Contreras},
  {Cortina}, {Cossio}, {Covino}, {da Vela}, {Dazzi}, {de Angelis}, {de Caneva},
  {de Cea Del Pozo}, {de Lotto}, {Delgado Mendez}, {Diago Ortega}, {Doert},
  {Dom{\'\i}nguez}, {Dominis Prester}, {Dorner}, {Doro}, {Eisenacher},
  {Elsaesser}, {Ferenc}, {Fonseca}, {Font}, {Fruck}, {Garc{\'\i}a L{\'o}pez},
  {Garczarczyk}, {Garrido}, {Giavitto}, {Godinovi{\'c}}, {Gozzini}, {Hadasch},
  {H{\"a}fner}, {Herrero}, {Hildebrand }, {H{\"o}hne-M{\"o}nch}, {Hose},
  {Hrupec}, {Huber}, {Jogler}, {Kadenius}, {Kellermann}, {Klepser},
  {Kr{\"a}henb{\"u}hl}, {Krause}, {La Barbera}, {Lelas}, {Leonardo}, {Lewand
  owska}, {Lindfors}, {Lombardi}, {L{\'o}pez}, {L{\'o}pez-Coto},
  {L{\'o}pez-Oramas}, {Lorenz}, {Makariev}, {Maneva}, {Mankuzhiyil},
  {Mannheim}, {Maraschi}, {Mariotti}, {Mart{\'\i}nez}, {Mazin}, {Meucci},
  {Mirand a}, {Mirzoyan}, {Mold{\'o}n}, {Moralejo}, {Munar-Adrover},
  {Niedzwiecki}, {Nieto}, {Nilsson}, {Nowak}, {Orito}, {Paiano}, {Paneque},
  {Paoletti}, {Pardo}, {Paredes}, {Partini}, {Perez-Torres}, {Persic},
  {Peruzzo}, {Pilia}, {Pochon}, {Prada}, {Prada Moroni}, {Prandini}, {Puerto
  Gimenez}, {Puljak}, {Reichardt}, {Reinthal}, {Rhode}, {Rib{\'o}}, {Rico},
  {R{\"u}gamer}, {Saggion}, {Saito}, {Saito}, {Salvati}, {Satalecka},
  {Scalzotto}, {Scapin}, {Schultz}, {Schweizer}, {Shayduk}, {Shore},
  {Sillanp{\"a}{\"a}}, {Sitarek}, {Snidaric}, {Sobczynska}, {Spanier}, {Spiro},
  {Stamatescu}, {Stamerra}, {Steinke}, {Storz}, {Strah}, {Sun}, {Suri{\'c}},
  {Takalo}, {Takami}, {Tavecchio}, {Temnikov}, {Terzi{\'c}}, {Tescaro},
  {Teshima}, {Tibolla}, {Torres}, {Treves}, {Uellenbeck}, {Vogler}, {Wagner},
  {Weitzel}, {Zabalza}, {Zandanel}, {Zanin}, {Pfrommer}, \&
  {Pinzke}}]{Aleksic2012}
{Aleksi{\'c}}, J., {Alvarez}, E.~A., {Antonelli}, L.~A., {et~al.} 2012, \aap,
  539, L2

\bibitem[{{Aleksi{\'c}} {et~al.}(2014){Aleksi{\'c}}, {Ansoldi}, {Antonelli},
  {Antoranz}, {Babic}, {Bangale}, {Barres de Almeida}, {Barrio}, {Becerra
  Gonz{\'a}lez}, {Bednarek}, {Berger}, {Bernardini}, {Biland}, {Blanch},
  {Bock}, {Bonnefoy}, {Bonnoli}, {Borracci}, {Bretz}, {Carmona}, {Carosi},
  {Carreto Fidalgo}, {Colin}, {Colombo}, {Contreras}, {Cortina}, {Covino}, {Da
  Vela}, {Dazzi}, {De Angelis}, {De Caneva}, {De Lotto}, {Delgado Mendez},
  {Doert}, {Dom{\'\i}nguez}, {Dominis Prester}, {Dorner}, {Doro}, {Einecke},
  {Eisenacher}, {Elsaesser}, {Farina}, {Ferenc}, {Fonseca}, {Font}, {Frantzen},
  {Fruck}, {Garc{\'\i}a L{\'o}pez}, {Garczarczyk}, {Garrido Terrats}, {Gaug},
  {Giavitto}, {Godinovi{\'c}}, {Gonz{\'a}lez Mu{\~n}oz}, {Gozzini}, {Hadamek},
  {Hadasch}, {Herrero}, {Hildebrand}, {Hose}, {Hrupec}, {Idec}, {Kadenius},
  {Kellermann}, {Knoetig}, {Krause}, {Kushida}, {La Barbera}, {Lelas},
  {Lewandowska}, {Lindfors}, {Lombardi}, {L{\'o}pez}, {L{\'o}pez-Coto},
  {L{\'o}pez-Oramas}, {Lorenz}, {Lozano}, {Makariev}, {Mallot}, {Maneva},
  {Mankuzhiyil}, {Mannheim}, {Maraschi}, {Marcote}, {Mariotti},
  {Mart{\'\i}nez}, {Mazin}, {Menzel}, {Meucci}, {Miranda}, {Mirzoyan},
  {Moralejo}, {Munar-Adrover}, {Nakajima}, {Niedzwiecki}, {Nilsson}, {Nowak},
  {Orito}, {Overkemping}, {Paiano}, {Palatiello}, {Paneque}, {Paoletti},
  {Paredes}, {Paredes-Fortuny}, {Partini}, {Persic}, {Prada}, {Prada Moroni},
  {Prand ini}, {Preziuso}, {Puljak}, {Reinthal}, {Rhode}, {Rib{\'o}}, {Rico},
  {Rodriguez Garcia}, {R{\"u}gamer}, {Saggion}, {Saito}, {Saito}, {Salvati},
  {Satalecka}, {Scalzotto}, {Scapin}, {Schultz}, {Schweizer}, {Shore},
  {Sillanp{\"a}{\"a}}, {Sitarek}, {Snidaric}, {Sobczynska}, {Spanier},
  {Stamatescu}, {Stamerra}, {Steinbring}, {Storz}, {Sun}, {Suri{\'c}},
  {Takalo}, {Tavecchio}, {Terzi{\'c}}, {Tescaro}, {Teshima}, {Thaele},
  {Tibolla}, {Torres}, {Toyama}, {Treves}, {Uellenbeck}, {Vogler}, {Wagner},
  {Zandanel}, {Zanin}, {MAGIC Collaboration}, {Balmaverde}, {Kataoka},
  {Rekola}, \& {Takahashi}}]{Aleksic2014}
{Aleksi{\'c}}, J., {Ansoldi}, S., {Antonelli}, L.~A., {et~al.} 2014, \aap, 564,
  A5

\bibitem[{{Asada} {et~al.}(2006){Asada}, {Kameno}, {Shen}, {Horiuchi},
  {Gabuzda}, \& {Inoue}}]{Asada2006}
{Asada}, K., {Kameno}, S., {Shen}, Z.-Q., {et~al.} 2006, \pasj, 58, 261

\bibitem[{{Asada} \& {Nakamura}(2012)}]{Asada2012}
{Asada}, K. \& {Nakamura}, M. 2012, \apjl, 745, L28

\bibitem[{{Astropy Collaboration} {et~al.}(2013){Astropy Collaboration},
  {Robitaille}, {Tollerud}, {Greenfield}, {Droettboom}, {Bray}, {Aldcroft},
  {Davis}, {Ginsburg}, {Price-Whelan}, {Kerzendorf}, {Conley}, {Crighton},
  {Barbary}, {Muna}, {Ferguson}, {Grollier}, {Parikh}, {Nair}, {Unther},
  {Deil}, {Woillez}, {Conseil}, {Kramer}, {Turner}, {Singer}, {Fox}, {Weaver},
  {Zabalza}, {Edwards}, {Azalee Bostroem}, {Burke}, {Casey}, {Crawford},
  {Dencheva}, {Ely}, {Jenness}, {Labrie}, {Lim}, {Pierfederici}, {Pontzen},
  {Ptak}, {Refsdal}, {Servillat}, \& {Streicher}}]{Astropy2013}
{Astropy Collaboration}, {Robitaille}, T.~P., {Tollerud}, E.~J., {et~al.} 2013,
  \aap, 558, A33

\bibitem[{{Begelman} \& {Cioffi}(1989)}]{Begelman1989}
{Begelman}, M.~C. \& {Cioffi}, D.~F. 1989, \apjl, 345, L21

\bibitem[{{Benbow} \& {VERITAS Collaboration}(2015)}]{Benbow2015}
{Benbow}, W. \& {VERITAS Collaboration}. 2015, in International Cosmic Ray
  Conference, Vol.~34, 34th International Cosmic Ray Conference (ICRC2015), 821

\bibitem[{{Beuchert} {et~al.}(2018){Beuchert}, {Kadler}, {Perucho},
  {Gro{\ss}berger}, {Schulz}, {Agudo}, {Casadio}, {G{\'o}mez}, {Gurwell},
  {Homan}, {Kovalev}, {Lister}, {Markoff}, {Molina}, {Pushkarev}, {Ros},
  {Savolainen}, {Steinbring}, {Thum}, \& {Wilms}}]{Beuchert2018}
{Beuchert}, T., {Kadler}, M., {Perucho}, M., {et~al.} 2018, \aap, 610, A32

\bibitem[{{Blandford} {et~al.}(2019){Blandford}, {Meier}, \&
  {Readhead}}]{Blandford2019}
{Blandford}, R., {Meier}, D., \& {Readhead}, A. 2019, \araa, 57, 467

\bibitem[{{Blandford} \& {Payne}(1982)}]{Blandford1982}
{Blandford}, R.~D. \& {Payne}, D.~G. 1982, \mnras, 199, 883

\bibitem[{{Blandford} \& {Znajek}(1977)}]{Blandford1977}
{Blandford}, R.~D. \& {Znajek}, R.~L. 1977, \mnras, 179, 433

\bibitem[{{Boccardi} {et~al.}(2016){Boccardi}, {Krichbaum}, {Bach}, {Bremer},
  \& {Zensus}}]{Boccardi2016}
{Boccardi}, B., {Krichbaum}, T.~P., {Bach}, U., {Bremer}, M., \& {Zensus},
  J.~A. 2016, \aap, 588, L9

\bibitem[{{Boehringer} {et~al.}(1993){Boehringer}, {Voges}, {Fabian}, {Edge},
  \& {Neumann}}]{Boehringer1993}
{Boehringer}, H., {Voges}, W., {Fabian}, A.~C., {Edge}, A.~C., \& {Neumann},
  D.~M. 1993, \mnras, 264, L25

\bibitem[{{Bruni} {et~al.}(2016){Bruni}, {Anderson}, {Alef}, {Rottmann},
  {Lobanov}, \& {Zensus}}]{Bruni2016}
{Bruni}, G., {Anderson}, J., {Alef}, W., {et~al.} 2016, Galaxies, 4, 55

\bibitem[{{Bruni} {et~al.}(2020){Bruni}, {Savolainen}, {G{\'o}mez}, {Lobanov},
  {Kovalev}, {RadioAstron AGN Imaging Team}, \& {KSP Team}}]{Bruni2020}
{Bruni}, G., {Savolainen}, T., {G{\'o}mez}, J.~L., {et~al.} 2020, Advances in
  Space Research, 65, 712

\bibitem[{{Chael} {et~al.}(2018){Chael}, {Johnson}, {Bouman}, {Blackburn},
  {Akiyama}, \& {Narayan}}]{Chael2018}
{Chael}, A.~A., {Johnson}, M.~D., {Bouman}, K.~L., {et~al.} 2018, \apj, 857, 23

\bibitem[{{Chael} {et~al.}(2016){Chael}, {Johnson}, {Narayan}, {Doeleman},
  {Wardle}, \& {Bouman}}]{Chael2016}
{Chael}, A.~A., {Johnson}, M.~D., {Narayan}, R., {et~al.} 2016, \apj, 829, 11

\bibitem[{{Childs}(2012)}]{Childs2012}
{Childs}, H. 2012, {High Performance Visualization --- Enabling Extreme-Scale
  Scientific Insight} ({Chapman \& Hall, CRC Computational Science}), 357

\bibitem[{{Clark}(1980)}]{Clark1980}
{Clark}, B.~G. 1980, \aap, 89, 377

\bibitem[{{Clark} {et~al.}(1968){Clark}, {Kellermann}, {Bare}, {Cohen}, \&
  {Jauncey}}]{Clark1968}
{Clark}, B.~G., {Kellermann}, K.~I., {Bare}, C.~C., {Cohen}, M.~H., \&
  {Jauncey}, D.~L. 1968, \apj, 153, 705

\bibitem[{{Deller} {et~al.}(2011){Deller}, {Brisken}, {Phillips}, {Morgan},
  {Alef}, {Cappallo}, {Middelberg}, {Romney}, {Rottmann}, {Tingay}, \&
  {Wayth}}]{Deller2011}
{Deller}, A.~T., {Brisken}, W.~F., {Phillips}, C.~J., {et~al.} 2011, \pasp,
  123, 275

\bibitem[{{Dent}(1966)}]{Dent1966}
{Dent}, W.~A. 1966, \apj, 144, 843

\bibitem[{{Dent} \& {Haddock}(1965)}]{Dent1965}
{Dent}, W.~A. \& {Haddock}, F.~T. 1965, \nat, 205, 487

\bibitem[{{Desai}(1998)}]{Desai1998}
{Desai}, K.~M. 1998, {The Calculation of SNR in KRING's FFT stage}, AIPS Memo
  101, NRAO

\bibitem[{{Dhawan} {et~al.}(1998){Dhawan}, {Kellermann}, \&
  {Romney}}]{Dhawan1998}
{Dhawan}, V., {Kellermann}, K.~I., \& {Romney}, J.~D. 1998, \apjl, 498, L111

\bibitem[{{Event Horizon Telescope Collaboration}
  {et~al.}(2019{\natexlab{a}}){Event Horizon Telescope Collaboration},
  {Akiyama}, {Alberdi}, {Alef}, {Asada}, {Azulay}, {Baczko}, {Ball},
  {Balokovi{\'c}}, {Barrett}, {Bintley}, {Blackburn}, {Boland}, {Bouman},
  {Bower}, {Bremer}, {Brinkerink}, {Brissenden}, {Britzen}, {Broderick},
  {Broguiere}, {Bronzwaer}, {Byun}, {Carlstrom}, {Chael}, {Chan}, {Chatterjee},
  {Chatterjee}, {Chen}, {Chen}, {Cho}, {Christian}, {Conway}, {Cordes}, {Crew},
  {Cui}, {Davelaar}, {De Laurentis}, {Deane}, {Dempsey}, {Desvignes}, {Dexter},
  {Doeleman}, {Eatough}, {Falcke}, {Fish}, {Fomalont}, {Fraga-Encinas},
  {Friberg}, {Fromm}, {G{\'o}mez}, {Galison}, {Gammie}, {Garc{\'\i}a},
  {Gentaz}, {Georgiev}, {Goddi}, {Gold}, {Gu}, {Gurwell}, {Hada}, {Hecht},
  {Hesper}, {Ho}, {Ho}, {Honma}, {Huang}, {Huang}, {Hughes}, {Ikeda}, {Inoue},
  {Issaoun}, {James}, {Jannuzi}, {Janssen}, {Jeter}, {Jiang}, {Johnson},
  {Jorstad}, {Jung}, {Karami}, {Karuppusamy}, {Kawashima}, {Keating},
  {Kettenis}, {Kim}, {Kim}, {Kim}, {Kino}, {Koay}, {Koch}, {Koyama}, {Kramer},
  {Kramer}, {Krichbaum}, {Kuo}, {Lauer}, {Lee}, {Li}, {Li}, {Lindqvist}, {Liu},
  {Liuzzo}, {Lo}, {Lobanov}, {Loinard}, {Lonsdale}, {Lu}, {MacDonald}, {Mao},
  {Markoff}, {Marrone}, {Marscher}, {Mart{\'\i}-Vidal}, {Matsushita},
  {Matthews}, {Medeiros}, {Menten}, {Mizuno}, {Mizuno}, {Moran}, {Moriyama},
  {Moscibrodzka}, {Mul{\ensuremath{\ddot{}}}ler}, {Nagai}, {Nagar}, {Nakamura},
  {Narayan}, {Narayanan}, {Natarajan}, {Neri}, {Ni}, {Noutsos}, {Okino},
  {Olivares}, {Oyama}, {{\"O}zel}, {Palumbo}, {Patel}, {Pen}, {Pesce},
  {Pi{\'e}tu}, {Plambeck}, {PopStefanija}, {Porth}, {Prather},
  {Preciado-L{\'o}pez}, {Psaltis}, {Pu}, {Ramakrishnan}, {Rao}, {Rawlings},
  {Raymond}, {Rezzolla}, {Ripperda}, {Roelofs}, {Rogers}, {Ros}, {Rose},
  {Roshanineshat}, {Rottmann}, {Roy}, {Ruszczyk}, {Ryan}, {Rygl},
  {S{\'a}nchez}, {S{\'a}nchez-Arguelles}, {Sasada}, {Savolainen}, {Schloerb},
  {Schuster}, {Shao}, {Shen}, {Small}, {Sohn}, {SooHoo}, {Tazaki}, {Tiede},
  {Tilanus}, {Titus}, {Toma}, {Torne}, {Trent}, {Trippe}, {Tsuda}, {van
  Bemmel}, {van Langevelde}, {van Rossum}, {Wagner}, {Wardle}, {Weintroub},
  {Wex}, {Wharton}, {Wielgus}, {Wong}, {Wu}, {Young}, {Young}, {Younsi},
  {Yuan}, {Yuan}, {Zensus}, {Zhao}, {Zhao}, {Zhu}, {Anczarski}, {Baganoff},
  {Eckart}, {Farah}, {Haggard}, {Meyer-Zhao}, {Michalik}, {Nadolski},
  {Neilsen}, {Nishioka}, {Nowak}, {Pradel}, {Primiani}, {Souccar},
  {Vertatschitsch}, {Yamaguchi}, \& {Zhang}}]{EHT2019e}
{Event Horizon Telescope Collaboration}, {Akiyama}, K., {Alberdi}, A., {et~al.}
  2019{\natexlab{a}}, \apjl, 875, L5

\bibitem[{{Event Horizon Telescope Collaboration}
  {et~al.}(2019{\natexlab{b}}){Event Horizon Telescope Collaboration},
  {Akiyama}, {Alberdi}, {Alef}, {Asada}, {Azulay}, {Baczko}, {Ball},
  {Balokovi{\'c}}, {Barrett}, {Bintley}, {Blackburn}, {Boland}, {Bouman},
  {Bower}, {Bremer}, {Brinkerink}, {Brissenden}, {Britzen}, {Broderick},
  {Broguiere}, {Bronzwaer}, {Byun}, {Carlstrom}, {Chael}, {Chan}, {Chatterjee},
  {Chatterjee}, {Chen}, {Chen}, {Cho}, {Christian}, {Conway}, {Cordes}, {Crew},
  {Cui}, {Davelaar}, {De Laurentis}, {Deane}, {Dempsey}, {Desvignes}, {Dexter},
  {Doeleman}, {Eatough}, {Falcke}, {Fish}, {Fomalont}, {Fraga-Encinas},
  {Freeman}, {Friberg}, {Fromm}, {G{\'o}mez}, {Galison}, {Gammie},
  {Garc{\'\i}a}, {Gentaz}, {Georgiev}, {Goddi}, {Gold}, {Gu}, {Gurwell},
  {Hada}, {Hecht}, {Hesper}, {Ho}, {Ho}, {Honma}, {Huang}, {Huang}, {Hughes},
  {Ikeda}, {Inoue}, {Issaoun}, {James}, {Jannuzi}, {Janssen}, {Jeter}, {Jiang},
  {Johnson}, {Jorstad}, {Jung}, {Karami}, {Karuppusamy}, {Kawashima},
  {Keating}, {Kettenis}, {Kim}, {Kim}, {Kim}, {Kino}, {Koay}, {Koch}, {Koyama},
  {Kramer}, {Kramer}, {Krichbaum}, {Kuo}, {Lauer}, {Lee}, {Li}, {Li},
  {Lindqvist}, {Liu}, {Liuzzo}, {Lo}, {Lobanov}, {Loinard}, {Lonsdale}, {Lu},
  {MacDonald}, {Mao}, {Markoff}, {Marrone}, {Marscher}, {Mart{\'\i}-Vidal},
  {Matsushita}, {Matthews}, {Medeiros}, {Menten}, {Mizuno}, {Mizuno}, {Moran},
  {Moriyama}, {Moscibrodzka}, {M{\"u}ller}, {Nagai}, {Nagar}, {Nakamura},
  {Narayan}, {Narayanan}, {Natarajan}, {Neri}, {Ni}, {Noutsos}, {Okino},
  {Olivares}, {Ortiz-Le{\'o}n}, {Oyama}, {{\"O}zel}, {Palumbo}, {Patel}, {Pen},
  {Pesce}, {Pi{\'e}tu}, {Plambeck}, {PopStefanija}, {Porth}, {Prather},
  {Preciado-L{\'o}pez}, {Psaltis}, {Pu}, {Ramakrishnan}, {Rao}, {Rawlings},
  {Raymond}, {Rezzolla}, {Ripperda}, {Roelofs}, {Rogers}, {Ros}, {Rose},
  {Roshanineshat}, {Rottmann}, {Roy}, {Ruszczyk}, {Ryan}, {Rygl},
  {S{\'a}nchez}, {S{\'a}nchez-Arguelles}, {Sasada}, {Savolainen}, {Schloerb},
  {Schuster}, {Shao}, {Shen}, {Small}, {Sohn}, {SooHoo}, {Tazaki}, {Tiede},
  {Tilanus}, {Titus}, {Toma}, {Torne}, {Trent}, {Trippe}, {Tsuda}, {van
  Bemmel}, {van Langevelde}, {van Rossum}, {Wagner}, {Wardle}, {Weintroub},
  {Wex}, {Wharton}, {Wielgus}, {Wong}, {Wu}, {Young}, {Young}, {Younsi},
  {Yuan}, {Yuan}, {Zensus}, {Zhao}, {Zhao}, {Zhu}, {Algaba}, {Allardi},
  {Amestica}, {Anczarski}, {Bach}, {Baganoff}, {Beaudoin}, {Benson},
  {Berthold}, {Blanchard}, {Blundell}, {Bustamente}, {Cappallo},
  {Castillo-Dom{\'\i}nguez}, {Chang}, {Chang}, {Chang}, {Chen}, {Chilson},
  {Chuter}, {C{\'o}rdova Rosado}, {Coulson}, {Crawford}, {Crowley}, {David},
  {Derome}, {Dexter}, {Dornbusch}, {Dudevoir}, {Dzib}, {Eckart}, {Eckert},
  {Erickson}, {Everett}, {Faber}, {Farah}, {Fath}, {Folkers}, {Forbes},
  {Freund}, {G{\'o}mez-Ruiz}, {Gale}, {Gao}, {Geertsema}, {Graham}, {Greer},
  {Grosslein}, {Gueth}, {Haggard}, {Halverson}, {Han}, {Han}, {Hao},
  {Hasegawa}, {Henning}, {Hern{\'a}ndez-G{\'o}mez}, {Herrero-Illana},
  {Heyminck}, {Hirota}, {Hoge}, {Huang}, {Impellizzeri}, {Jiang}, {Kamble},
  {Keisler}, {Kimura}, {Kono}, {Kubo}, {Kuroda}, {Lacasse}, {Laing}, {Leitch},
  {Li}, {Lin}, {Liu}, {Liu}, {Lu}, {Marson}, {Martin-Cocher}, {Massingill},
  {Matulonis}, {McColl}, {McWhirter}, {Messias}, {Meyer-Zhao}, {Michalik},
  {Monta{\~n}a}, {Montgomerie}, {Mora-Klein}, {Muders}, {Nadolski}, {Navarro},
  {Neilsen}, {Nguyen}, {Nishioka}, {Norton}, {Nowak}, {Nystrom}, {Ogawa},
  {Oshiro}, {Oyama}, {Parsons}, {Paine}, {Pe{\~n}alver}, {Phillips}, {Poirier},
  {Pradel}, {Primiani}, {Raffin}, {Rahlin}, {Reiland}, {Risacher}, {Ruiz},
  {S{\'a}ez-Mada{\'\i}n}, {Sassella}, {Schellart}, {Shaw}, {Silva}, {Shiokawa},
  {Smith}, {Snow}, {Souccar}, {Sousa}, {Sridharan}, {Srinivasan}, {Stahm},
  {Stark}, {Story}, {Timmer}, {Vertatschitsch}, {Walther}, {Wei}, {Whitehorn},
  {Whitney}, {Woody}, {Wouterloot}, {Wright}, {Yamaguchi}, {Yu}, {Zeballos},
  {Zhang}, \& {Ziurys}}]{EHT2019a}
{Event Horizon Telescope Collaboration}, {Akiyama}, K., {Alberdi}, A., {et~al.}
  2019{\natexlab{b}}, \apjl, 875, L1

\bibitem[{{Event Horizon Telescope Collaboration}
  {et~al.}(2019{\natexlab{c}}){Event Horizon Telescope Collaboration},
  {Akiyama}, {Alberdi}, {Alef}, {Asada}, {Azulay}, {Baczko}, {Ball},
  {Balokovi{\'c}}, {Barrett}, {Bintley}, {Blackburn}, {Boland}, {Bouman},
  {Bower}, {Bremer}, {Brinkerink}, {Brissenden}, {Britzen}, {Broderick},
  {Broguiere}, {Bronzwaer}, {Byun}, {Carlstrom}, {Chael}, {Chan}, {Chatterjee},
  {Chatterjee}, {Chen}, {Chen}, {Cho}, {Christian}, {Conway}, {Cordes}, {Crew},
  {Cui}, {Davelaar}, {De Laurentis}, {Deane}, {Dempsey}, {Desvignes}, {Dexter},
  {Doeleman}, {Eatough}, {Falcke}, {Fish}, {Fomalont}, {Fraga-Encinas},
  {Freeman}, {Friberg}, {Fromm}, {G{\'o}mez}, {Galison}, {Gammie},
  {Garc{\'\i}a}, {Gentaz}, {Georgiev}, {Goddi}, {Gold}, {Gu}, {Gurwell},
  {Hada}, {Hecht}, {Hesper}, {Ho}, {Ho}, {Honma}, {Huang}, {Huang}, {Hughes},
  {Ikeda}, {Inoue}, {Issaoun}, {James}, {Jannuzi}, {Janssen}, {Jeter}, {Jiang},
  {Johnson}, {Jorstad}, {Jung}, {Karami}, {Karuppusamy}, {Kawashima},
  {Keating}, {Kettenis}, {Kim}, {Kim}, {Kim}, {Kino}, {Koay}, {Koch}, {Koyama},
  {Kramer}, {Kramer}, {Krichbaum}, {Kuo}, {Lauer}, {Lee}, {Li}, {Li},
  {Lindqvist}, {Liu}, {Liuzzo}, {Lo}, {Lobanov}, {Loinard}, {Lonsdale}, {Lu},
  {MacDonald}, {Mao}, {Markoff}, {Marrone}, {Marscher}, {Mart{\'\i}-Vidal},
  {Matsushita}, {Matthews}, {Medeiros}, {Menten}, {Mizuno}, {Mizuno}, {Moran},
  {Moriyama}, {Moscibrodzka}, {M{\"u}ller}, {Nagai}, {Nagar}, {Nakamura},
  {Narayan}, {Narayanan}, {Natarajan}, {Neri}, {Ni}, {Noutsos}, {Okino},
  {Olivares}, {Oyama}, {{\"O}zel}, {Palumbo}, {Patel}, {Pen}, {Pesce},
  {Pi{\'e}tu}, {Plambeck}, {PopStefanija}, {Porth}, {Prather},
  {Preciado-L{\'o}pez}, {Psaltis}, {Pu}, {Ramakrishnan}, {Rao}, {Rawlings},
  {Raymond}, {Rezzolla}, {Ripperda}, {Roelofs}, {Rogers}, {Ros}, {Rose},
  {Roshanineshat}, {Rottmann}, {Roy}, {Ruszczyk}, {Ryan}, {Rygl},
  {S{\'a}nchez}, {S{\'a}nchez-Arguelles}, {Sasada}, {Savolainen}, {Schloerb},
  {Schuster}, {Shao}, {Shen}, {Small}, {Sohn}, {SooHoo}, {Tazaki}, {Tiede},
  {Tilanus}, {Titus}, {Toma}, {Torne}, {Trent}, {Trippe}, {Tsuda}, {van
  Bemmel}, {van Langevelde}, {van Rossum}, {Wagner}, {Wardle}, {Weintroub},
  {Wex}, {Wharton}, {Wielgus}, {Wong}, {Wu}, {Young}, {Young}, {Younsi},
  {Yuan}, {Yuan}, {Zensus}, {Zhao}, {Zhao}, {Zhu}, {Farah}, {Meyer-Zhao},
  {Michalik}, {Nadolski}, {Nishioka}, {Pradel}, {Primiani}, {Souccar},
  {Vertatschitsch}, \& {Yamaguchi}}]{EHT2019d}
{Event Horizon Telescope Collaboration}, {Akiyama}, K., {Alberdi}, A., {et~al.}
  2019{\natexlab{c}}, \apjl, 875, L4

\bibitem[{{Fabian} {et~al.}(2003){Fabian}, {Sanders}, {Allen}, {Crawford},
  {Iwasawa}, {Johnstone}, {Schmidt}, \& {Taylor}}]{Fabian2003}
{Fabian}, A.~C., {Sanders}, J.~S., {Allen}, S.~W., {et~al.} 2003, \mnras, 344,
  L43

\bibitem[{{Fabian} {et~al.}(2000){Fabian}, {Sanders}, {Ettori}, {Taylor},
  {Allen}, {Crawford}, {Iwasawa}, {Johnstone}, \& {Ogle}}]{Fabian2000}
{Fabian}, A.~C., {Sanders}, J.~S., {Ettori}, S., {et~al.} 2000, \mnras, 318,
  L65

\bibitem[{{Ford} {et~al.}(2014){Ford}, {Anderson}, {Belousov}, {Brandt},
  {Ford}, {Kanevsky}, {Kovalenko}, {Kovalev}, {Maddalena}, {Sergeev},
  {Smirnov}, {Watts}, \& {Weadon}}]{Ford2014}
{Ford}, H.~A., {Anderson}, R., {Belousov}, K., {et~al.} 2014, in \procspie,
  Vol. 9145, Ground-based and Airborne Telescopes V, 91450B

\bibitem[{{Forman} {et~al.}(1972){Forman}, {Kellogg}, {Gursky}, {Tananbaum}, \&
  {Giacconi}}]{Forman1972}
{Forman}, W., {Kellogg}, E., {Gursky}, H., {Tananbaum}, H., \& {Giacconi}, R.
  1972, \apj, 178, 309

\bibitem[{{Fujita} {et~al.}(2016){Fujita}, {Kawakatu}, {Shlosman}, \&
  {Ito}}]{Fujita2016}
{Fujita}, Y., {Kawakatu}, N., {Shlosman}, I., \& {Ito}, H. 2016, \mnras, 455,
  2289

\bibitem[{{Fujita} \& {Nagai}(2017)}]{Fujita2017}
{Fujita}, Y. \& {Nagai}, H. 2017, \mnras, 465, L94

\bibitem[{{Giovannini} {et~al.}(2018){Giovannini}, {Savolainen}, {Orienti},
  {Nakamura}, {Nagai}, {Kino}, {Giroletti}, {Hada}, {Bruni}, {Kovalev},
  {Anderson}, {D'Ammando}, {Hodgson}, {Honma}, {Krichbaum}, {Lee}, {Lico},
  {Lisakov}, {Lobanov}, {Petrov}, {Sohn}, {Sokolovsky}, {Voitsik}, {Zensus}, \&
  {Tingay}}]{Giovannini2018}
{Giovannini}, G., {Savolainen}, T., {Orienti}, M., {et~al.} 2018, NatAs, 2, 472

\bibitem[{{Hada} {et~al.}(2016){Hada}, {Kino}, {Doi}, {Nagai}, {Honma},
  {Akiyama}, {Tazaki}, {Lico}, {Giroletti}, {Giovannini}, {Orienti}, \&
  {Hagiwara}}]{Hada2016}
{Hada}, K., {Kino}, M., {Doi}, A., {et~al.} 2016, \apj, 817, 131

\bibitem[{{Hiura} {et~al.}(2018){Hiura}, {Nagai}, {Kino}, {Niinuma}, {Sorai},
  {Chida}, {Akiyama}, {D'Ammando}, {Giovannini}, {Giroletti}, {Hada}, {Honma},
  {Koyama}, {Orienti}, {Orosz}, \& {Sawada-Satoh}}]{Hiura2018}
{Hiura}, K., {Nagai}, H., {Kino}, M., {et~al.} 2018, \pasj, 70, 83

\bibitem[{{H{\"o}gbom}(1974)}]{Hogbom1974}
{H{\"o}gbom}, J.~A. 1974, \aaps, 15, 417

\bibitem[{{Janssen} {et~al.}(2021){Janssen}, {Falcke}, {Kadler}, {Ros},
  {Wielgus}, {Akiyama}, {Balokovi{\'c}}, {Blackburn}, {Bouman}, {Chael},
  {Chan}, {Chatterjee}, {Davelaar}, {Edwards}, {Fromm}, {G{\'o}mez}, {Goddi},
  {Issaoun}, {Johnson}, {Kim}, {Koay}, {Krichbaum}, {Liu}, {Liuzzo}, {Markoff},
  {Markowitz}, {Marrone}, {Mizuno}, {M{\"u}ller}, {Ni}, {Pesce},
  {Ramakrishnan}, {Roelofs}, {Rygl}, {van Bemmel}, {Event Horizon Telescope
  Collaboration}, {Alberdi}, {Alef}, {Algaba}, {Anantua}, {Asada}, {Azulay},
  {Baczko}, {Ball}, {Ball}, {Barrett}, {Benson}, {Bintley}, {Bintley},
  {Blundell}, {Boland}, {Boland}, {Bower}, {Boyce}, {Bremer}, {Brinkerink},
  {Brissenden}, {Britzen}, {Broderick}, {Broguiere}, {Bronzwaer}, {Byun},
  {Carlstrom}, {Carlstrom}, {Carlstrom}, {Carlstrom}, {Chatterjee}, {Chen},
  {Chen}, {Chesler}, {Cho}, {Christian}, {Conway}, {Cordes}, {Crawford},
  {Crew}, {Cruz-Osorio}, {Cui}, {Cui}, {De Laurentis}, {Deane}, {Dempsey},
  {Desvignes}, {Dexter}, {Doeleman}, {Eatough}, {Farah}, {Farah}, {Fish},
  {Fomalont}, {Ford}, {Fraga-Encinas}, {Friberg}, {Friberg}, {Fuentes},
  {Galison}, {Gammie}, {Garc{\'\i}a}, {Gelles}, {Gentaz}, {Georgiev},
  {Georgiev}, {Gold}, {Gold}, {G{\'o}mez-Ruiz}, {Gu}, {Gurwell}, {Hada},
  {Haggard}, {Hecht}, {Hesper}, {Himwich}, {Ho}, {Ho}, {Honma}, {Huang},
  {Huang}, {Hughes}, {Ikeda}, {Inoue}, {James}, {Jannuzi}, {Jeter}, {Jiang},
  {Jimenez-Rosales}, {Jorstad}, {Jung}, {Karami}, {Karuppusamy}, {Kawashima},
  {Keating}, {Kettenis}, {Kim}, {Kim}, {Kim}, {Kino}, {Kino}, {Kofuji},
  {Koyama}, {Kramer}, {Kramer}, {Kuo}, {Lauer}, {Lee}, {Levis}, {Li}, {Li},
  {Lindqvist}, {Lico}, {Lindahl}, {Lindahl}, {Liu}, {Lo}, {Lobanov}, {Loinard},
  {Lonsdale}, {Lu}, {MacDonald}, {Mao}, {Marchili}, {Marscher},
  {Mart{\'\i}-Vidal}, {Matsushita}, {Matthews}, {Medeiros}, {Menten}, {Mizuno},
  {Moran}, {Moriyama}, {Moscibrodzka}, {Musoke}, {Mej{\'\i}as}, {Nagai},
  {Nagar}, {Nakamura}, {Narayan}, {Narayanan}, {Natarajan}, {Nathanail},
  {Neilsen}, {Neri}, {Noutsos}, {Nowak}, {Okino}, {Olivares}, {Ortiz-Le{\'o}n},
  {Oyama}, {{\"O}zel}, {Palumbo}, {Park}, {Patel}, {Pen}, {Pi{\'e}tu},
  {Plambeck}, {PopStefanija}, {Porth}, {P{\"o}tzl}, {Prather},
  {Preciado-L{\'o}pez}, {Psaltis}, {Pu}, {Rao}, {Rawlings}, {Raymond},
  {Rezzolla}, {Ricarte}, {Ripperda}, {Rogers}, {Rose}, {Roshanineshat},
  {Rottmann}, {Roy}, {Ruszczyk}, {S{\'a}nchez}, {S{\'a}nchez-Arguelles},
  {Sasada}, {Savolainen}, {Schloerb}, {Schuster}, {Shao}, {Shen}, {Small},
  {Sohn}, {SooHoo}, {Sun}, {Tazaki}, {Tetarenko}, {Tiede}, {Tilanus}, {Titus},
  {Torne}, {Trent}, {Traianou}, {Trippe}, {van Bemmel}, {van Langevelde}, {van
  Rossum}, {Wagner}, {Ward-Thompson}, {Wardle}, {Weintroub}, {Wex}, {Wharton},
  {Wharton}, {Wong}, {Wu}, {Yoon}, {Young}, {Young}, {Younsi}, {Yuan}, {Yuan},
  {Zensus}, {Zhao}, \& {Zhao}}]{Janssen2021}
{Janssen}, M., {Falcke}, H., {Kadler}, M., {et~al.} 2021, NatAs, 5, 1017

\bibitem[{{Kardashev} {et~al.}(2013){Kardashev}, {Khartov}, {Abramov},
  {Avdeev}, {Alakoz}, {Aleksandrov}, {Ananthakrishnan}, {Andreyanov},
  {Andrianov}, {Antonov}, {Artyukhov}, {Arkhipov}, {Baan}, {Babakin},
  {Babyshkin}, {Bartel'}, {Belousov}, {Belyaev}, {Berulis}, {Burke},
  {Biryukov}, {Bubnov}, {Burgin}, {Busca}, {Bykadorov}, {Bychkova},
  {Vasil'kov}, {Wellington}, {Vinogradov}, {Wietfeldt}, {Voitsik},
  {Gvamichava}, {Girin}, {Gurvits}, {Dagkesamanskii}, {D'Addario},
  {Giovannini}, {Jauncey}, {Dewdney}, {D'yakov}, {Zharov}, {Zhuravlev},
  {Zaslavskii}, {Zakhvatkin}, {Zinov'ev}, {Ilinen}, {Ipatov}, {Kanevskii},
  {Knorin}, {Casse}, {Kellermann}, {Kovalev}, {Kovalev}, {Kovalenko}, {Kogan},
  {Komaev}, {Konovalenko}, {Kopelyanskii}, {Korneev}, {Kostenko}, {Kotik},
  {Kreisman}, {Kukushkin}, {Kulishenko}, {Cooper}, {Kut'kin}, {Cannon},
  {Larionov}, {Lisakov}, {Litvinenko}, {Likhachev}, {Likhacheva}, {Lobanov},
  {Logvinenko}, {Langston}, {McCracken}, {Medvedev}, {Melekhin}, {Menderov},
  {Murphy}, {Mizyakina}, {Mozgovoi}, {Nikolaev}, {Novikov}, {Novikov},
  {Oreshko}, {Pavlenko}, {Pashchenko}, {Ponomarev}, {Popov}, {Pravin-Kumar},
  {Preston}, {Pyshnov}, {Rakhimov}, {Rozhkov}, {Romney}, {Rocha}, {Rudakov},
  {R{\"a}is{\"a}nen}, {Sazankov}, {Sakharov}, {Semenov}, {Serebrennikov},
  {Schilizzi}, {Skulachev}, {Slysh}, {Smirnov}, {Smith}, {Soglasnov},
  {Sokolovskii}, {Sondaar}, {Stepan'yants}, {Turygin}, {Turygin}, {Tuchin},
  {Urpo}, {Fedorchuk}, {Finkel'shtein}, {Fomalont}, {Fejes}, {Fomina},
  {Khapin}, {Tsarevskii}, {Zensus}, {Chuprikov}, {Shatskaya}, {Shapirovskaya},
  {Sheikhet}, {Shirshakov}, {Schmidt}, {Shnyreva}, {Shpilevskii}, {Ekers}, \&
  {Yakimov}}]{Kardashev2013}
{Kardashev}, N.~S., {Khartov}, V.~V., {Abramov}, V.~V., {et~al.} 2013,
  Astronomy Reports, 57, 153

\bibitem[{{Kettenis} {et~al.}(2006){Kettenis}, {van Langevelde}, {Reynolds}, \&
  {Cotton}}]{Kettenis2006}
{Kettenis}, M., {van Langevelde}, H.~J., {Reynolds}, C., \& {Cotton}, B. 2006,
  in Astronomical Society of the Pacific Conference Series, Vol. 351,
  Astronomical Data Analysis Software and Systems XV, ed. C.~{Gabriel},
  C.~{Arviset}, D.~{Ponz}, \& S.~{Enrique}, 497

\bibitem[{{Kim} {et~al.}(2018){Kim}, {Krichbaum}, {Lu}, {Ros}, {Bach},
  {Bremer}, {de Vicente}, {Lindqvist}, \& {Zensus}}]{Kim2018}
{Kim}, J.~Y., {Krichbaum}, T.~P., {Lu}, R.~S., {et~al.} 2018, \aap, 616, A188

\bibitem[{{Kim} {et~al.}(2019){Kim}, {Krichbaum}, {Marscher}, {Jorstad},
  {Agudo}, {Thum}, {Hodgson}, {MacDonald}, {Ros}, {Lu}, {Bremer}, {de Vicente},
  {Lindqvist}, {Trippe}, \& {Zensus}}]{Kim2019}
{Kim}, J.~Y., {Krichbaum}, T.~P., {Marscher}, A.~P., {et~al.} 2019, \aap, 622,
  A196

\bibitem[{Kino {et~al.}(2021)Kino, Niinuma, Kawakatu, Nagai, Giovannini,
  Orienti, Wajima, D'Ammando, Hada, Giroletti, \& Gurwell}]{Kino2021}
Kino, M., Niinuma, K., Kawakatu, N., {et~al.} 2021, \apj, 920, L24

\bibitem[{{Kino} {et~al.}(2018){Kino}, {Wajima}, {Kawakatu}, {Nagai},
  {Orienti}, {Giovannini}, {Hada}, {Niinuma}, \& {Giroletti}}]{Kino2018}
{Kino}, M., {Wajima}, K., {Kawakatu}, N., {et~al.} 2018, \apj, 864, 118

\bibitem[{{Kogan}(1996)}]{Kogan1996}
{Kogan}, L. 1996, {Global Ground VLBI Network as a Tied Array for Space VLBI},
  VLBA Scientific Memo~13, NRAO

\bibitem[{{Komissarov} \& {Porth}(2021)}]{Komissarov2021}
{Komissarov}, S. \& {Porth}, O. 2021, \nar, 92, 101610

\bibitem[{{Komissarov} \& {Falle}(1998)}]{Komissarov1998}
{Komissarov}, S.~S. \& {Falle}, S.~A.~E.~G. 1998, \mnras, 297, 1087

\bibitem[{{Kovalev} {et~al.}(2014){Kovalev}, {Vasil'kov}, {Popov}, {Soglasnov},
  {Voitsik}, {Lisakov}, {Kut'kin}, {Nikolaev}, {Nizhel'skii}, {Zhekanis}, \&
  {Tsybulev}}]{Kovalev2014}
{Kovalev}, Y.~A., {Vasil'kov}, V.~I., {Popov}, M.~V., {et~al.} 2014, Cosmic
  Research, 52, 393

\bibitem[{{Kovalev} {et~al.}(2020){Kovalev}, {Kardashev}, {Sokolovsky},
  {Voitsik}, {An}, {Anderson}, {Andrianov}, {Avdeev}, {Bartel}, {Bignall},
  {Burgin}, {Edwards}, {Ellingsen}, {Frey}, {Garc{\'\i}a-Mir{\'o}},
  {Gawro{\'n}ski}, {Ghigo}, {Ghosh}, {Giovannini}, {Girin}, {Giroletti},
  {Gurvits}, {Jauncey}, {Horiuchi}, {Ivanov}, {Kharinov}, {Koay}, {Kostenko},
  {Kovalenko}, {Kovalev}, {Kravchenko}, {Kunert-Bajraszewska}, {Kutkin},
  {Likhachev}, {Lisakov}, {Litovchenko}, {McCallum}, {Melis}, {Melnikov},
  {Migoni}, {Nair}, {Pashchenko}, {Phillips}, {Polatidis}, {Pushkarev},
  {Quick}, {Rakhimov}, {Reynolds}, {Rizzo}, {Rudnitskiy}, {Savolainen},
  {Shakhvorostova}, {Shatskaya}, {Shen}, {Shchurov}, {Vermeulen}, {de Vicente},
  {Wolak}, {Zensus}, \& {Zuga}}]{2020AdSpR..65..705K}
{Kovalev}, Y.~Y., {Kardashev}, N.~S., {Sokolovsky}, K.~V., {et~al.} 2020,
  Advances in Space Research, 65, 705

\bibitem[{{Lewis} \& {Austin}(2002)}]{Lewis2002}
{Lewis}, G.~M. \& {Austin}, P.~H. 2002, in 11th Conference on Atmospheric
  Radiation, American Meteorological Society Conference Series, ed. G.~{Smith}
  \& J.~{Brodie}, 123--126

\bibitem[{{Lister}(2001)}]{Lister2001}
{Lister}, M.~L. 2001, \apj, 562, 208

\bibitem[{{Lister} {et~al.}(2009{\natexlab{a}}){Lister}, {Aller}, {Aller},
  {Cohen}, {Homan}, {Kadler}, {Kellermann}, {Kovalev}, {Ros}, {Savolainen},
  {Zensus}, \& {Vermeulen}}]{Lister2009a}
{Lister}, M.~L., {Aller}, H.~D., {Aller}, M.~F., {et~al.} 2009{\natexlab{a}},
  \aj, 137, 3718

\bibitem[{{Lister} {et~al.}(2009{\natexlab{b}}){Lister}, {Cohen}, {Homan},
  {Kadler}, {Kellermann}, {Kovalev}, {Ros}, {Savolainen}, \&
  {Zensus}}]{Lister2009b}
{Lister}, M.~L., {Cohen}, M.~H., {Homan}, D.~C., {et~al.} 2009{\natexlab{b}},
  \aj, 138, 1874

\bibitem[{{Lister} {et~al.}(2020){Lister}, {Homan}, {Kovalev}, {Mandal},
  {Pushkarev}, \& {Siemiginowska}}]{Lister2020}
{Lister}, M.~L., {Homan}, D.~C., {Kovalev}, Y.~Y., {et~al.} 2020, \apj, 899,
  141

\bibitem[{{Lobanov}(2005)}]{Lobanov2005}
{Lobanov}, A.~P. 2005, ArXiv Astrophysics e-prints [\eprint{astro-ph/0503225}]

\bibitem[{{MAGIC Collaboration} {et~al.}(2018){MAGIC Collaboration}, {Ansoldi},
  {Antonelli}, {Arcaro}, {Baack}, {Babi{\'c}}, {Banerjee}, {Bangale}, {Barres
  de Almeida}, {Barrio}, {Becerra Gonz{\'a}lez}, {Bednarek}, {Bernardini},
  {Berse}, {Berti}, {Bhattacharyya}, {Bigongiari}, {Biland}, {Blanch},
  {Bonnoli}, {Carosi}, {Ceribella}, {Chatterjee}, {Colak}, {Colin}, {Colombo},
  {Contreras}, {Cortina}, {Covino}, {Cumani}, {D'Elia}, {da Vela}, {Dazzi}, {de
  Angelis}, {de Lotto}, {Delfino}, {Delgado}, {di Pierro}, {Dom{\'\i}nguez},
  {Dominis Prester}, {Dorner}, {Doro}, {Einecke}, {Elsaesser}, {Fallah
  Ramazani}, {Fattorini}, {Fern{\'a}ndez-Barral}, {Ferrara}, {Fidalgo},
  {Foffano}, {Fonseca}, {Font}, {Fruck}, {Galindo}, {Gallozzi}, {Garc{\'\i}a
  L{\'o}pez}, {Garczarczyk}, {Gaug}, {Giammaria}, {Godinovi{\'c}}, {Gora},
  {Guberman}, {Hadasch}, {Hahn}, {Hassan}, {Hayashida}, {Herrera}, {Hoang},
  {Hose}, {Hrupec}, {Ishio}, {Konno}, {Kubo}, {Kushida}, {Lamastra}, {Lelas},
  {Leone}, {Lindfors}, {Lombardi}, {Longo}, {L{\'o}pez}, {Maggio}, {Majumdar},
  {Makariev}, {Maneva}, {Manganaro}, {Mannheim}, {Maraschi}, {Mariotti},
  {Mart{\'\i}nez}, {Masuda}, {Mazin}, {Mielke}, {Minev}, {Miranda}, {Mirzoyan},
  {Moralejo}, {Moreno}, {Moretti}, {Nagayoshi}, {Neustroev}, {Niedzwiecki},
  {Nievas Rosillo}, {Nigro}, {Nilsson}, {Ninci}, {Nishijima}, {Noda},
  {Nogu{\'e}s}, {Paiano}, {Palacio}, {Paneque}, {Paoletti}, {Paredes},
  {Pedaletti}, {Pe{\~n}il}, {Peresano}, {Persic}, {Pfrang}, {Prada Moroni},
  {Prandini}, {Puljak}, {Garcia}, {Reichardt}, {Rhode}, {Rib{\'o}}, {Rico},
  {Righi}, {Rugliancich}, {Saha}, {Saito}, {Satalecka}, {Schweizer}, {Sitarek},
  {{\v{S}}nidari{\'c}}, {Sobczynska}, {Stamerra}, {Strzys}, {Suri{\'c}},
  {Takahashi}, {Tavecchio}, {Temnikov}, {Terzi{\'c}}, {Teshima},
  {Torres-Alb{\`a}}, {Tsujimoto}, {Vanzo}, {Vazquez Acosta}, {Vovk}, {Ward},
  {Will}, {Zari{\'c}}, {Glawion}, {Takalo}, \& {Jormanainen}}]{MAGIC2018}
{MAGIC Collaboration}, {Ansoldi}, S., {Antonelli}, L.~A., {et~al.} 2018, \aap,
  617, A91

\bibitem[{{Mart{\'\i}} {et~al.}(1997){Mart{\'\i}}, {M{\"u}ller}, {Font},
  {Ib{\'a}{\~n}ez}, \& {Marquina}}]{Marti1997}
{Mart{\'\i}}, J.~M., {M{\"u}ller}, E., {Font}, J.~A., {Ib{\'a}{\~n}ez},
  J.~M.~Z., \& {Marquina}, A. 1997, \apj, 479, 151

\bibitem[{{Mart{\'{\i}}-Vidal} \& {Marcaide}(2008)}]{MartiVidal2008}
{Mart{\'{\i}}-Vidal}, I. \& {Marcaide}, J.~M. 2008, \aap, 480, 289

\bibitem[{{McNamara} \& {Nulsen}(2007)}]{MacNamara2007}
{McNamara}, B.~R. \& {Nulsen}, P.~E.~J. 2007, \araa, 45, 117

\bibitem[{{McNamara} \& {Nulsen}(2012)}]{McNamara2012}
{McNamara}, B.~R. \& {Nulsen}, P.~E.~J. 2012, New Journal of Physics, 14,
  055023

\bibitem[{{Morganti} {et~al.}(2013){Morganti}, {Fogasy}, {Paragi}, {Oosterloo},
  \& {Orienti}}]{Morganti2013}
{Morganti}, R., {Fogasy}, J., {Paragi}, Z., {Oosterloo}, T., \& {Orienti}, M.
  2013, Science, 341, 1082

\bibitem[{{Morganti} {et~al.}(2005){Morganti}, {Tadhunter}, \&
  {Oosterloo}}]{Morganti2005}
{Morganti}, R., {Tadhunter}, C.~N., \& {Oosterloo}, T.~A. 2005, \aap, 444, L9

\bibitem[{{Mukherjee} {et~al.}(2018){Mukherjee}, {Bicknell}, {Wagner},
  {Sutherland}, \& {Silk}}]{Mukherjee2018}
{Mukherjee}, D., {Bicknell}, G.~V., {Wagner}, A.~Y., {Sutherland}, R.~S., \&
  {Silk}, J. 2018, \mnras, 479, 5544

\bibitem[{{Murphy}(2000)}]{Murphy2000}
{Murphy}, D.~W. 2000, Advances in Space Research, 26, 609

\bibitem[{{Nagai} {et~al.}(2017){Nagai}, {Fujita}, {Nakamura}, {Orienti},
  {Kino}, {Asada}, \& {Giovannini}}]{Nagai2017}
{Nagai}, H., {Fujita}, Y., {Nakamura}, M., {et~al.} 2017, \apj, 849, 52

\bibitem[{{Nagai} {et~al.}(2014){Nagai}, {Haga}, {Giovannini}, {Doi},
  {Orienti}, {D'Ammando}, {Kino}, {Nakamura}, {Asada}, {Hada}, \&
  {Giroletti}}]{Nagai2014}
{Nagai}, H., {Haga}, T., {Giovannini}, G., {et~al.} 2014, \apj, 785, 53

\bibitem[{{Nagai} {et~al.}(2019){Nagai}, {Onishi}, {Kawakatu}, {Fujita},
  {Kino}, {Fukazawa}, {Lim}, {Forman}, {Vrtilek}, {Nakanishi}, {Noda}, {Asada},
  {Wajima}, {Ohyama}, {David}, \& {Daikuhara}}]{Nagai2019}
{Nagai}, H., {Onishi}, K., {Kawakatu}, N., {et~al.} 2019, \apj, 883, 193

\bibitem[{{Nagai} {et~al.}(2010){Nagai}, {Suzuki}, {Asada}, {Kino}, {Kameno},
  {Doi}, {Inoue}, {Kataoka}, {Bach}, {Hirota}, {Matsumoto}, {Honma},
  {Kobayashi}, \& {Fujisawa}}]{Nagai2010}
{Nagai}, H., {Suzuki}, K., {Asada}, K., {et~al.} 2010, \pasj, 62, L11

\bibitem[{{Nesterov} {et~al.}(1995){Nesterov}, {Lyuty}, \&
  {Valtaoja}}]{Nesterov1995}
{Nesterov}, N.~S., {Lyuty}, V.~M., \& {Valtaoja}, E. 1995, \aap, 296, 628

\bibitem[{{Pacholczyk}(1970)}]{Pacholcyk1970}
{Pacholczyk}, A.~G. 1970, {Radio astrophysics. Nonthermal processes in galactic
  and extragalactic sources} ({San~Francisco}: {W.~H.~Freeman and Company})

\bibitem[{{Paraschos} {et~al.}(2021){Paraschos}, {Kim}, {Krichbaum}, \&
  {Zensus}}]{Paraschos2021}
{Paraschos}, G.~F., {Kim}, J.~Y., {Krichbaum}, T.~P., \& {Zensus}, J.~A. 2021,
  \aap, 650, L18

\bibitem[{{Paraschos} {et~al.}(2023){Paraschos}, {Mpisketzis}, {Kim}, {Witzel},
  {Krichbaum}, {Zensus}, {Gurwell}, {L{\"a}hteenm{\"a}ki}, {Tornikoski},
  {Kiehlmann}, \& {Readhead}}]{Paraschos2022}
{Paraschos}, G.~F., {Mpisketzis}, V., {Kim}, J.~Y., {et~al.} 2023, \aap, 669,
  A32

\bibitem[{{Pauliny-Toth} \& {Kellermann}(1966)}]{PaulinyToth1966}
{Pauliny-Toth}, I.~I.~K. \& {Kellermann}, K.~I. 1966, \apj, 146, 634

\bibitem[{{Pedlar} {et~al.}(1990){Pedlar}, {Ghataure}, {Davies}, {Harrison},
  {Perley}, {Crane}, \& {Unger}}]{Pedlar1990}
{Pedlar}, A., {Ghataure}, H.~S., {Davies}, R.~D., {et~al.} 1990, \mnras, 246,
  477

\bibitem[{{Perucho} {et~al.}(2021){Perucho}, {L{\'o}pez-Miralles}, {Reynaldi},
  \& {Labiano}}]{Perucho2021}
{Perucho}, M., {L{\'o}pez-Miralles}, J., {Reynaldi}, V., \& {Labiano}, {\'A}.
  2021, Astronomische Nachrichten, 342, 1171

\bibitem[{{Perucho} \& {Mart{\'\i}}(2007)}]{Perucho2007b}
{Perucho}, M. \& {Mart{\'\i}}, J.~M. 2007, \mnras, 382, 526

\bibitem[{{Perucho} {et~al.}(2010){Perucho}, {Mart{\'\i}}, {Cela}, {Hanasz},
  {de La Cruz}, \& {Rubio}}]{Perucho2010}
{Perucho}, M., {Mart{\'\i}}, J.~M., {Cela}, J.~M., {et~al.} 2010, \aap, 519,
  A41

\bibitem[{{Perucho} {et~al.}(2019){Perucho}, {Mart{\'\i}}, \&
  {Quilis}}]{Perucho2019}
{Perucho}, M., {Mart{\'\i}}, J.-M., \& {Quilis}, V. 2019, \mnras, 482, 3718

\bibitem[{{Perucho} {et~al.}(2017){Perucho}, {Mart{\'\i}}, {Quilis}, \&
  {Borja-Lloret}}]{Perucho2017}
{Perucho}, M., {Mart{\'\i}}, J.-M., {Quilis}, V., \& {Borja-Lloret}, M. 2017,
  \mnras, 471, L120

\bibitem[{{Perucho} {et~al.}(2014){Perucho}, {Mart{\'\i}}, {Quilis}, \&
  {Ricciardelli}}]{Perucho2014}
{Perucho}, M., {Mart{\'\i}}, J.-M., {Quilis}, V., \& {Ricciardelli}, E. 2014,
  \mnras, 445, 1462

\bibitem[{{Perucho} {et~al.}(2011){Perucho}, {Quilis}, \&
  {Mart{\'\i}}}]{Perucho2011}
{Perucho}, M., {Quilis}, V., \& {Mart{\'\i}}, J.-M. 2011, \apj, 743, 42

\bibitem[{{Petrov}(2014)}]{Petrov2015}
{Petrov}, L. 2014, in Proceedings of the 12th European VLBI Network Symposium
  and Users Meeting (EVN 2014). 7-10 October 2014. Cagliari, 35

\bibitem[{{Petrov} {et~al.}(2011){Petrov}, {Kovalev}, {Fomalont}, \&
  {Gordon}}]{Petrov2011}
{Petrov}, L., {Kovalev}, Y.~Y., {Fomalont}, E.~B., \& {Gordon}, D. 2011, \aj,
  142, 35

\bibitem[{{Pushkarev} {et~al.}(2012){Pushkarev}, {Hovatta}, {Kovalev},
  {Lister}, {Lobanov}, {Savolainen}, \& {Zensus}}]{Pushkarev2012}
{Pushkarev}, A.~B., {Hovatta}, T., {Kovalev}, Y.~Y., {et~al.} 2012, \aap, 545,
  A113

\bibitem[{{Rafferty} {et~al.}(2006){Rafferty}, {McNamara}, {Nulsen}, \&
  {Wise}}]{Rafferty2006}
{Rafferty}, D.~A., {McNamara}, B.~R., {Nulsen}, P.~E.~J., \& {Wise}, M.~W.
  2006, \apj, 652, 216

\bibitem[{{Readhead} {et~al.}(1983){Readhead}, {Mason}, {Mofett}, {Pearson},
  {Seielstad}, {Woody}, {Backer}, {Plambeck}, {Welch}, {Wright}, {Rogers},
  {Webber}, {Shapiro}, {Moran}, {Goldsmith}, {Predmore}, {Baath}, \&
  {Ronnang}}]{Readhead1983}
{Readhead}, A.~C.~S., {Mason}, C.~R., {Mofett}, A.~T., {et~al.} 1983, \nat,
  303, 504

\bibitem[{{Rieger}(2019)}]{Rieger2019}
{Rieger}, F.~M. 2019, Galaxies, 7, 78

\bibitem[{{Rienecker} {et~al.}(2018){Rienecker}, {Suarez}, {Todling},
  {Bacmeister}, {Takacs}, {Liu}, {Sienkiewicz}, {Koster}, {Gelaro}, I., \&
  E.}]{Rienecker2018}
{Rienecker}, M., {Suarez}, M., {Todling}, R., {et~al.} 2018, NASA Technical
  Memorandum, 104606, 1

\bibitem[{{Schulz} {et~al.}(2020){Schulz}, {Kadler}, {Ros}, {Perucho},
  {Krichbaum}, {Agudo}, {Beuchert}, {Lindqvist}, {Mannheim}, {Wilms}, \&
  {Zensus}}]{Schulz2020}
{Schulz}, R., {Kadler}, M., {Ros}, E., {et~al.} 2020, \aap, 644, A85

\bibitem[{{Schwab} \& {Cotton}(1983)}]{Schwab1983}
{Schwab}, F.~R. \& {Cotton}, W.~D. 1983, \aj, 88, 688

\bibitem[{{Shepherd}(1997)}]{Shepherd1997}
{Shepherd}, M.~C. 1997, in Astronomical Society of the Pacific Conference
  Series, Vol. 125, Astronomical Data Analysis Software and Systems VI, ed.
  G.~{Hunt} \& H.~{Payne}, 77

\bibitem[{{Shklovsky}(1966)}]{Shklovsky1966}
{Shklovsky}, I.~S. 1966, \sovast, 9, 683

\bibitem[{{Strauss} {et~al.}(1992){Strauss}, {Huchra}, {Davis}, {Yahil},
  {Fisher}, \& {Tonry}}]{Strauss1992}
{Strauss}, M.~A., {Huchra}, J.~P., {Davis}, M., {et~al.} 1992, \apjs, 83, 29

\bibitem[{{Suzuki} {et~al.}(2012){Suzuki}, {Nagai}, {Kino}, {Kataoka}, {Asada},
  {Doi}, {Inoue}, {Orienti}, {Giovannini}, {Giroletti}, {L{\"a}hteenm{\"a}ki},
  {Tornikoski}, {Le{\'o}n-Tavares}, {Bach}, {Kameno}, \&
  {Kobayashi}}]{Suzuki2012}
{Suzuki}, K., {Nagai}, H., {Kino}, M., {et~al.} 2012, \apj, 746, 140

\bibitem[{{Tavecchio} \& {Ghisellini}(2014)}]{Tavecchio2014}
{Tavecchio}, F. \& {Ghisellini}, G. 2014, \mnras, 443, 1224

\bibitem[{{Tchekhovskoy} {et~al.}(2011){Tchekhovskoy}, {Narayan}, \&
  {McKinney}}]{Tchekhovskoy2011}
{Tchekhovskoy}, A., {Narayan}, R., \& {McKinney}, J.~C. 2011, \mnras, 418, L79

\bibitem[{{Thompson} {et~al.}(2007){Thompson}, {Moran}, \&
  {Swenson}}]{Thompson2007}
{Thompson}, A.~R., {Moran}, J.~M., \& {Swenson}, G.~W. 2007, {Interferometry
  and Synthesis in Radio Astronomy} ({New~York}: John Wiley \& Sons)

\bibitem[{{Vaidya} {et~al.}(2015){Vaidya}, {Mignone}, {Bodo}, \&
  {Massaglia}}]{Vaidya2015}
{Vaidya}, B., {Mignone}, A., {Bodo}, G., \& {Massaglia}, S. 2015, \aap, 580,
  A110

\bibitem[{{Wagner} \& {Bicknell}(2011)}]{Wagner2011}
{Wagner}, A.~Y. \& {Bicknell}, G.~V. 2011, \apj, 728, 29

\bibitem[{{Wagner} {et~al.}(2012){Wagner}, {Bicknell}, \&
  {Umemura}}]{Wagner2012}
{Wagner}, A.~Y., {Bicknell}, G.~V., \& {Umemura}, M. 2012, \apj, 757, 136

\bibitem[{{Walker} {et~al.}(2000){Walker}, {Dhawan}, {Romney}, {Kellermann}, \&
  {Vermeulen}}]{Walker2000}
{Walker}, R.~C., {Dhawan}, V., {Romney}, J.~D., {Kellermann}, K.~I., \&
  {Vermeulen}, R.~C. 2000, \apj, 530, 233

\bibitem[{{Walker} {et~al.}(1994){Walker}, {Romney}, \& {Benson}}]{Walker1994}
{Walker}, R.~C., {Romney}, J.~D., \& {Benson}, J.~M. 1994, \apjl, 430, L45

\bibitem[{{Zakhvatkin} {et~al.}(2020){Zakhvatkin}, {Andrianov}, {Avdeev},
  {Kostenko}, {Kovalev}, {Likhachev}, {Litovchenko}, {Litvinov}, {Rudnitskiy},
  {Shchurov}, {Sokolovsky}, {Stepanyants}, {Tuchin}, {Voitsik}, {Zaslavskiy},
  {Zharov}, \& {Zuga}}]{2020AdSpR..65..798Z}
{Zakhvatkin}, M.~V., {Andrianov}, A.~S., {Avdeev}, V.~Y., {et~al.} 2020,
  Advances in Space Research, 65, 798

\bibitem[{{Zakhvatkin} {et~al.}(2014){Zakhvatkin}, {Ponomarev}, {Stepan'yants},
  {Tuchin}, \& {Zaslavskiy}}]{2014CosRe..52..342Z}
{Zakhvatkin}, M.~V., {Ponomarev}, Y.~N., {Stepan'yants}, V.~A., {Tuchin},
  A.~G., \& {Zaslavskiy}, G.~S. 2014, Cosmic Research, 52, 342

\bibitem[{{Zamaninasab} {et~al.}(2014){Zamaninasab}, {Clausen-Brown},
  {Savolainen}, \& {Tchekhovskoy}}]{Zamaninasab2014}
{Zamaninasab}, M., {Clausen-Brown}, E., {Savolainen}, T., \& {Tchekhovskoy}, A.
  2014, \nat, 510, 126

\end{thebibliography}

\begin{appendix}

\section{Probability of false detection in the fringe search} \label{false_det}

\begin{figure}
\centering
\includegraphics[width=\columnwidth]{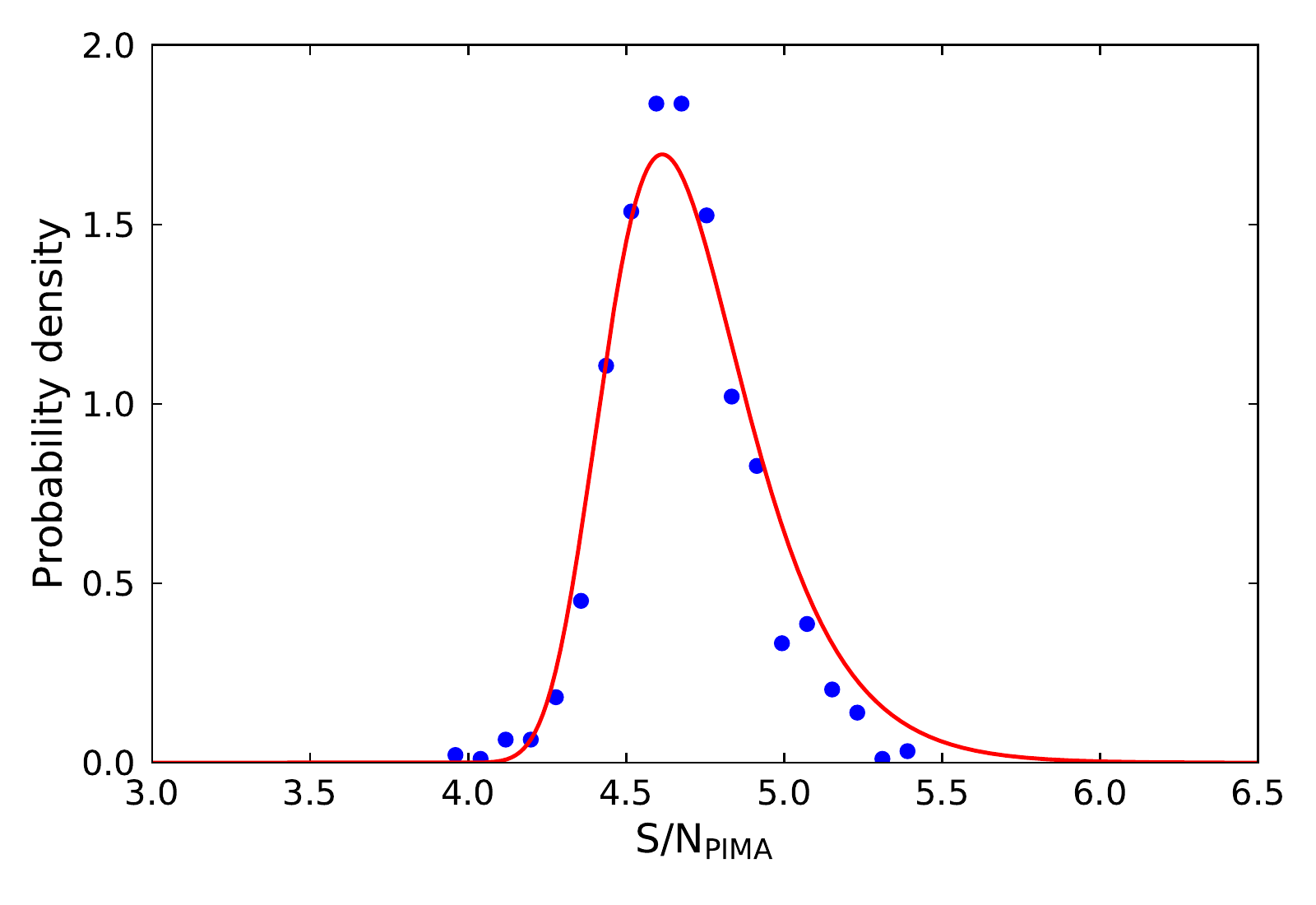}\\
\includegraphics[width=\columnwidth]{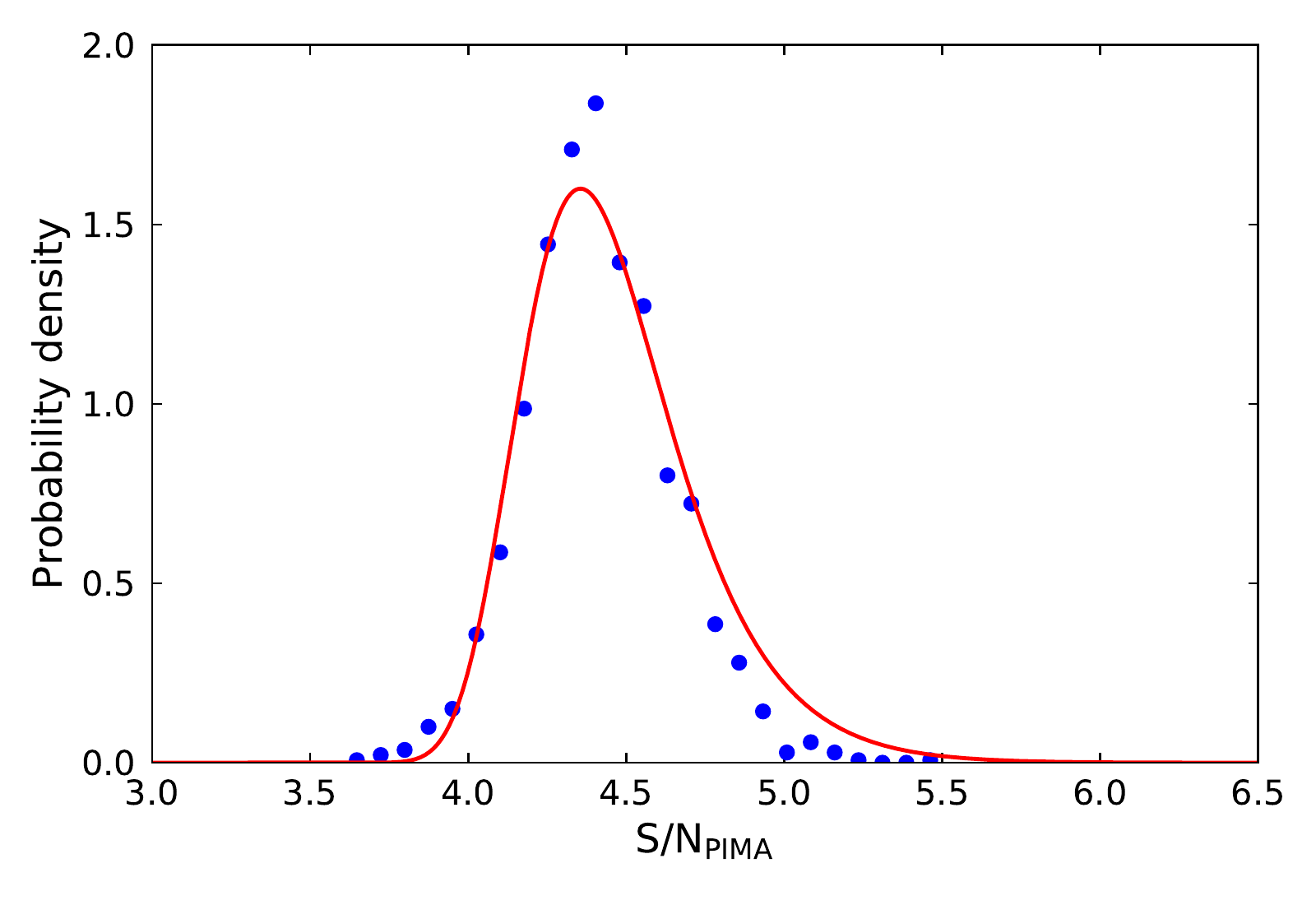}
\caption{Empirical distribution of the maximum $\mathrm{S/N}_\mathrm{PIMA}$ of the fringe amplitude from \textsc{PIMA} fringe-fitting data on the SRT baselines n the C band (\emph{top}) and K band (\emph{bottom}) when no signal was present in the chosen search region (blue points). The red curve shows the fitted theoretical probability distribution. The shift in the maximum position is mainly due to the different size of the search space: C band fringe-fitting was carried out first and therefore a larger search space was used than in the K band.}
\label{false_snr_fig}
\end{figure}

In order to calculate the false detection rates in the fringe search on space baselines when using baseline-based fringe search implemented in \textsc{PIMA}, we estimated the probability density distribution of the maximum S/N, $\mathrm{S/N}_\mathrm{PIMA}$, for the fringe amplitude in the case of no signal by following the method described in \citet{Petrov2011}. Figure~\ref{false_snr_fig} shows the empirical distribution of $\mathrm{S/N}_\mathrm{PIMA}$ of the fringe peak found by \textsc{PIMA} on the baselines to the SRT in the C and K bands, when the search region was set far from the known signal location \citep[compare it with the results of the \textit{RadioAstron} AGN survey statistics in][]{2020AdSpR..65..705K}. The size of the search space was set to be the same as the one used in the actual fringe searches: $\pm 2$\,$\mu$s in delay, $\pm 2 \times 10^{-11}$\,s/s in rate and $\pm 1 \times 10^{-13}$\,s/s$^2$ in acceleration in the C band and $\pm 0.6$\,$\mu$s in delay, $\pm 5 \times 10^{-12}$\,s/s in rate and $\pm 1 \times 10^{-13}$\,s/s$^2$ in acceleration in the K band. This difference in the size of the search space results in a difference in the $\mathrm{S/N}_\mathrm{PIMA}$ distributions between the bands. The figure also shows a fitted theoretical probability distribution of:
\begin{equation}
p(s) = \frac{n_\mathrm{eff}}{\sigma_\mathrm{eff}} s e^{ - \frac{s^2}{2} } \Big( 1-e^{ - \frac{s^2}{2} }  \Big)^{n_\mathrm{eff}-1} 
\end{equation}
where $s$ is the $\mathrm{S/N}_\mathrm{PIMA}$ of the maximum fringe amplitude. The parameters $\sigma_\mathrm{eff}$ and $n_\mathrm{eff}$ are the effective rms noise of the correlator output and the effective number of spectrum points in the search region. If all the $n$ samples in the delay-rate--acceleration space were statistically independent, $n_\mathrm{eff}$ would be equal to $n$. This, however, is not generally the case \citep{Petrov2011}, and we fit for $\sigma_\mathrm{eff}$ and $n_\mathrm{eff}$. With these estimates, we calculated the false detection probabilities in this experiment for the above-mentioned search spaces (see Table~\ref{false_snr_tab}).

\begin{figure}
\centering
\includegraphics[width=\columnwidth]{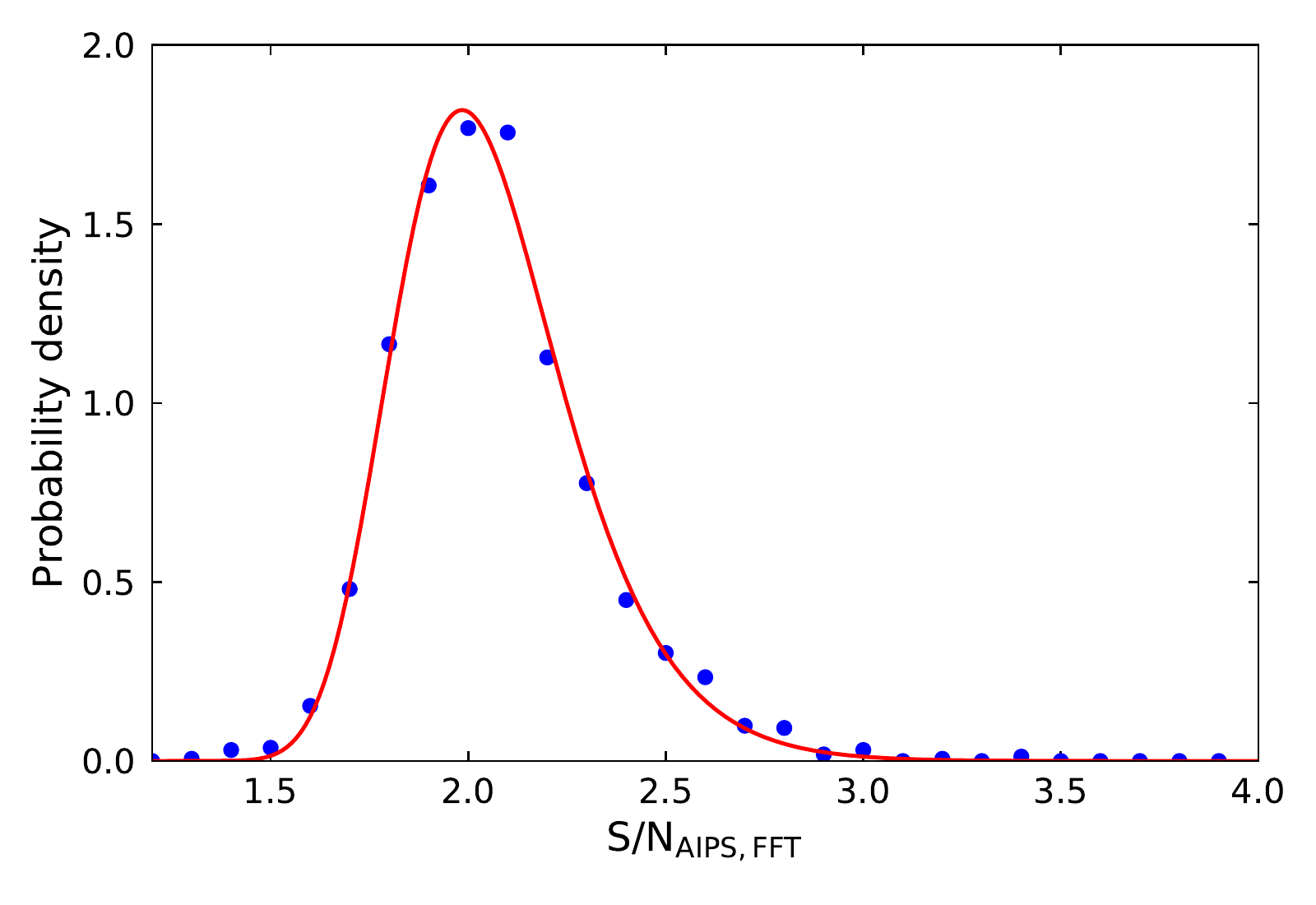}
\includegraphics[width=\columnwidth]{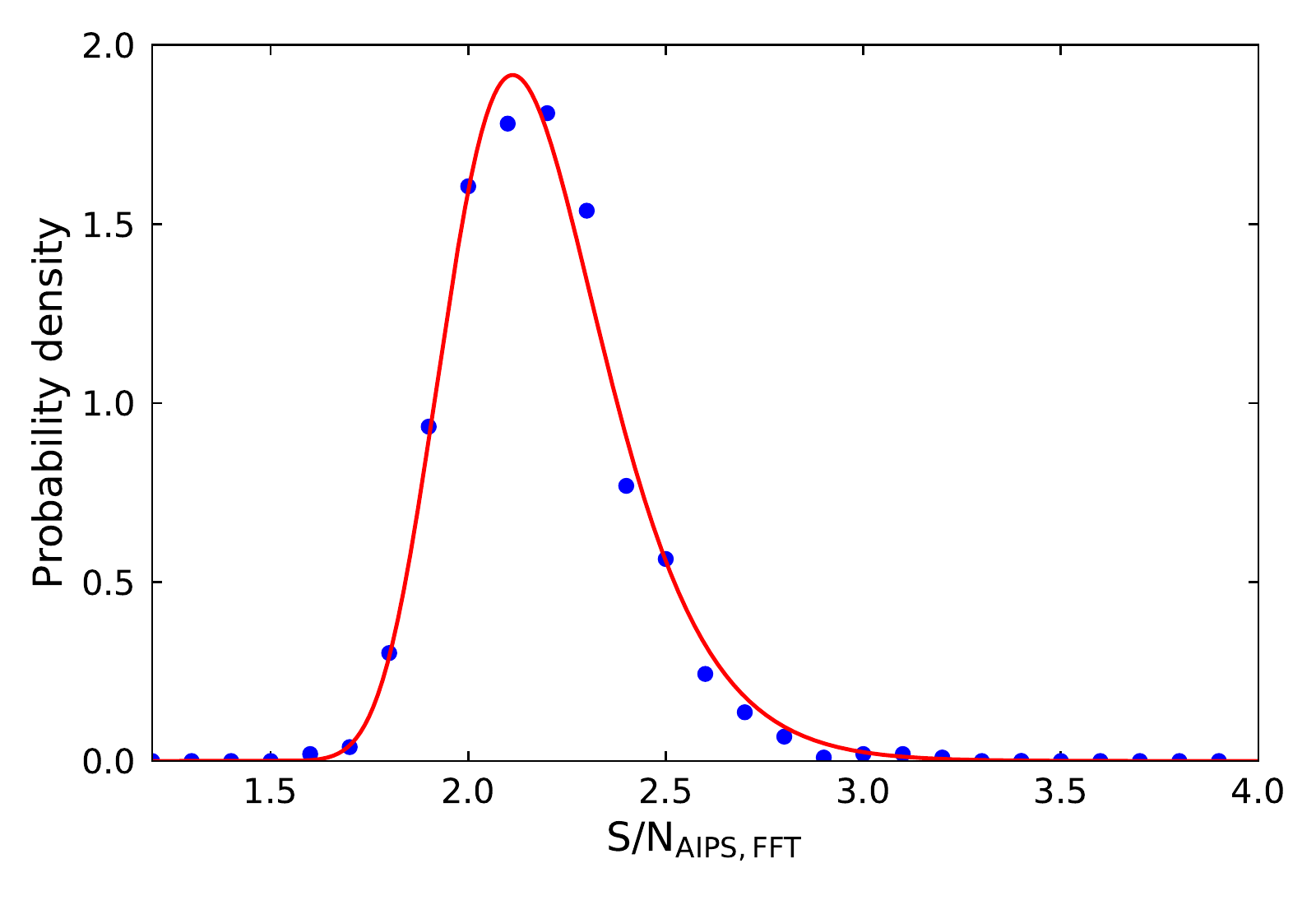}
\caption{Empirical distribution of the maximum $\mathrm{S/N}_\mathrm{AIPS,FFT}$ of the fringe amplitude in the FFT stage of the \textsc{AIPS} fringe-fitting task \textsc{fring} in the C band (\textit{top}) and K band (\textit{bottom}) when no signal was present in the chosen search region (blue points). The search window is constrained to be $\pm 100$\,ns in delay and $\pm 25$\,mHz (C band) or $\pm 50$\,mHz (K band) in rate. The red curve shows the fitted theoretical probability distribution. Note: in addition to a very large difference in the search space size between Figs.~\ref{false_snr_fig} and \ref{false_snr_fig_aips}, $\mathrm{S/N}_\mathrm{PIMA}$ and $\mathrm{S/N}_\mathrm{AIPS,FFT}$ are also defined differently.}
\label{false_snr_fig_aips}
\end{figure}

A similar technique is used to calculate the false detection probabilities for different S/N thresholds, $\mathrm{S/N}_\mathrm{AIPS,FFT}$, in the initial FFT fringe search stage of the \textsc{AIPS} task \textsc{fring} (now without the acceleration term). Figure~\ref{false_snr_fig_aips} shows the results. The location of the peak in Fig.~\ref{false_snr_fig_aips} differs from that in Fig.~\ref{false_snr_fig} for several reasons and the two should not be compared directly. First, the size of the search space is vastly larger in Fig.~\ref{false_snr_fig} than in Fig.~\ref{false_snr_fig_aips}. Secondly, S/N is defined differently in the \textsc{AIPS} task \textsc{fring} and in \textsc{PIMA} \citep{Petrov2011}. And thirdly, $\mathrm{S/N}_\mathrm{PIMA}$ is calculated for the fringe amplitude after the least-squares solution \citep{Petrov2011}. Furthermore, \textsc{fring} is known to underestimate the $\mathrm{S/N}_\mathrm{AIPS,FFT}$ of the FFT step at the low S/N limit \citep{Desai1998} and, therefore, we fit a following modified probability distribution to the data:
\begin{equation}
p(s) = \frac{n_\mathrm{eff}}{\sigma_\mathrm{eff}} f_\mathrm{S/N}^2 s e^{ - \frac{(f_\mathrm{S/N} s)^2}{2} } \Big( 1-e^{ - \frac{(f_\mathrm{S/N} s)^2}{2} }  \Big) ^{n_\mathrm{eff}-1} . 
\end{equation}
Here, $f_\mathrm{S/N}$ is the scaling factor of the $\mathrm{S/N}_\mathrm{AIPS,FFT}$, which is left as a free parameter. Table~\ref{false_snr_tab_aips} lists the false detection probabilities calculated from these fits.

\begin{table}
\caption{Probability of false detection in the \textsc{PIMA} fringe search}
\label{false_snr_tab}
\centering
\begin{tabular}{lll}
\hline\hline
$P_e$     & \multicolumn{2}{c}{$\mathrm{S/N}_\mathrm{PIMA}$}     \\
          & C-band\tablefootmark{a}  & K-band\tablefootmark{b} \\
\hline
0.1       & 5.07       & 4.84 \\  
0.01      & 5.52       & 5.30 \\
0.001     & 5.92       & 5.72 \\
0.0001    & 6.30       & 6.11 \\
$10^{-5}$  & 6.65       & 6.48 \\
$10^{-6}$  & 6.99       & 6.82 \\
\hline
\end{tabular}
\tablefoot{
\tablefoottext{a}{Search space: $\pm 2$\,$\mu$s in delay, $\pm 2 \times 10^{-11}$\,s/s in rate, and $\pm1 \times 10^{-13}$\,s/s$^2$ in acceleration.}
\tablefoottext{b}{Search space: $\pm 0.6$\,$\mu$s in delay, $\pm 5 \times 10^{-12}$\,s/s in rate, and $\pm1 \times 10^{-13}$\,s/s$^2$ in acceleration.}
}
\end{table}

\begin{table}
\caption{Probability of false detection in the \textsc{AIPS} narrow window fringe search}
\label{false_snr_tab_aips}
\centering
\begin{tabular}{lll}
\hline\hline
$\mathrm{S/N}_\mathrm{AIPS,FFT}$   & \multicolumn{2}{c}{$P_e$}     \\
          & C band\tablefootmark{a}  & K band\tablefootmark{b} \\
\hline
3.0       & 0.002   &  0.004  \\  
3.1       & 0.0008  &  0.002  \\
3.2       & 0.0004  &  0.0008 \\
3.3       & 0.0002  &  0.0004 \\
3.4       & $7\times10^{-5}$   &  0.0002 \\
3.5       & $3\times10^{-5}$   &  $7\times10^{-5}$ \\
3.6       & $2\times10^{-5}$   &  $3\times10^{-5}$ \\
3.7       & $5\times10^{-6}$   &  $1\times10^{-5}$ \\
3.8       & $2\times10^{-6}$   &  $5\times10^{-6}$ \\
3.9       & $8\times10^{-7}$   &  $2\times10^{-6}$ \\
\hline
\end{tabular}
\tablefoot{
\tablefoottext{a}{Search window: $\pm 100$\,ns in delay and $\pm 25$\,mHz in rate.}
\tablefoottext{b}{Search window: $\pm 100$\,ns in delay and $\pm 50$\,mHz in rate.}
}
\end{table}

\clearpage

\section{Coherence time on baselines to the Space Radio Telescope} \label{apxcoh}

Since atmospheric phase fluctuations do not affect the signal recorded by the Space Radio Telescope (SRT), it is expected that the coherence time on the space-to-ground baselines is longer than on the ground-to-ground baselines by a factor of $\lesssim$$\sqrt{2}$, if phase changes due to potential residual acceleration of the spacecraft have been corrected for. This is potentially important in the K band, where the space-VLBI data suffers from the poor sensitivity of the SRT (SEFD of 46700\,Jy). Furthermore, correcting for the phase drifts in the ground array data by self-calibration before searching for the fringes to the SRT is expected to allow integration times in the SRT fringe fitting that are longer than the atmospheric coherence time. In order to maximise the sensitivity in the global fringe search at the K band with AIPS, we estimated the maximum integration time in \textsc{fring} which still improves the S/N.

\begin{figure}
\centering
\includegraphics[angle=-90,width=\columnwidth]{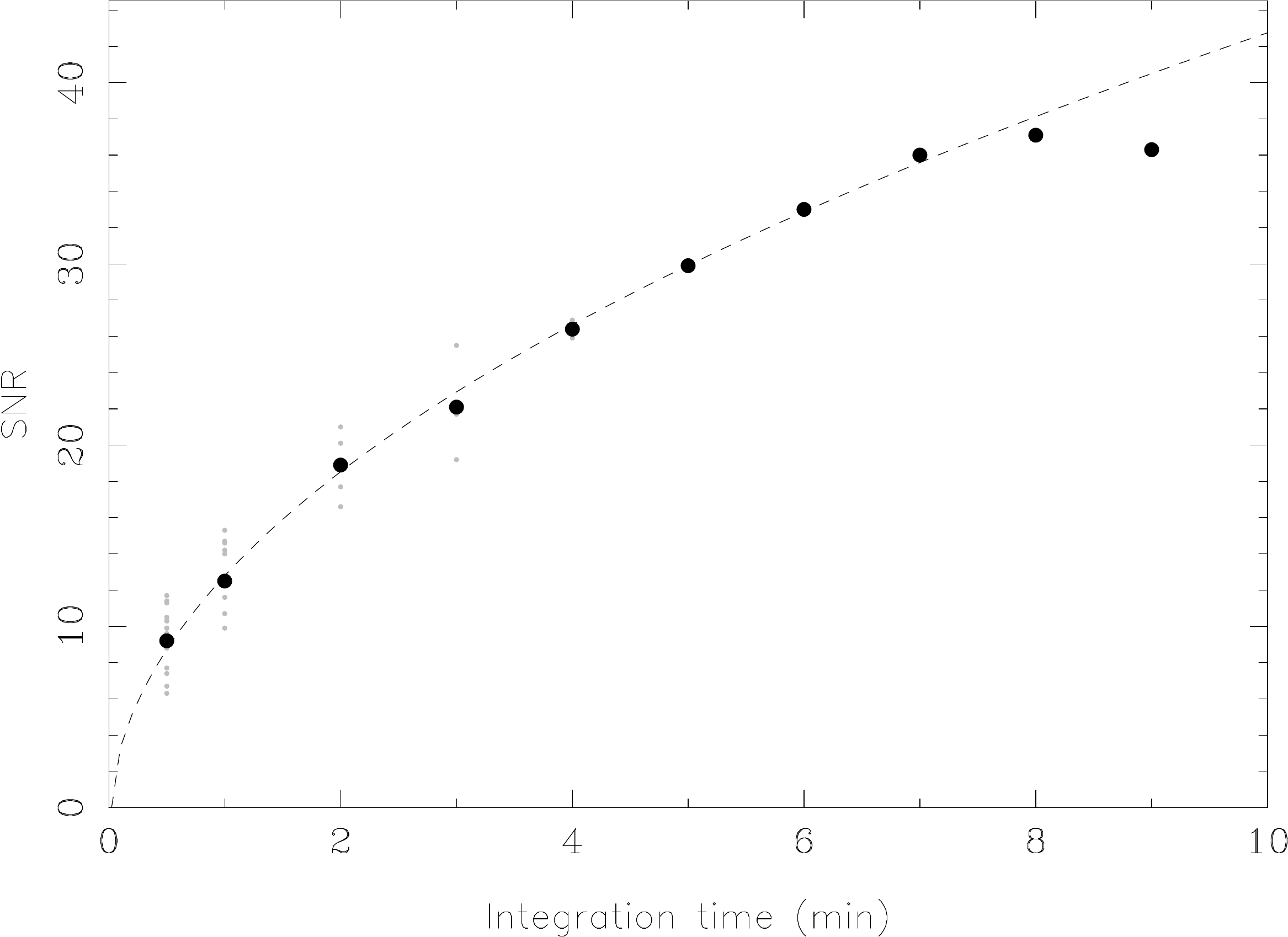}
\includegraphics[angle=-90,width=\columnwidth]{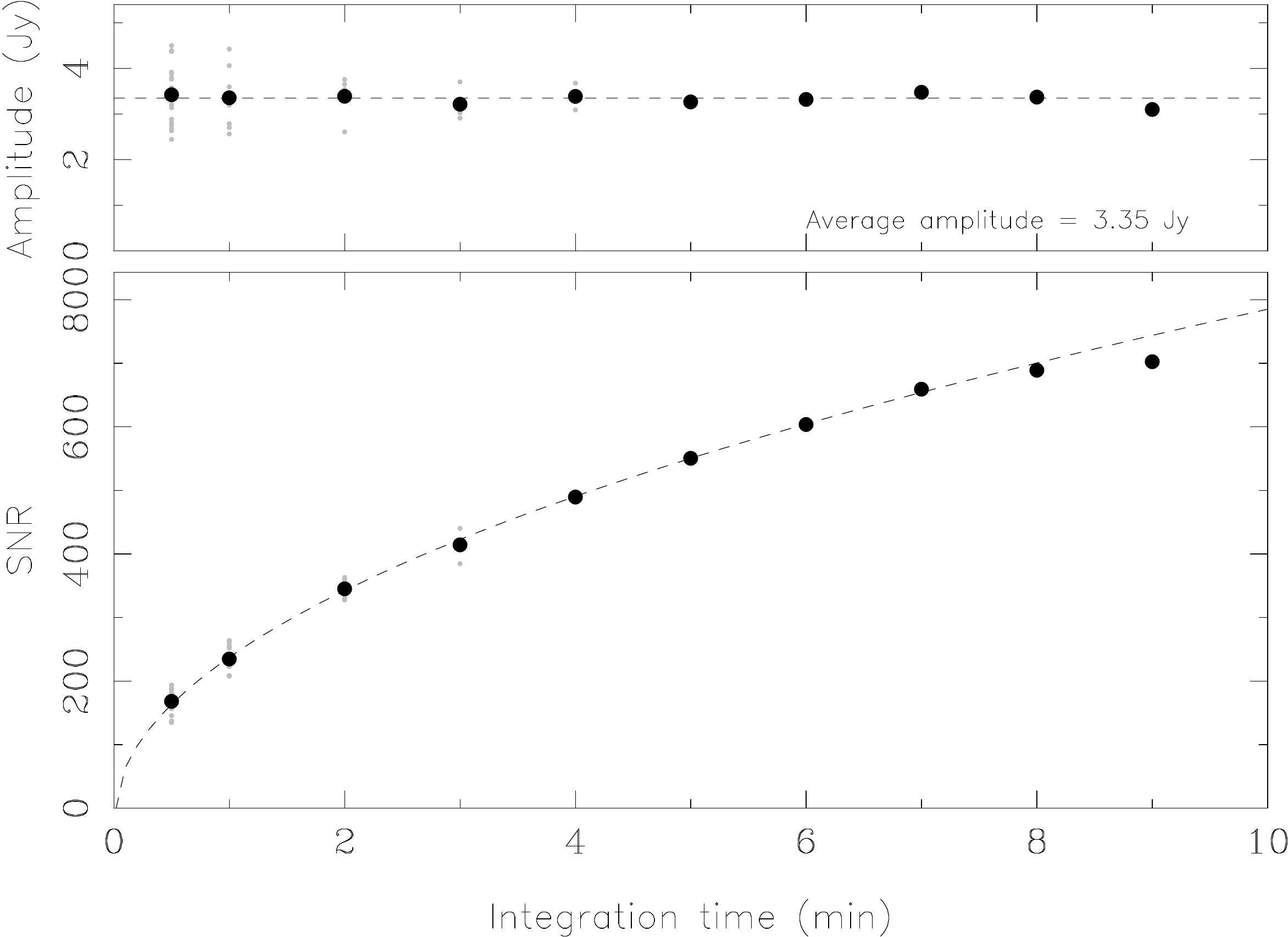}
\caption{Coherence plots for the Space Radio Telescope at 22\,GHz. The data shown are for the time interval 17:23:00$-$17:33:00\,UT. \textit{Top panel:} S/N in the FFT fringe search step of a global fringe fitting algorithm as a function of integration time. Grey dots show the S/N of individual solutions, while large black dots indicate the averages. The dashed line shows the expected $\sqrt{\Delta t}$ -dependence of the S/N. The data starts to deviate from the dashed line at $\sim8$\,min integration time, which marks the maximum averaging time that still improves the S/N. Longer averaging times do not improve the S/N because of increased coherence losses due to a drifting visibility phase. \textit{Middle panel:}  Average visibility amplitude (after applying the corresponding fringe fit solutions). \textit{Bottom panel:} S/N after the least-squares minimisation step of a global fringe fitting algorithm as a function of integration time. Symbols are the same as in the top panel.}
\label{coherence_fft}
\end{figure}

The maximum integration time was measured by fringe fitting the SRT data with different solution intervals ranging from 0.5 to 9 minutes. We selected a time interval of 17:23:00$-$17:33:00\,UT when the projected baselines to the SRT were short ($\sim0.2$\,$D_\oplus$) and, therefore, had a high-enough S/N to allow detections down to 0.5\,min integrations. Before the test, we corrected for the residual acceleration term, the instrumental phase offset between the IFs, and the single-band delays of the SRT. The ground array data were imaged and fully self-calibrated before the test, so that baseline stacking was possible. In the test we ran the \textsc{AIPS} task \textsc{fring} using the image from the ground array as an input model, stacked the baselines in the initial FFT step, and searched solutions only for the SRT. The results are shown in Fig.~\ref{coherence_fft}. The S/N keeps increasing as $\sqrt{\Delta t}$ up to $\sim 8$\,min. Coherence losses show up only at the integration time of 9\,min, for which the vector averaged visibility amplitude -- after applying the fringe fit solutions -- drops to 94\% of the vector averaged visibility amplitude determined from the $0.5-2$\,min solution interval results. This integration time is significantly longer than the expected 22\,GHz atmospheric coherence times on ground-ground baselines, which by conventional wisdom are thought to range from tens of seconds in bad weather conditions to a few minutes in good weather conditions. However, we would like to point out that this conventional wisdom appears to be rather conservative and in good weather conditions coherence times as long as about 8~min can be achieved even on the ground at 22~GHz. This was found by statistical analysis of the large amount of \textit{RadioAstron} AGN survey data (Kovalev al., in~prep.).

We used the KVN-Tamna station as the reference station in the test described above. In order to have an idea of the weather conditions at this station during the observation, we computed the wet path delay, opacity and brightness temperature at Tamna using the output of the NASA numerical weather model GEOS-FPIT \citep{Rienecker2018} via a direct integration of the equations of wave propagation in the heterogeneous media \citep{Petrov2015}. We obtained a wet path delay of 687\,ps, opacity of 0.24, and sky brightness temperature of 60\,K in the direction of observations at 22\,GHz. For comparisons, these quantities are approximately two times greater than at the low-altitude site of the Mets\"ahovi Radio Observatory in Finland for the same time of year. This illustrates that the weather conditions at the reference station on the ground were not particularly good, and yet it was still possible to use rather long integration times in the ground-to-space fringe search.

\clearpage

\section{Imaging parameter survey with the \textsc{eht-imaging} library} \label{appendix:eht}
\begin{figure*}[b]
\centering
\includegraphics[angle=-90,width=0.24\textwidth]{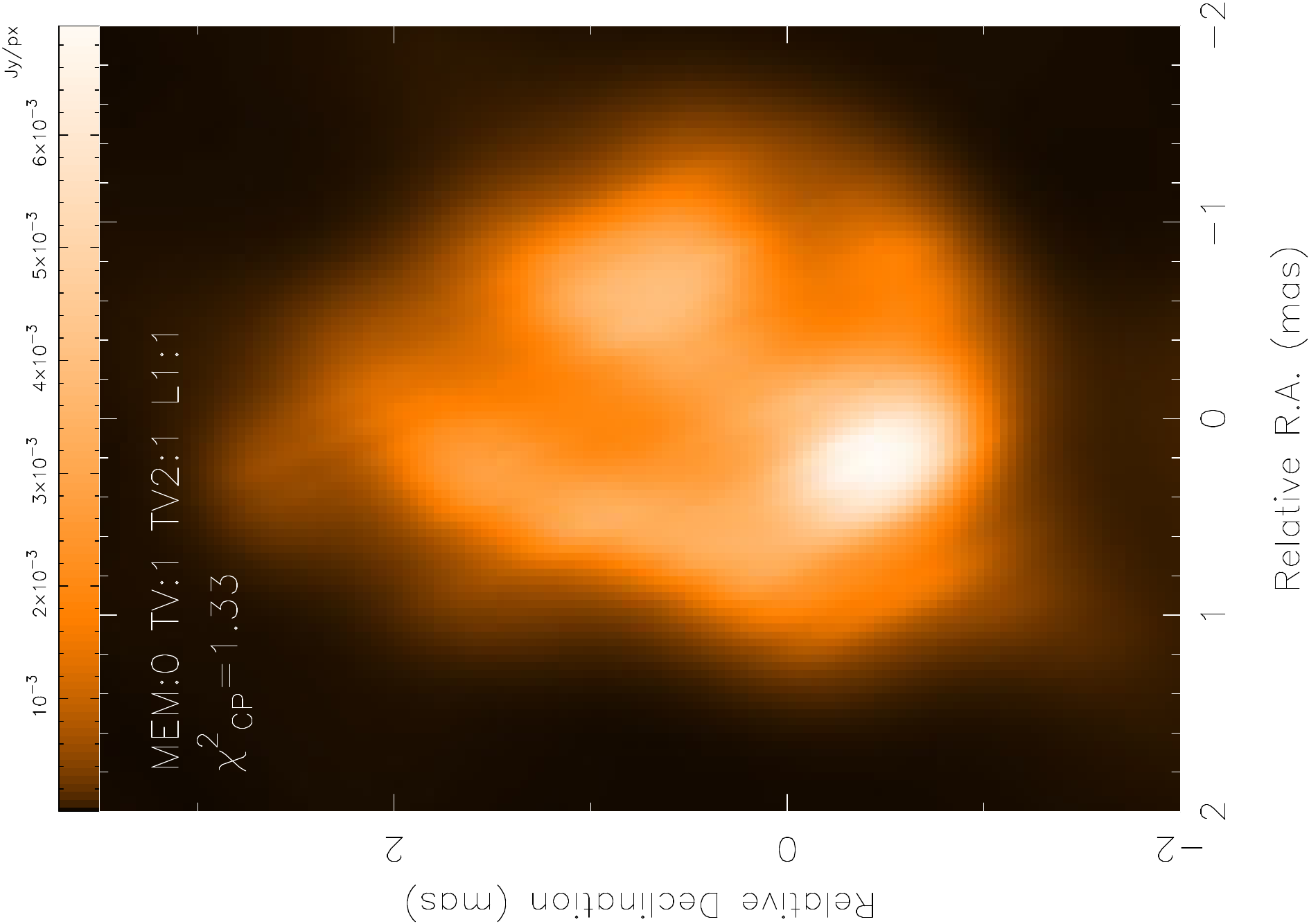}
\includegraphics[angle=-90,width=0.24\textwidth]{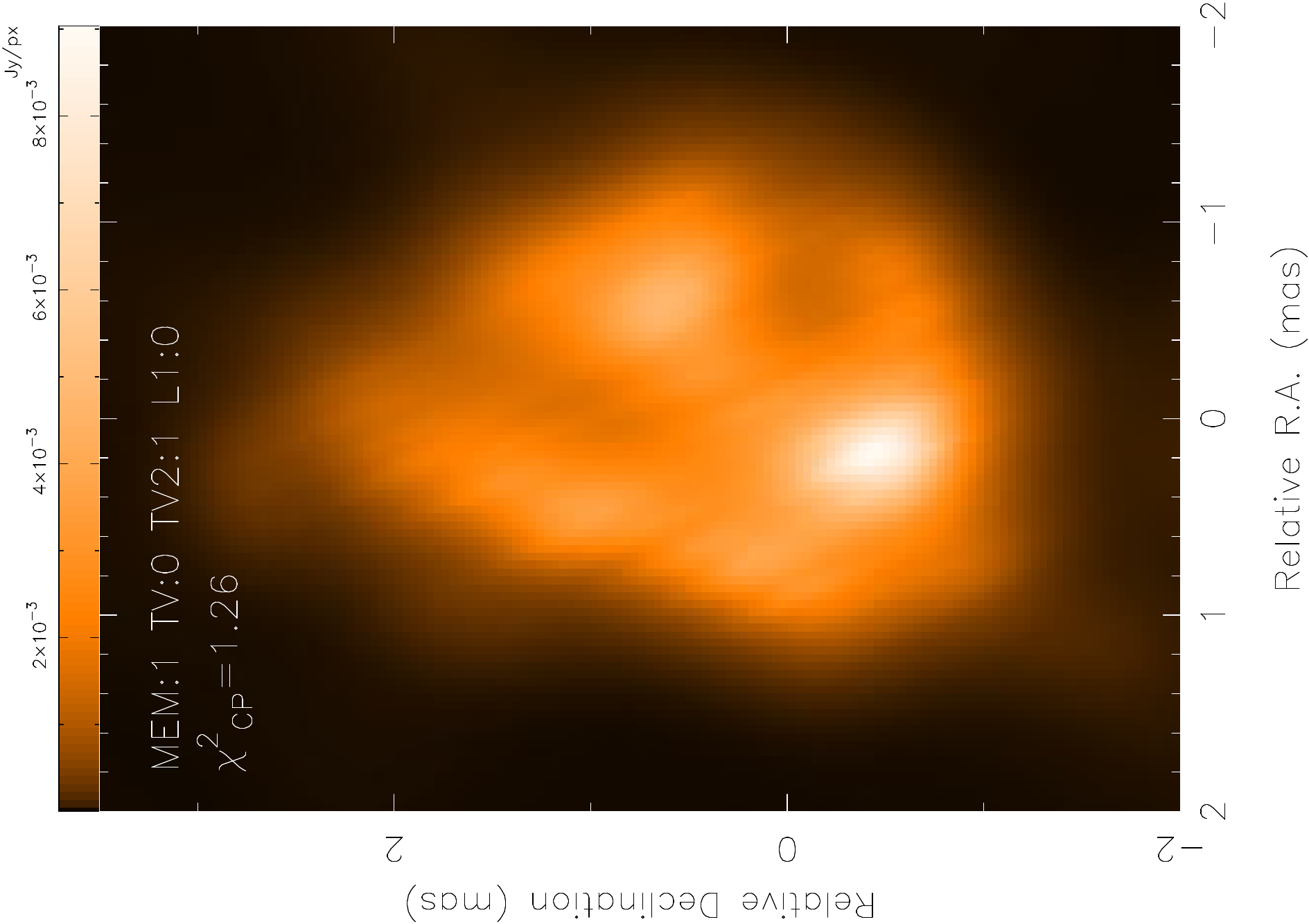}
\includegraphics[angle=-90,width=0.24\textwidth]{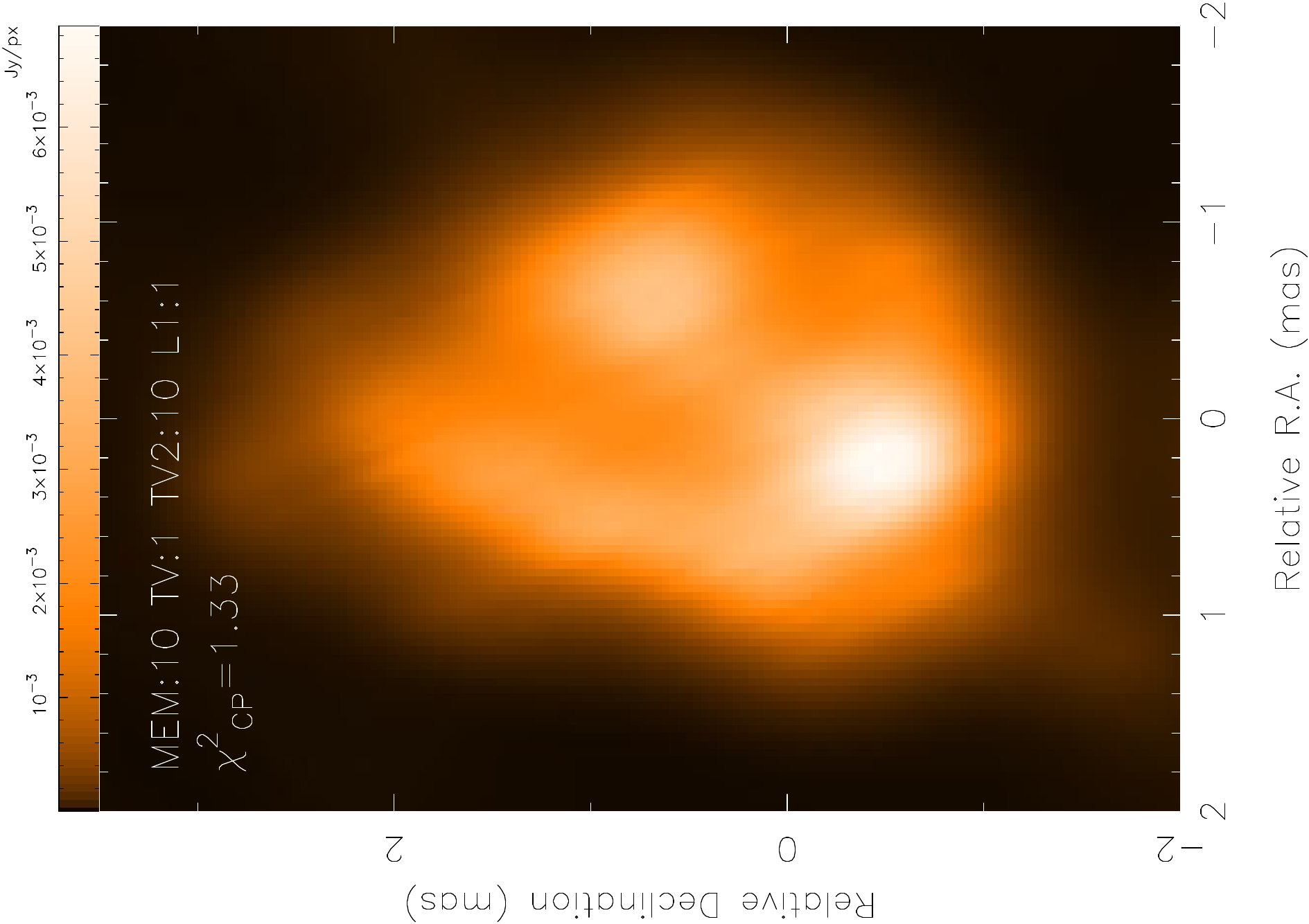}
\includegraphics[angle=-90,width=0.24\textwidth]{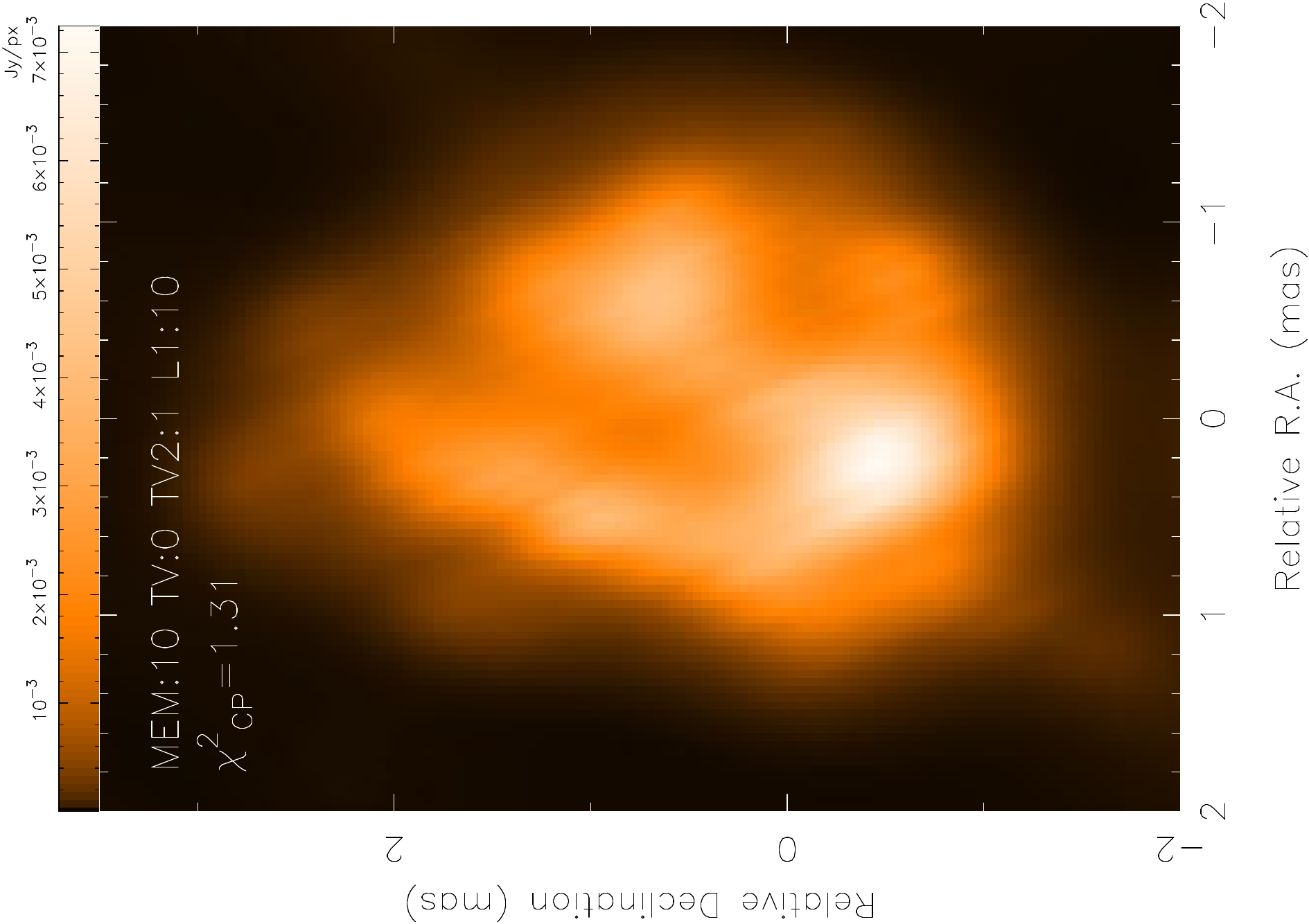} \\
\includegraphics[angle=-90,width=0.24\textwidth]{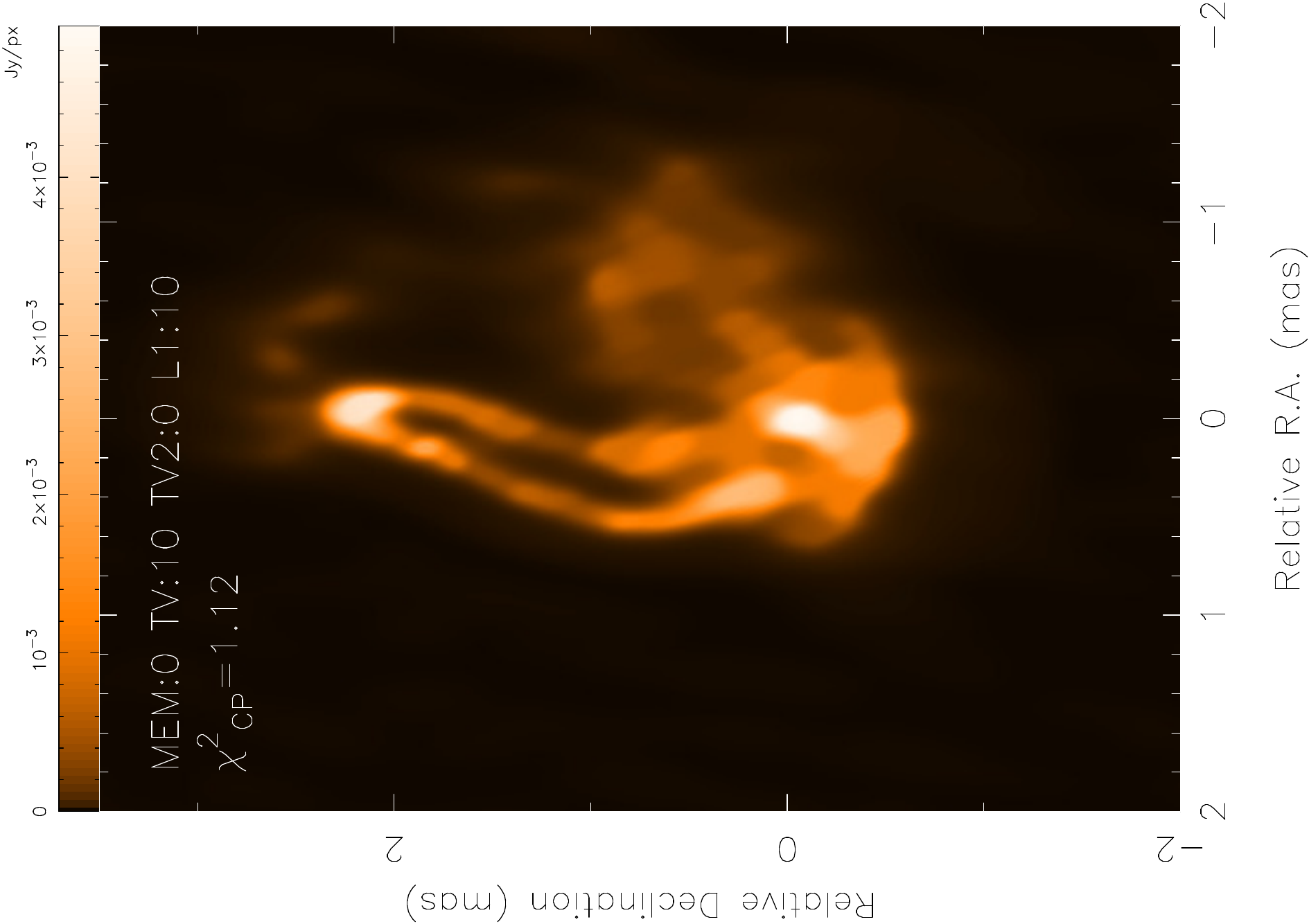}
\includegraphics[angle=-90,width=0.24\textwidth]{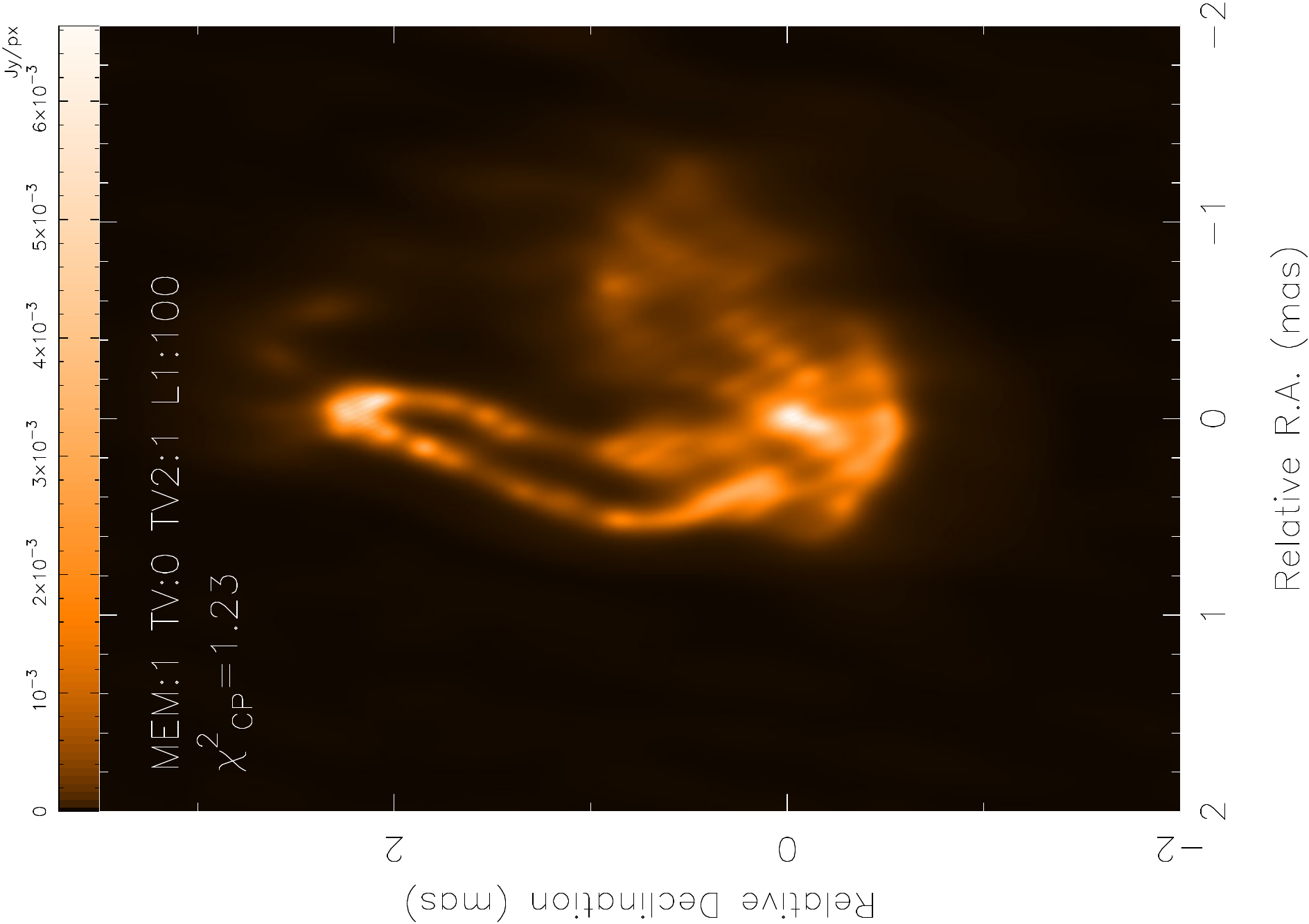}
\includegraphics[angle=-90,width=0.24\textwidth]{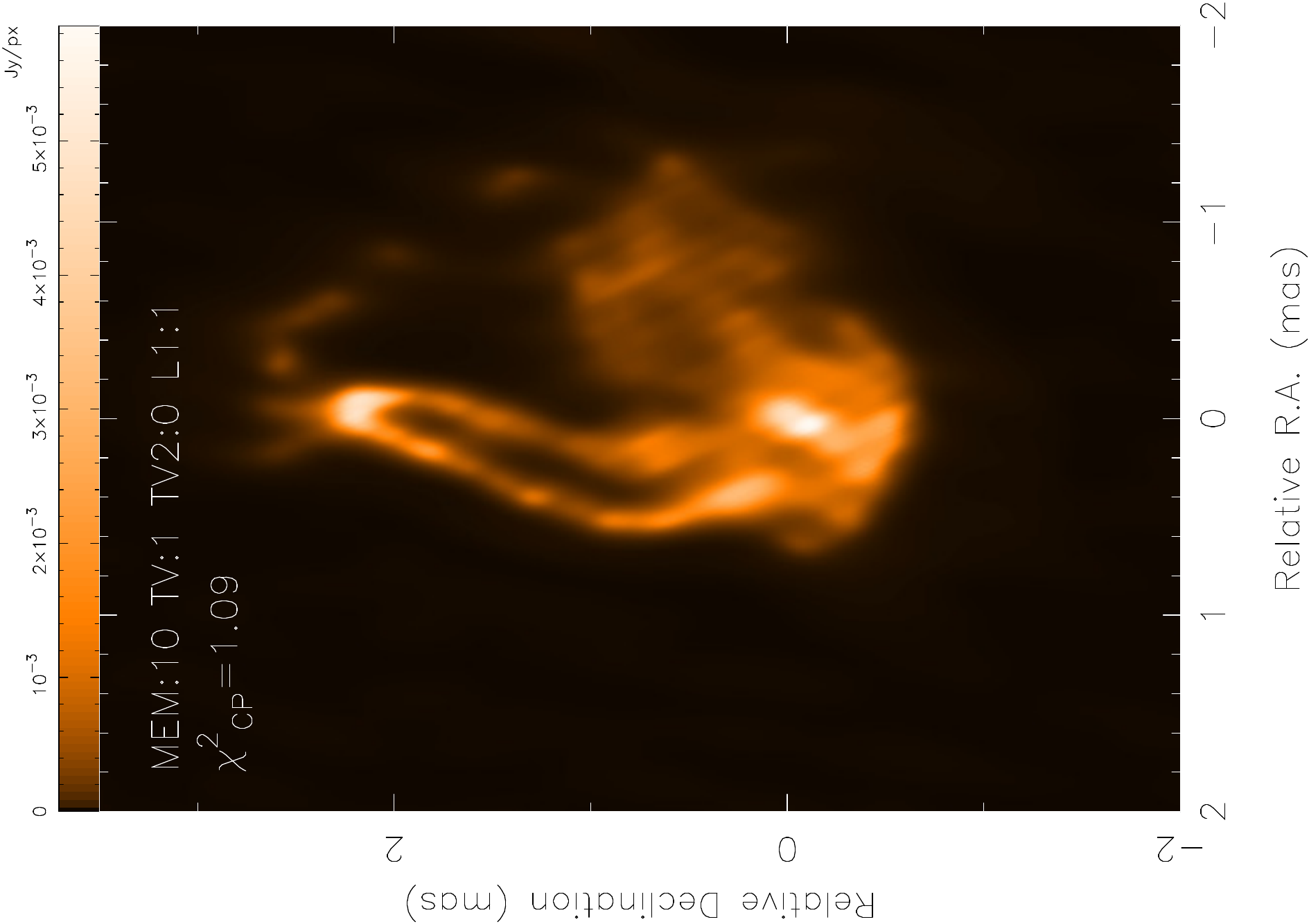}
\includegraphics[angle=-90,width=0.24\textwidth]{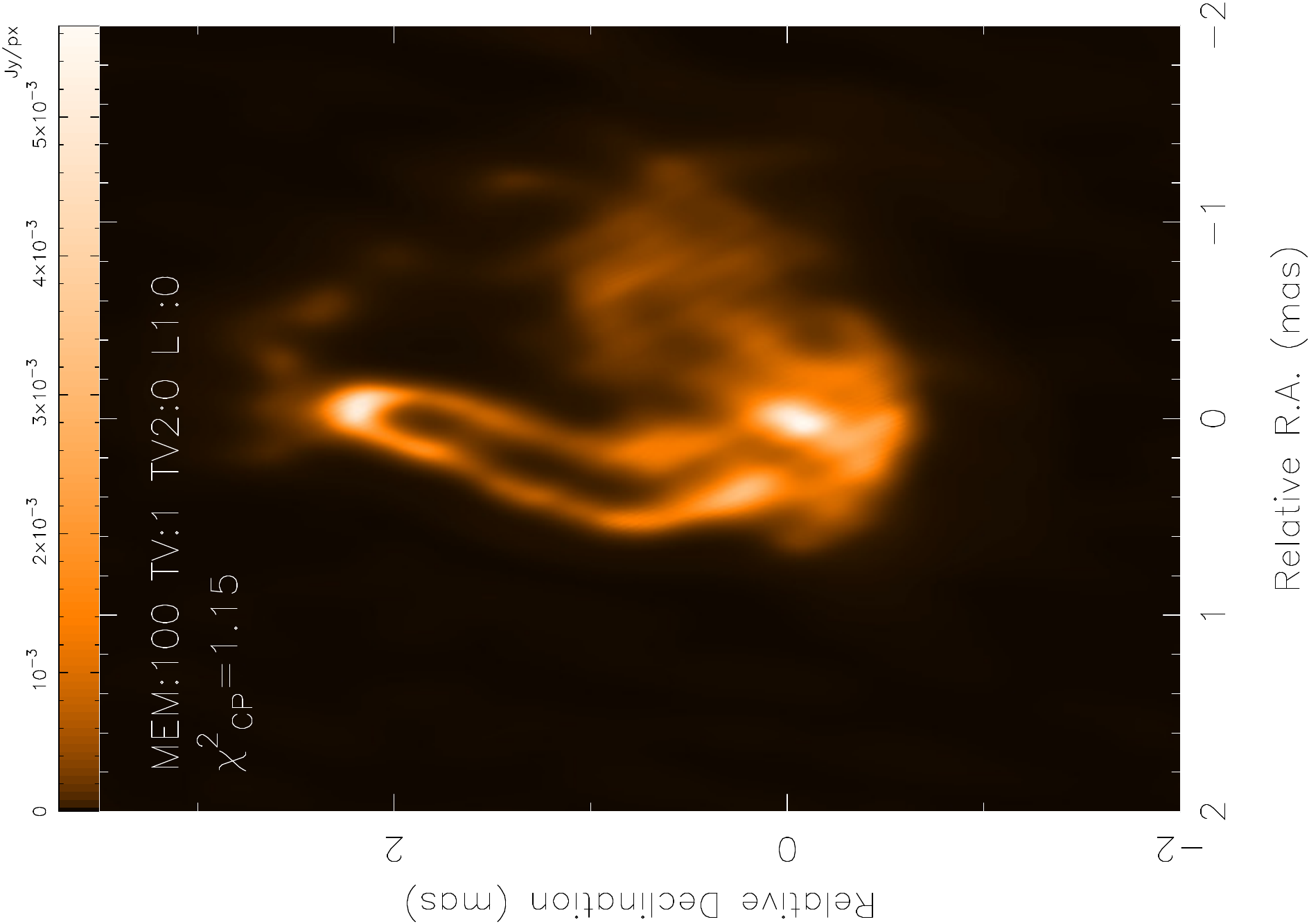}
\caption{Set of RML space-VLBI images of 3C\,84 from the parameter survey carried out with \textsc{ehtim}. \emph{Top row:} Four examples of 5\,GHz images. \emph{Bottom row:} Four examples of 22\,GHz images. The hyperparameter values are indicated in the images together with the $\chi^2$ value of the fit to closure phases.}
\label{RA_ehtim}
\end{figure*}

As described in Sect.~\ref{sec:imaging}, we imaged the space-VLBI data also with a regularised maximum likelihood method (RML) implemented in the \textsc{eht-imaging} library \citep{Chael2016, Chael2018}. Since the output of the RML methods depends on the choice of the regularisers and their weights, we have explored the effect of the hyperparameters $\beta_R$ by carrying out a small imaging parameter survey. We varied $\beta_R$ between 0 and 100 for four regulariser terms: relative entropy (MEM), total variation (TV), total squared variation (TV2), and the $l_1$ norm. Furthermore, the expected total flux density in the image was used as a regulariser with $\beta_\mathrm{flux} = 1000$ in all the imaging trials. The best-fitting images from the parameter survey have mostly small differences in their reduced $\chi^2$ of the fit to closure phases and from that point of view majority of them could be considered valid image reconstructions (only setting TV regulariser weight to $\beta_\mathrm{TV} = 100$ consistently gives $\chi^2_\mathrm{CP} > 2.0$). However, several of them show obvious signs of imaging artefacts in the form of noticeable striping when inspected by eye. After discarding the images most strongly affected by such artefacts, we chose four representative reconstructions at both 5\,GHz and 22\,GHz that are shown in Fig.~\ref{RA_ehtim}. These reconstructions confirm the fidelity of the main emission structures visible in the CLEAN images. \balance

\clearpage

\section{Alternative CLEAN images} \label{appendix:clean}

Here, we present a set of \emph{RadioAstron} CLEAN images at 5\,GHz and 22\,GHz made with different visibility weighting schemes (super-uniform, uniform, and natural). Furthermore, we show the super-uniformly weighted images with several different restoring beams (for details, see image captions). 

\begin{figure*}[b]
\centering
\includegraphics[angle=-90,width=0.32\textwidth]{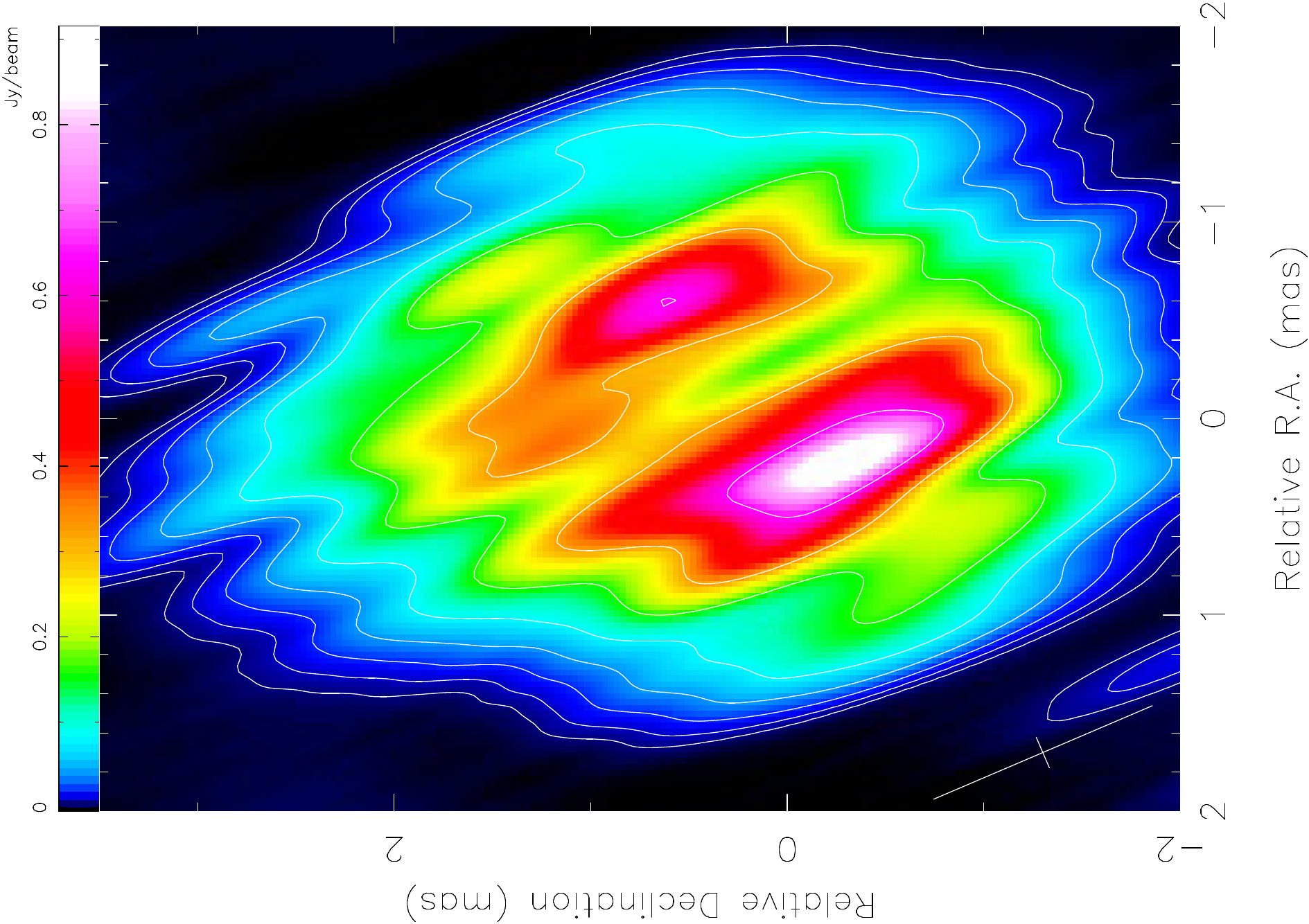}
\includegraphics[angle=-90,width=0.32\textwidth]{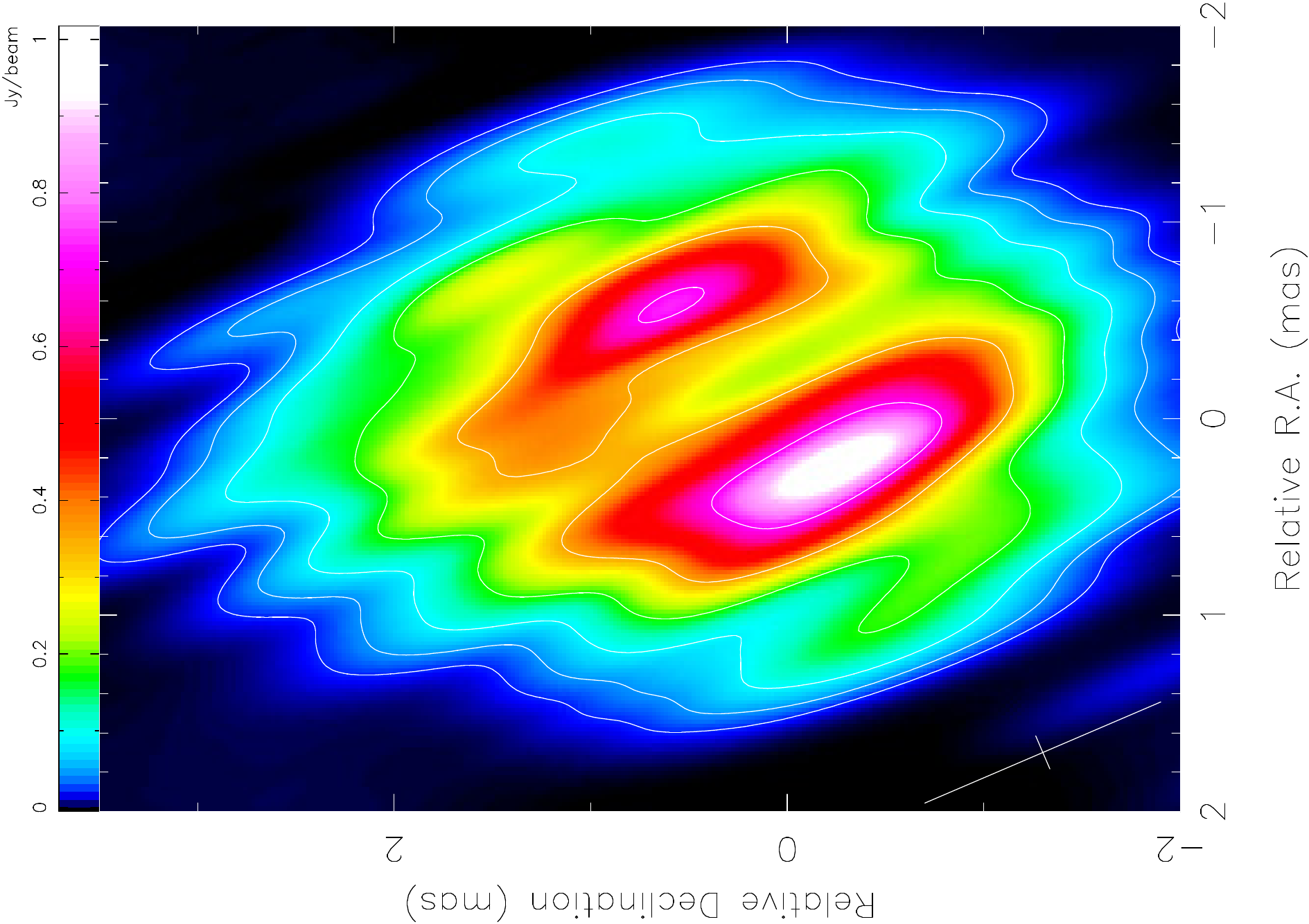}
\includegraphics[angle=-90,width=0.32\textwidth]{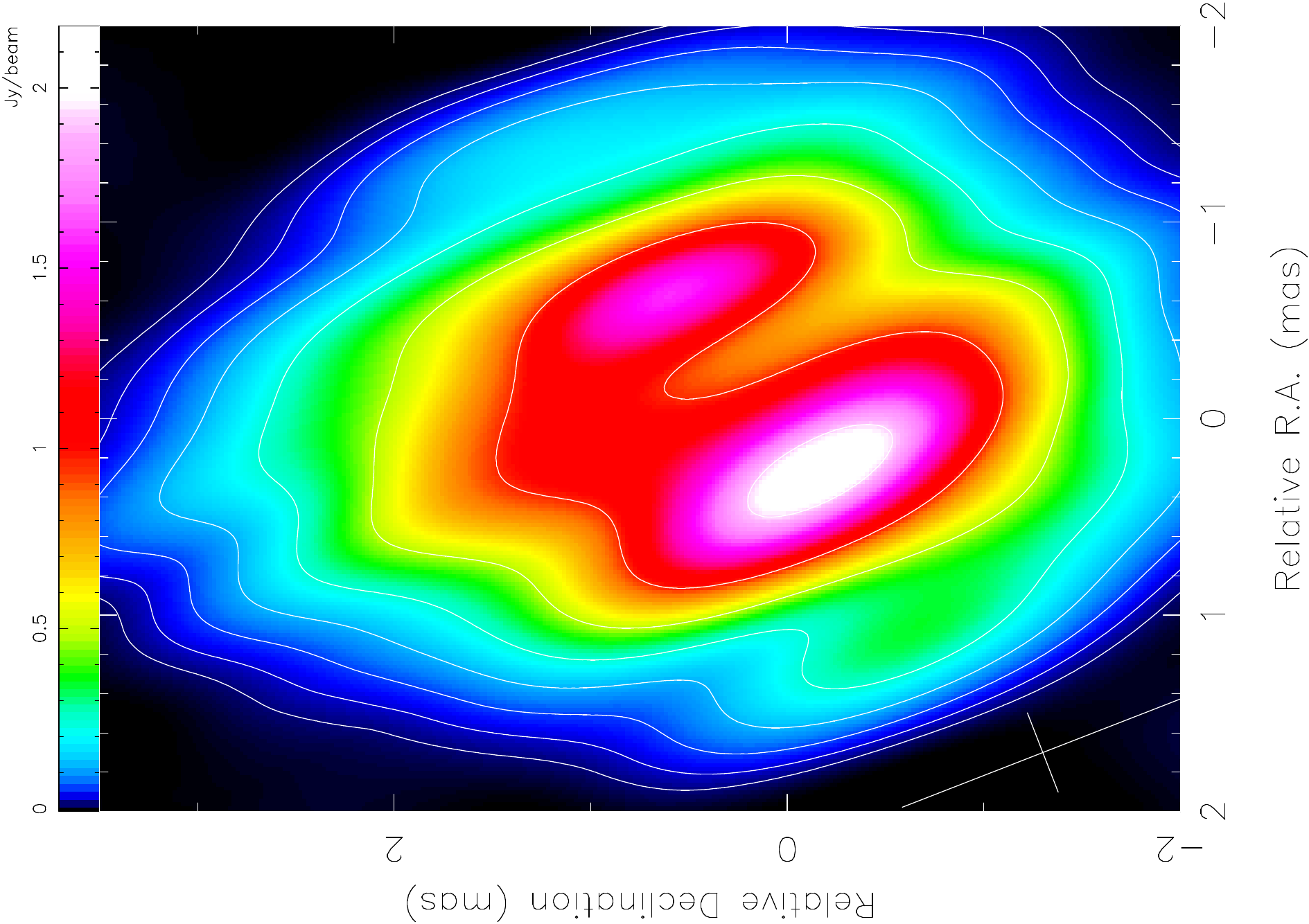}\\
\includegraphics[angle=-90,width=0.32\textwidth]{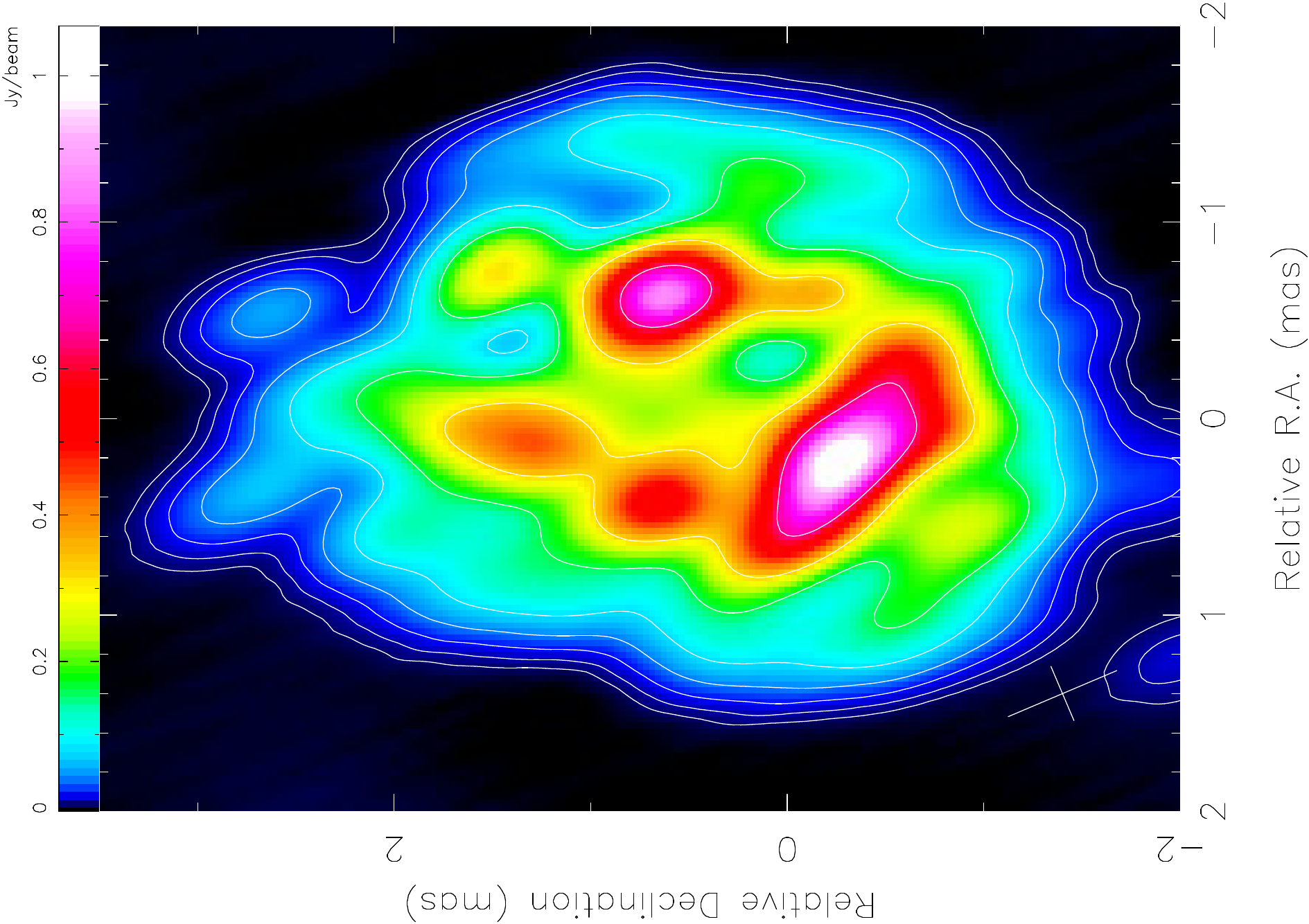}
\includegraphics[angle=-90,width=0.32\textwidth]{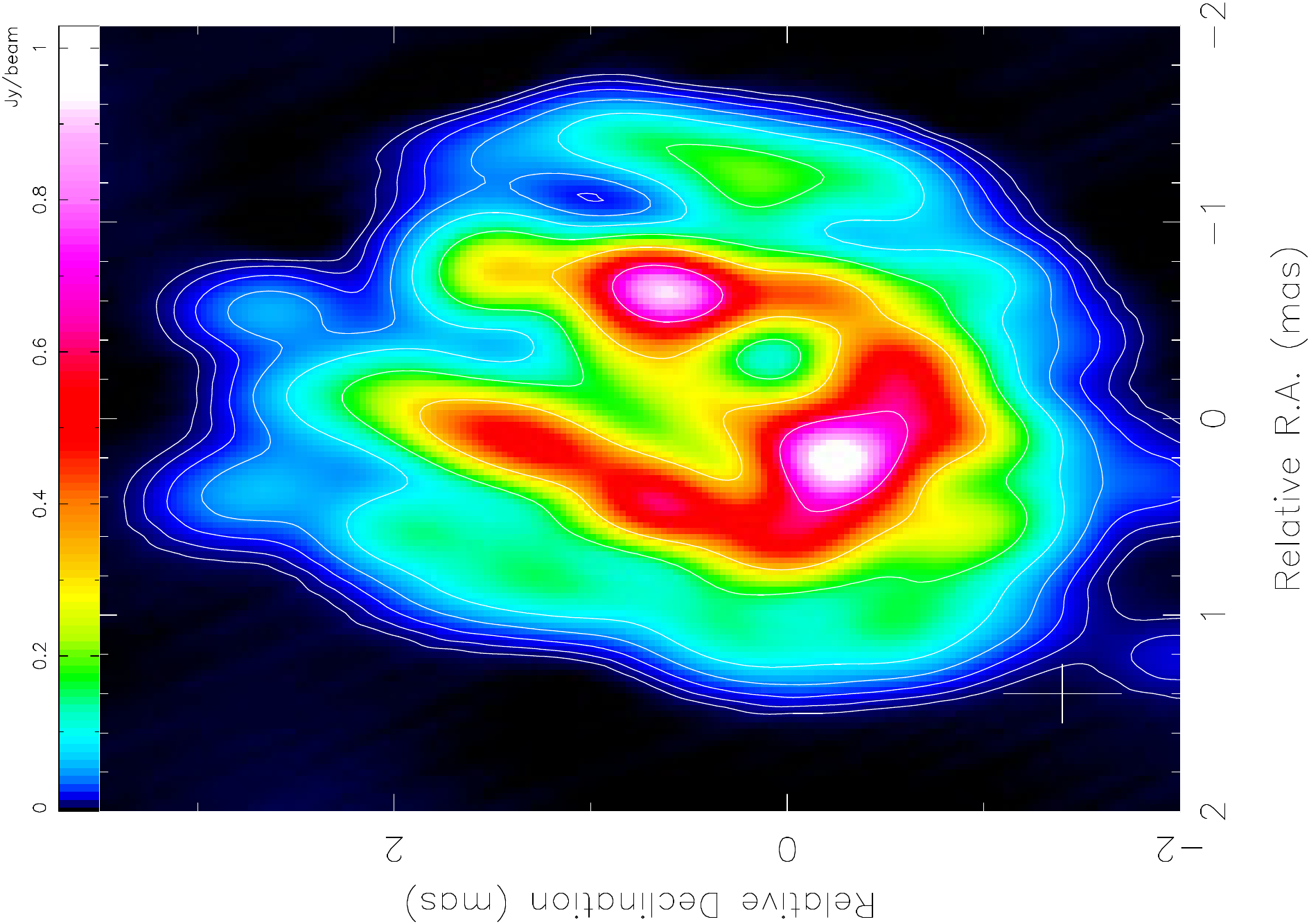} 
\includegraphics[angle=-90,width=0.32\textwidth]{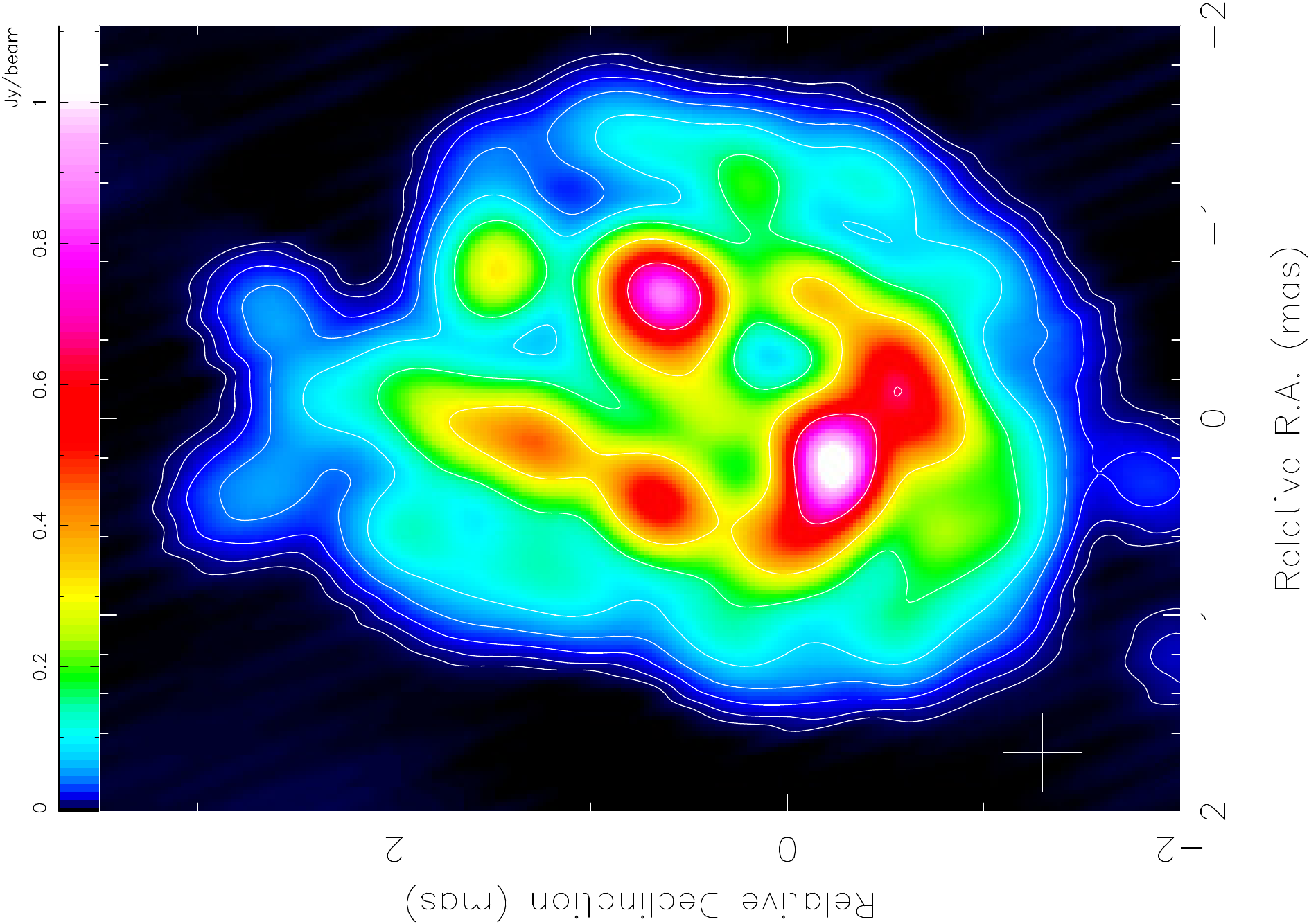}
\caption{Set of 5\,GHz CLEAN space-VLBI images of 3C\,84. \emph{Top row:} Images with different weighting schemes. From left to right: super-uniform (\textsc{uvweight} 5,$-$1 in \textsc{Difmap}; see Sect.~\ref{sec:imaging}), uniform (\textsc{uvweight} 2,$-$1), and natural (\textsc{uvweight} 0,$-$1). The restoring beam sizes are $1.22 \times 0.17$\,mas at PA=23$^\circ$ (\emph{left}), $1.31 \times 0.18$\,mas at PA=23$^\circ$ (\emph{middle}), and $1.53 \times 0.43$\,mas at PA=21$^\circ$ (\emph{right}). Peak flux densities from left to right are 916\,mJy/beam, 1018\,mJy/beam, and 2172\,mJy/beam. \emph{Bottom row:} Super-uniformly weighted images with different restoring beams. From left to right: $0.6 \times 0.3$\,mas at PA=23$^\circ$, $0.6 \times 0.3$\,mas at PA=0$^\circ$, and $0.40 \times 0.40$\,mas. Peak flux densities from left to right are 1060\,mJy/beam, 1029\,mJy/beam, and 1109\,mJy/beam. In all images contours start from 1\% of the peak flux density and increase in steps by factors of 2. Note: the colour scale is different in each panel.}
\label{RA_C_multi_beam}
\end{figure*}

\begin{figure*}[t]
\centering
\includegraphics[angle=-90,width=0.32\textwidth]{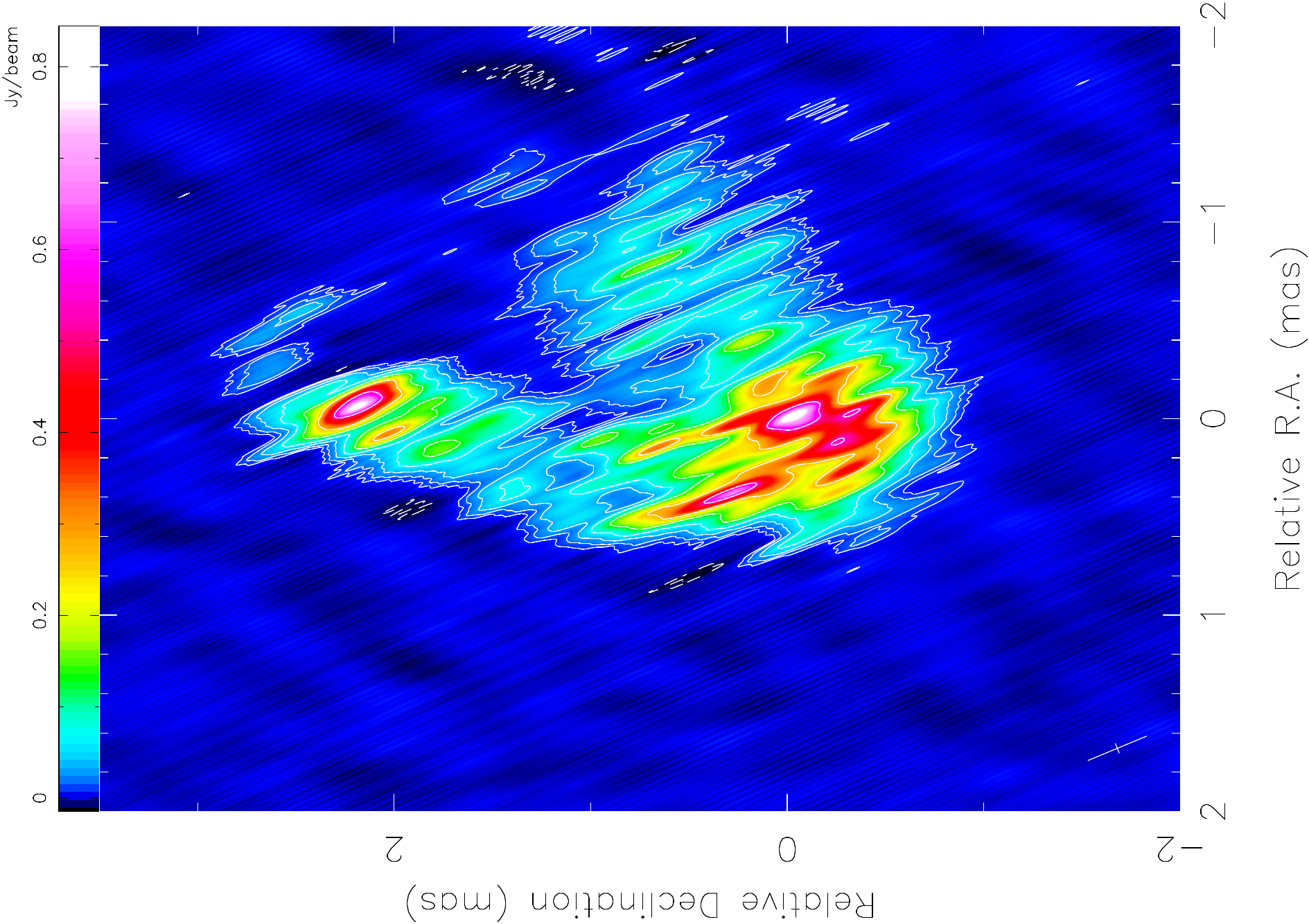}
\includegraphics[angle=-90,width=0.32\textwidth]{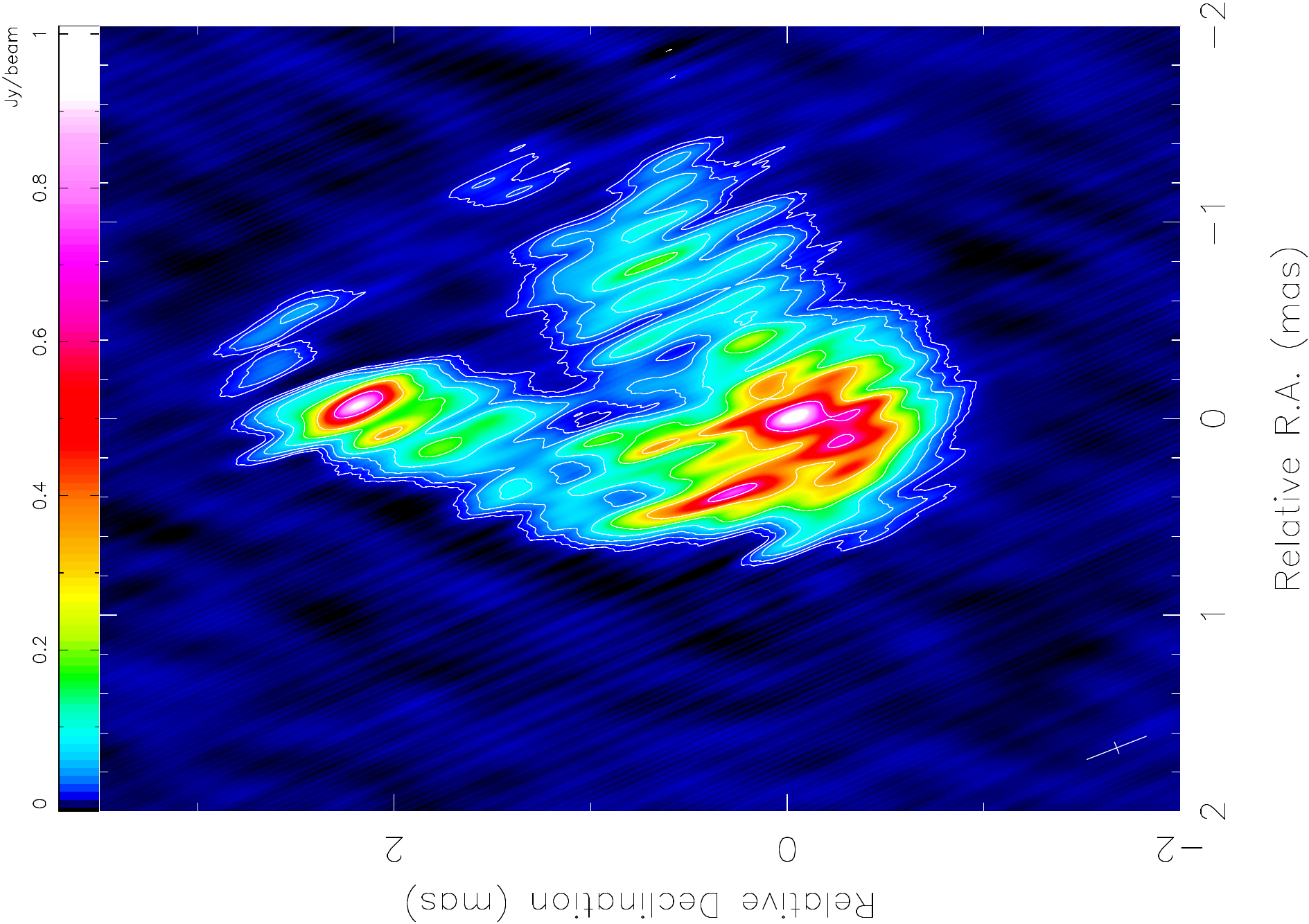}
\includegraphics[angle=-90,width=0.32\textwidth]{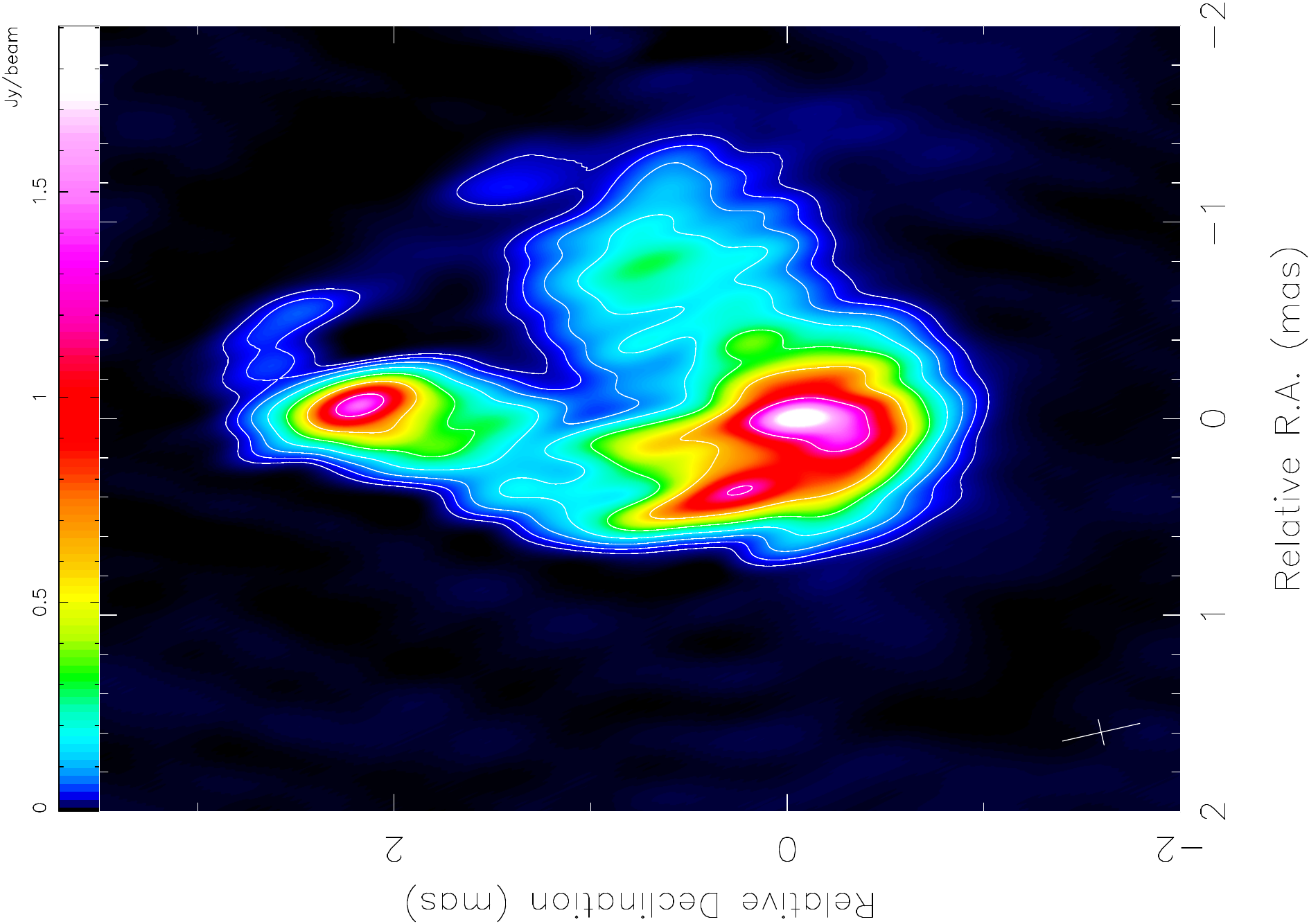} \\
\includegraphics[angle=-90,width=0.32\textwidth]{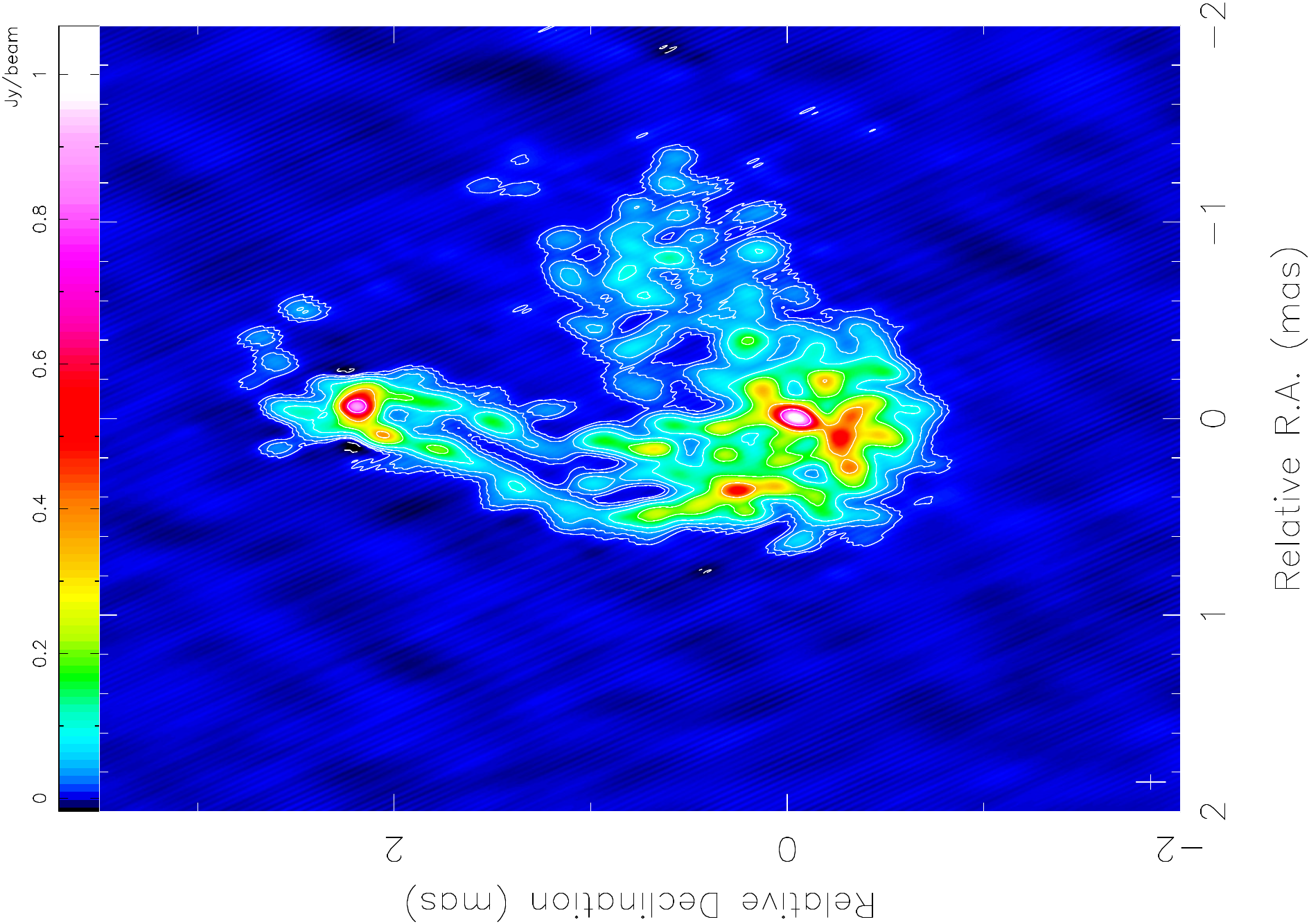} 
\includegraphics[angle=-90,width=0.32\textwidth]{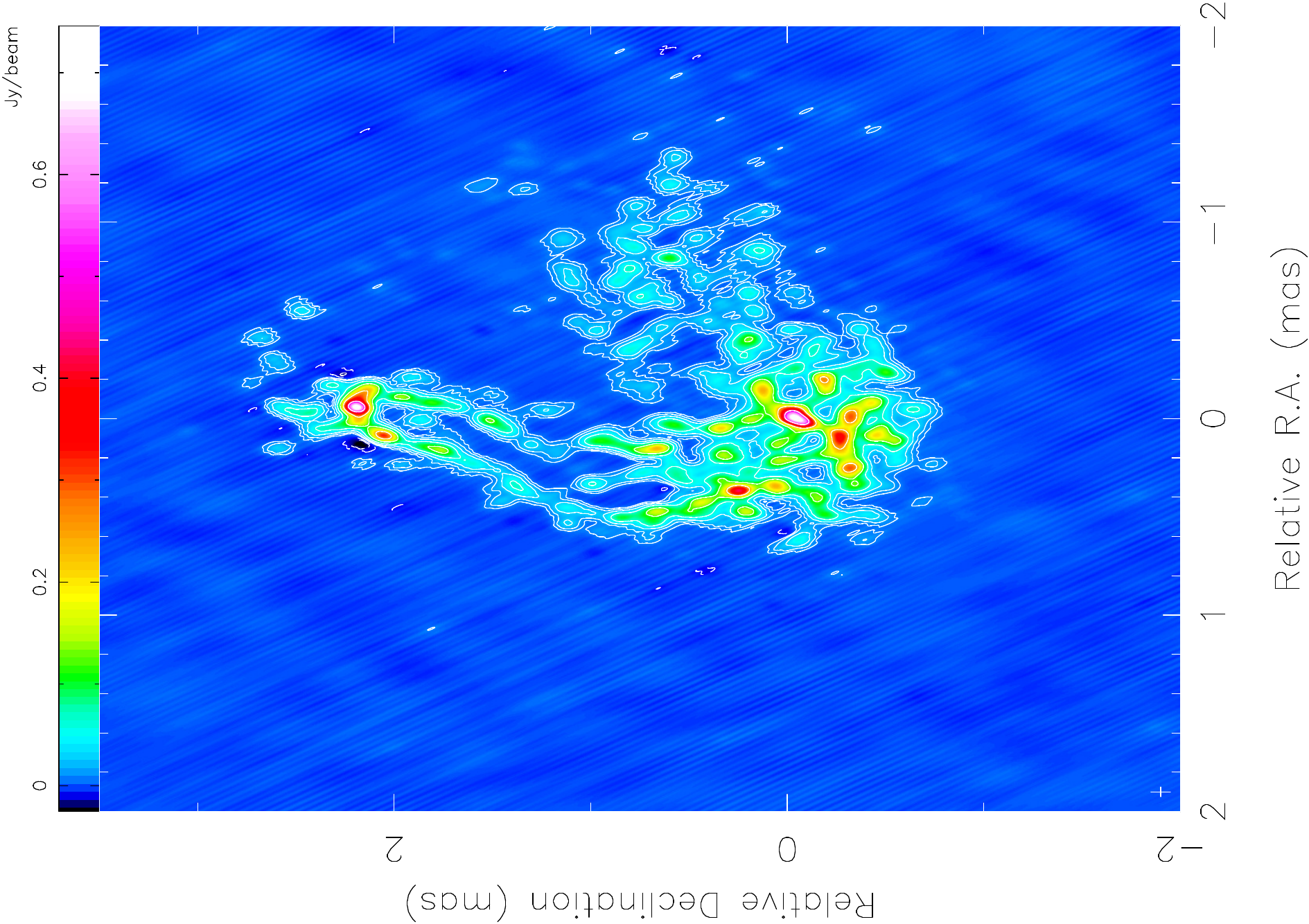} 
\includegraphics[angle=-90,width=0.32\textwidth]{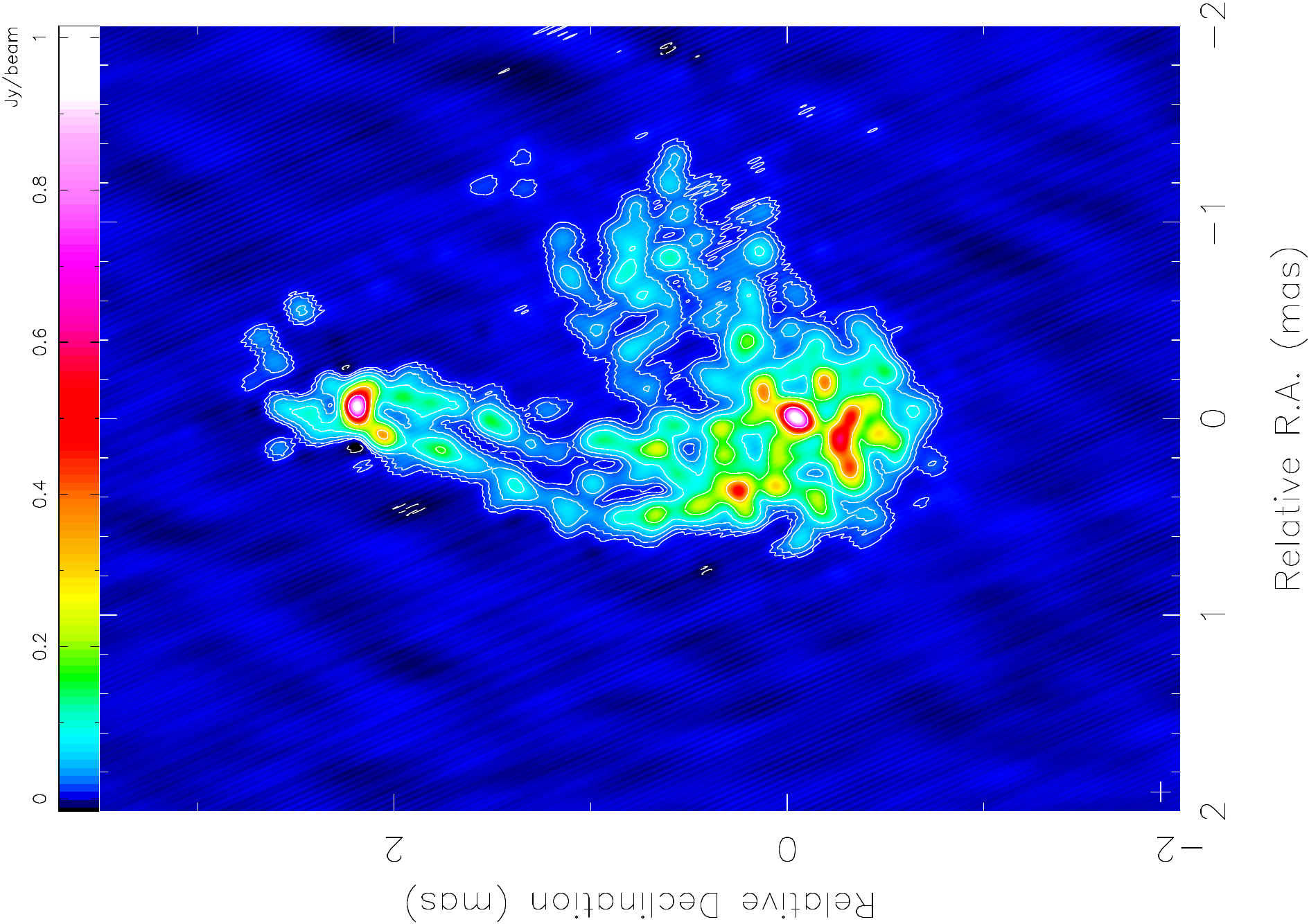}
\caption{Set of 22\,GHz CLEAN space-VLBI images of 3C\,84. \emph{Top row:} Images with different weighting schemes. From left to right: super-uniform (\textsc{uvweight} 5,$-$1 in \textsc{Difmap}; see Sect.~\ref{sec:imaging}), uniform (\textsc{uvweight} 2,$-$1), and natural (\textsc{uvweight} 0,$-$1). The restoring beam sizes are $0.32 \times 0.05$\,mas at PA=22$^\circ$ (\emph{left}), $0.33 \times 0.07$\,mas at PA=21$^\circ$ (\emph{middle}), and $0.40 \times 0.13$\,mas at PA=13$^\circ$ (\emph{right}). Peak flux densities from left to right are 845\,mJy/beam, 1011\,mJy/beam, and 1905\,mJy/beam.
\emph{Bottom row:} Super-uniformly weighted images with different restoring beams. From left to right: $0.15 \times 0.075$\,mas at PA=0$^\circ$, 0.10\,$\times$\,0.05\,mas at PA=0$^\circ$, and $0.10 \times 0.10$\,mas. Peak flux densities from left to right are 1067\,mJy/beam, 746\,mJy/beam, and 1016\,mJy/beam. In all images contours start from 1\% of the peak flux density and increase in steps by factors of 2. Note: the colour scale is different in each panel.}
\label{RA_K_multi_beam}
\end{figure*}

\begin{figure*}[t]
\centering
\includegraphics[angle=-90,width=0.6\textwidth]{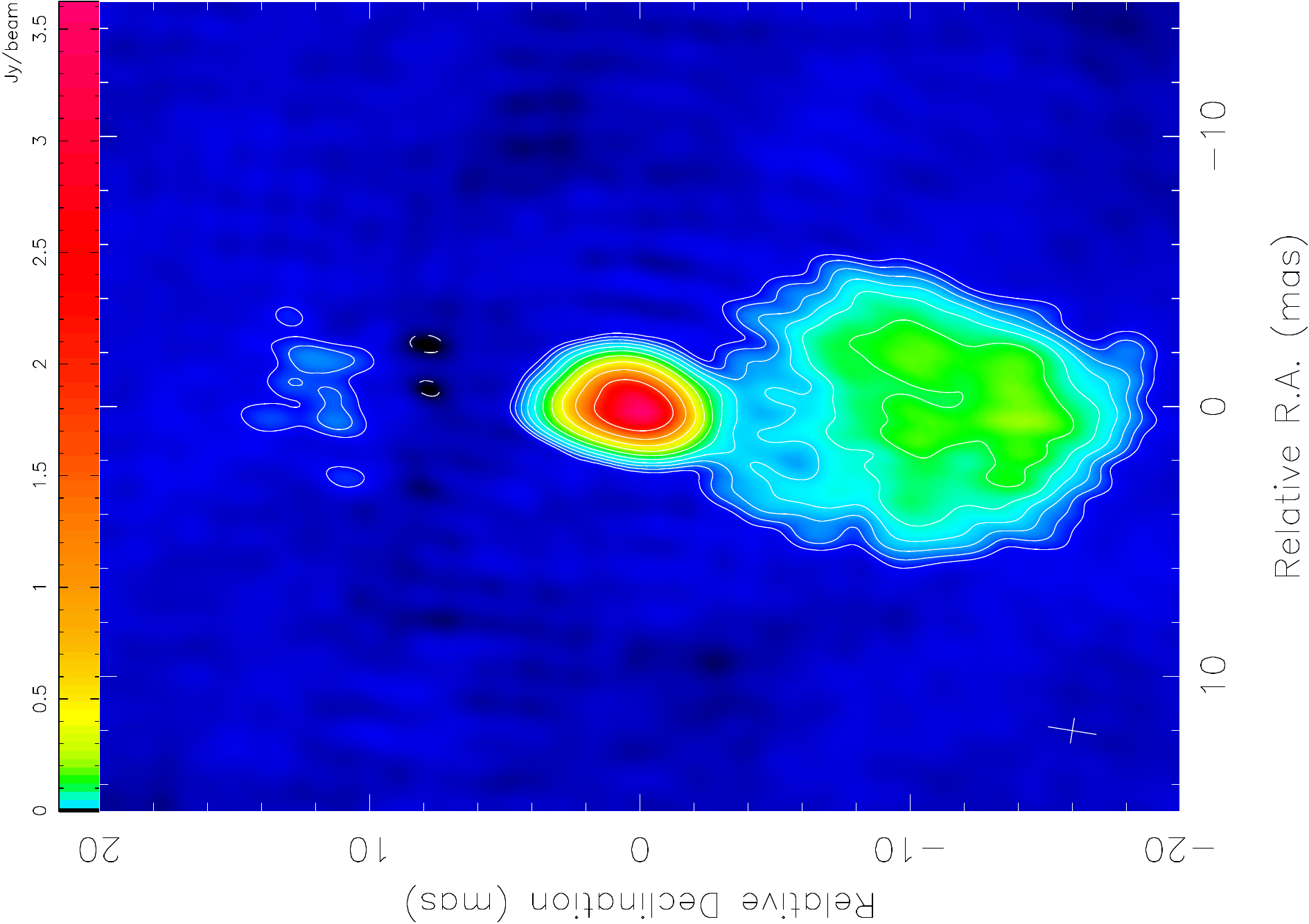}
\caption{Ground-only 5\,GHz CLEAN image with intensity contours overlaid. The image parameters are the same as in the left panel of Fig.~\ref{RAcombined} except that here a logarithmic colour scale is used. Contours start from 0.1\% of the peak flux density and increase in steps by factors of 2.}
\label{GR_C_log}
\end{figure*}

\end{appendix}

\end{document}